\documentclass[useAMS,usenatbib]{mn2e}
\usepackage{epsf}
\usepackage{graphicx}
\usepackage{psfrag}
\usepackage[usenames]{color} 

\newcommand{\ifm}[1]{\relax\ifmmode#1\else$\mathsurround=0pt #1$\fi}

\newcommand{\pc}{\,{\rm pc}}
\newcommand{\msun}{M_\odot}
\newcommand{\Fig}[1]{Figure~\ref{fig:#1}}
\newcommand{\kpc}{\,{\rm kpc}}

\def\CD{CDB}
\def\Mv{M_{\rm v}}
\def\Rv{R_{\rm v}}
\def\cmms{\,{\rm cm}^{-2}}
\def\cmmc{\,{\rm cm}^{-3}}
\def\lya{${\rm Ly}\alpha$\ }
\def\la{${\rm Ly}\alpha$}

\def\EW{EW}
\def\s10{S10}
\def\f11{F11}
\def\HI{H\,\textsc{i}}
\def\CII{C\,\textsc{ii}}
\def\CIV{C\,\textsc{iv}}
\def\OI{O\,\textsc{i}}
\def\SiII{Si\,\textsc{ii}}
\def\SiIV{Si\,\textsc{iv}}
\def\MgII{Mg\,\textsc{ii}}
\def\FeII{Fe\,\textsc{ii}}
\def\OII{O\,\textsc{ii}}
\def\OIII{O\,\textsc{iii}}

\def\gtrsim{\lower.5ex\hbox{$\; \buildrel > \over \sim \;$}}
\def\lsim{\lower.5ex\hbox{$\; \buildrel < \over \sim \;$}}

\usepackage{color}

\title[Absorption in cold streams feeding high-$z$ galaxies]
{Detectability of cold streams into high-redshift galaxies by absorption lines}

\author[Tobias Goerdt et al.]
{\parbox[t]{\textwidth}{Tobias Goerdt$^{1}$\thanks{tobias.goerdt@uam.es},
Avishai Dekel$^2$\thanks{dekel@phys.huji.ac.il}, 
Amiel Sternberg$^3$\thanks{amiel@wise.tau.ac.il},
Orly Gnat$^2$\thanks{orlig@phys.huji.ac.il} and \\
Daniel Ceverino$^{1}$\thanks{daniel.ceverino@uam.es}}\\  \vspace*{3pt} \\
$^1$Departamento de F\'isica Te\'orica, Universidad Aut\'onoma de Madrid,
28049 Madrid, Espa\~na\\
$^2$Racah Institute of Physics, The Hebrew University, Jerusalem 91904,
Israel\\
$^3$The Raymond and Beverly Sackler School of Physics and Astronomy, Tel Aviv
University, Tel Aviv 69978, Israel}

\date{Draft version \today}
\pagerange{\pageref{firstpage}--\pageref{lastpage}}
\pubyear{2012}

\begin{document}

\maketitle

\label{firstpage}

\begin{abstract}
Cold gas streaming along the dark-matter filaments of the cosmic web is
predicted to be the major provider of resources for disc buildup, violent disk
instability and star formation in massive galaxies in the early universe. We
study to what extent these cold streams are traceable in the extended
circum-galactic environment of galaxies via {\la} absorption and selected low
ionisation metal absorption lines. We model the expected absorption signatures
using high resolution zoom-in AMR cosmological simulations. In the
postprocessing, we distinguish between self-shielded gas and unshielded gas. In
the self-shielded gas, which is optically thick to Lyman continuum radiation,
we assume pure collisional ionisation for species with an ionisation potential
greater than 13.6 eV. In the optically thin, unshielded gas these species are
also photoionised by the meta-galactic radiation. In addition to absorption of
radiation from background quasars, we compute the absorption line profiles of
radiation emitted by the galaxy at the centre of the same halo. We predict the
strength of the absorption signal for individual galaxies without stacking. We
find that the {\la} absorption profiles produced by the streams are consistent
with observations of absorption and emission {\la} profiles in high redshift
galaxies. Due to the low metallicities in the streams, and their low covering
factors, the metal absorption features are weak and difficult to detect.
\end{abstract}


\section{Introduction}
Cold gas is thought to flow into massive haloes $\sim 10^{12} \msun$ at $z=2-3$
along filaments with velocities of$\gtrsim 200\,{\rm km\,s^{-1}}$. This
phenomenon is predicted by simulations and theoretical analysis, where high-z
massive galaxies are continuously fed by narrow, cold, intense, partly clumpy,
gaseous streams that penetrate through the shock-heated halo gas into the inner
galaxy \citep{bd03, keresa, db06, ocvirk, keresb, nature, peter}. They form a
dense, unstable, turbulent disc with a bulge and trigger rapid star formation
\citep{dsc, oscara, cd, andi, oscarb, c12, marcello, genel12, genzel11}. N-body
simulations suggest that about half the mass in dark-matter halos is built-up
smoothly, suggesting that the baryons are also accreted semi-continuously as
the galaxies grow \citep{genel10}. Indeed, hydrodynamical cosmological
simulations reveal that the rather smooth gas components, including mini-minor
mergers with mass ratio smaller than 1:10, brings in about two thirds of the
mass \citep{nature}. The massive, clumpy and star-forming discs observed at
$z\sim 2$ \citep{genzel08, genel, foerster2, foerster3} may have been formed
primarily via the smooth and steady accretion provided by the cold streams,
with a smaller contribution by major merger events \citep{oscarb, cd}.

\citet{pettini02, adelberger03, shapley03} explored the kinematics of the
intergalactic medium in the early universe in great dept using high-resolution
spectroscopy. \citet{steidel96} proposed that the predicted inflowing gas could
show up as redshifted metal absorption lines in the spectra of Lyman-break
galaxies (LBGs). However, it seems that only a small fraction of the LBGs
exhibit redshifted absorption features, which was interpreted as an indication
for the absence of cold streams in haloes with $4\times10^{11} < \Mv < 10^{12}
M_{\odot}$ \citep[][hereafter {\s10}]{steidel}. This has to be reconciled with
the robust theoretical prediction that the predicted filaments are ubiquitous
in haloes in the early universe. The goal of our paper is to predict the
absorption-line signatures of the cold flows, for a detailed comparison with
observations like {\s10}.

\citet{mich} used cosmological hydrodynamical AMR simulations to predict the
characteristics of {\la} emission from the cold gas streams. The {\la}
luminosity in their simulations is powered by the release of gravitational
energy as the gas is flowing with a rather constant velocity down the potential
gradient toward the halo centre. The simulated {\la}-blobs (LABs) are similar
in many ways to the observed LABs. Some of the observed LABs may thus be
regarded as direct detections of the cold streams that drove galaxy evolution
at high $z$. Observations seem to support this picture \citep{rauch, erb}. On
the other hand, \citet{faucher}, using SPH simulations, predict lower {\la}
luminosities than \citet{mich}, and find it difficult to explain the observed
LAB luminosities with inflow-driven cooling radiation alone. This is partly
because \citet{faucher} exclude emission from very dense gas, and partly
because the SPH simulations underestimate the energy dissipation of the
inflowing gas as it interacts with itself and with the hot medium. They make
the point that the gas temperature should also be taken into account when
distinguishing between the optically thin and thick regimes. New AMR
simulations incorporating radiative transfer seem to confirm the \citet{mich}
model \citep{joki, dan}. In particular, \citet{dan} analysing high-resolution
cosmological AMR-hydro simulations with full radiative transport for the
ionising radiation from background and from stars and for the {\la} scattering
find that about half the spatially extended {\la} luminosity originates from
fluorescence due to ionising radiation from new stars, and the other half from
cooling emission due to the gravitational energy gain, yielding a total
luminosity that is comparable to the findings of \citet{mich} based on crude
approximations for the radiative transport and without the fluorescence. In a
pioneering observation, \citet{sebastiano} detected extended {\la} emission
from circum-galactic filaments of cold gas that seem to largely arise from
fluorescence excited by the UV radiation from an AGN. They detected dark
galaxies and circum-galactic filaments that are fluorescently illuminated by a
quasar. The properties of these {\la} sources all indicate that the sample is
consistent with having much of the {\la} emission originating in fluorescent
reprocessing of quasar radiation.

\citet{serena} investigated the nature and detectability of redshifted
rest-frame UV line emission from the IGM at $2<z<5$. They used simulations to
create maps of a set of strong rest-frame UV emission lines. They conclude that
several species (e.g., C~{\small III}, C~{\small IV}, Si~{\small III},
Si~{\small IV}, and O~{\small VI}) of those from the high-redshift IGM will
become detectable in the near future. They also claim that lower ionisation
lines provide us with tools to image cold accretion flows as well as cold
outflowing clouds.

\begin{figure}
\begin{center}
\includegraphics[width=8.45cm]{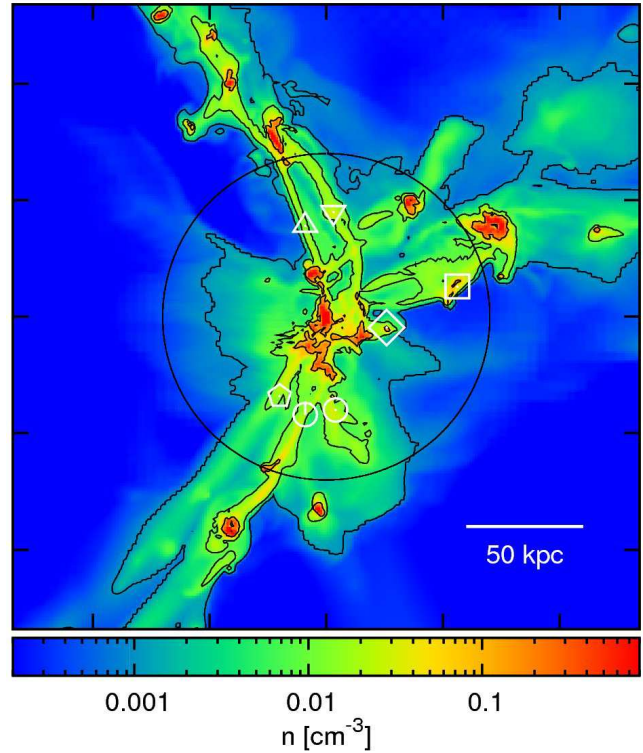}
\end{center}
\caption{Gas density in a simulated galaxy (resolution $70\pc$) at $z=2.3$ with
$\Mv = 3.5 \times 10^{11}\msun$. The colour refers to the maximum density along
the line of sight. The contours mark $n=0.1$, 0.01 and $0.001 \cmmc$,
respectively. The circle shows the virial radius which is 74 kpc. The inflow is
dominated by three cold narrow streams that are partly clumpy. The density in
the streams is $n=0.003-0.1 \cmmc$, with the clump cores reaching $n \sim 1
\cmmc$. The white open boxes tag main features such as streams. The same type
of points is used to identify these features in figure \ref{fig:hammer}.}
\label{fig:denmap}
\end{figure}

\citet{freeke}, using SPH simulations, find that nearly all of the {\HI}
absorption arises in gas that has remained fairly cold, at least while it was
extragalactic. In addition, the majority of the {\HI} is rapidly falling
towards a nearby galaxy, with non-negligible contributions from outflowing and
static gas. They identify carefully the environment of the {\HI} absorbers and
conclude with this information that cold accretion flows are critical for the
success of simulations in reproducing the observed rate of incidence of damped
{\la} and particularly that of Lyman limit systems. They propose that cold
accretion flows exist and have already been detected in the form of high column
density {\HI} absorbers.

\begin{figure*}
\begin{center}
\psfrag{XHI}[B][B][1][0] {$X_{\rm HI}$}
\psfrag{nHcm-3}[B][B][1][0] {$n_{\rm H}$ [cm$^{-3}$]}
\psfrag{UVX}[Bl][Bl][1][0] {UVX}
\psfrag{UVX + local sources}[Bl][Bl][1][0] {UVX + local sources}
\psfrag{radiative transfer}[Bl][Bl][1][0] {radiative transfer}
\psfrag{shielded/thin}[Bl][Bl][1][0] {shielded/thin}
\psfrag{density criterion}[Bl][Bl][1][0] {density criterion}
\psfrag{(this paper)}[Bl][Bl][1][0] {(this paper)}
\includegraphics[width=6.09cm]{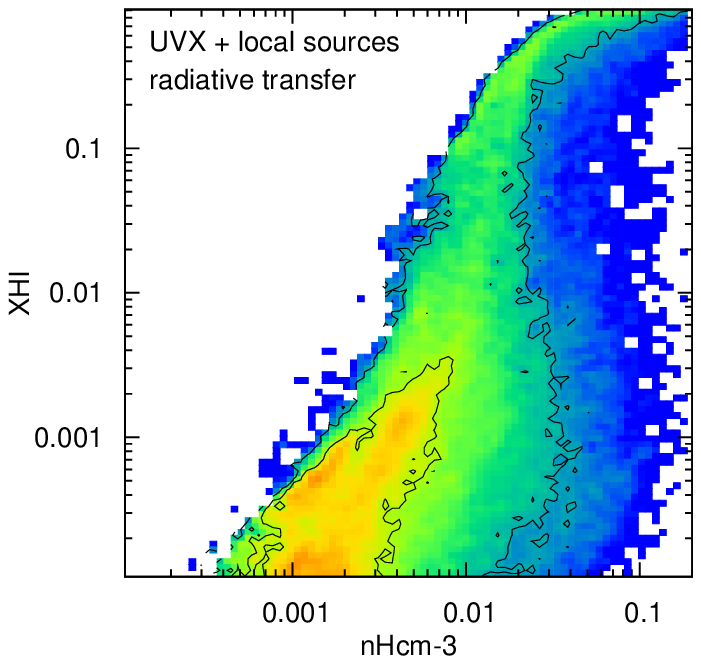}
\includegraphics[width=5.05cm]{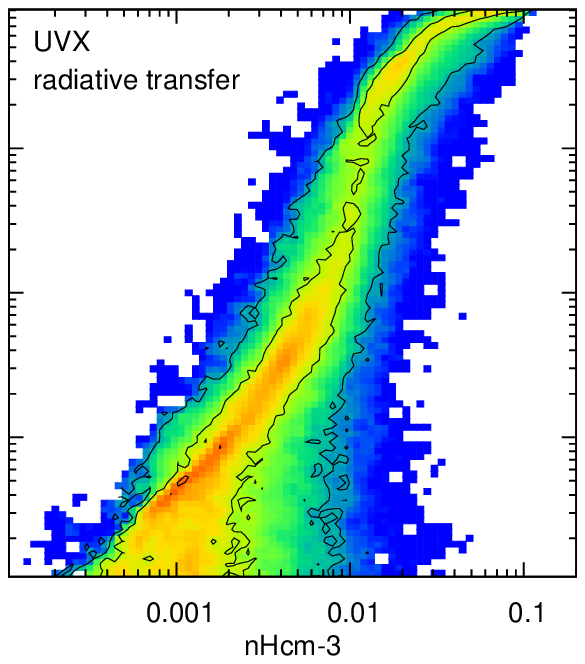}
\includegraphics[width=6.39cm]{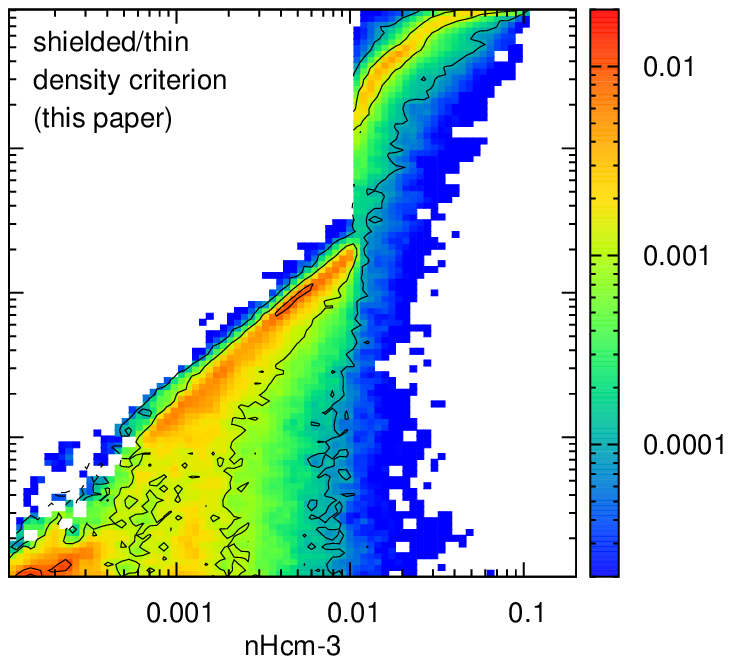}
\end{center}
\caption{Distribution by volume of the neutral hydrogen fractions $x_{\rm HI}$
and total hydrogen gas densities $n_{\rm H}$ in the circumgalactic environment
of our simulated galaxy (see text). The left panel shows the distributions for
the full radiative transfer calculation including the metagalactic UVX field
plus internal sources presented by {\f11}. The middle panel is for the
radiative transfer calculation, but with UVX only. The right panel are results
employing a pure density criterion of $n_{\rm H, shield} = 0.01$\,cm$^{-3}$ for
the boundary between self-shielded and optically thin cells. The units of the
colour-bar axis are unity per unit area in the (log$_{10}\, X_{\rm HI}$,
log$_{10} \, n_{\rm H}$) plane. Contour lines are at 0.0001, 0.001 and 0.01
respectively. Our simplified ionisation model provides a useful crude
approximation for the UVX radiative transfer results (see text).}
\label{fig:danieltest}
\end{figure*}

\citet{kacprzak} analysed synthetic {\MgII} absorption spectra from simulated
galaxies at $z = 1$ and they tentatively concluded that most of the {\MgII}
absorption arises in filaments and tidal streams of the galactic halo. This
material is infalling with velocities around 200 km s$^{-1}$. They claim that
large equivalent-width (\EW) {\MgII} are associated with outflows.

\citet{faucher2}, \citet{kimm}, \citet{kyle} and \citet[][hereafter
\f11]{michele} estimated from simulations the covering factor of the cold
streams at $z=2-4$ for absorption from a background source. They found that it
is small and decreasing with time. \citet{kyle} and {\f11} argue that as a
galaxy switches from cold- to hot-mode accretion with decreasing redshift the
reduced cold gas accretion naturally results in a suppression of cool
circumgalactic gas. This transition occurs when the halo is massive enough to
support a stable shock near the virial radius \citep{bd03,keresa,db06}. {\f11}
discover that within $\sim$ 500 Myr of reaching this threshold mass, the
covering fraction drops. Because the cold-stream covering factor is small
compared to the order unity covering fraction expected for galactic winds, the
cold streams are generally overwhelmed by outflows in absorption spectra.
However, {\f11} show that there is enough cross section in the cold flows that
could make them detectable in {\la} absorption in surveys of background
sources. Note the differences with the observations of galaxies at zero impact
parameter, i.e. observations in the transverse direction, using e.g. quasars.
\citet{freeke} and {\f11} show that in fact there is a significant contribution
to the population of absorption line systems that comes from the streams. That
is, outflows are prominent, but inflows should be visible as well.

\citet{kimm} and {\f11} investigated whether the cold gas streams are expected
to be detectable in low-ionisation metal absorption lines, such as  {\CII}
(1334 \AA) or {\SiII} (1260 \AA). They find that the metal absorption signal by
the interstellar medium of the central galaxy itself is so deep and so broad in
velocity space that it swamps the metal signal from the filamentary gas, which
is rather metal poor. A cold filament might be detectable in metal lines if it
lies precisely along the line-of-sight, but this would be rare.

\citet{shen12} used SPH simulations adopting a blastwave scheme for supernova
feedback which, in combination with a high gas density threshold for star
formation, has been shown to generate large-scale galactic outflows. Their
strong outflows avoid the incoming dense narrow streams, and find their way out
in a wide solid angle through the dilute medium between the streams. Showing
that the cold inflows are not affected by the outflows. They generate synthetic
spectra by drawing sightlines through the simulated CGM at different
galactocentric impact parameters, and compared the theoretical interstellar
absorption line strengths with the observations. They find that they are in
broad agreement with those observed at high-redshift by {\s10}.

\begin{figure*}
\begin{center}
\includegraphics[width=17.73cm]{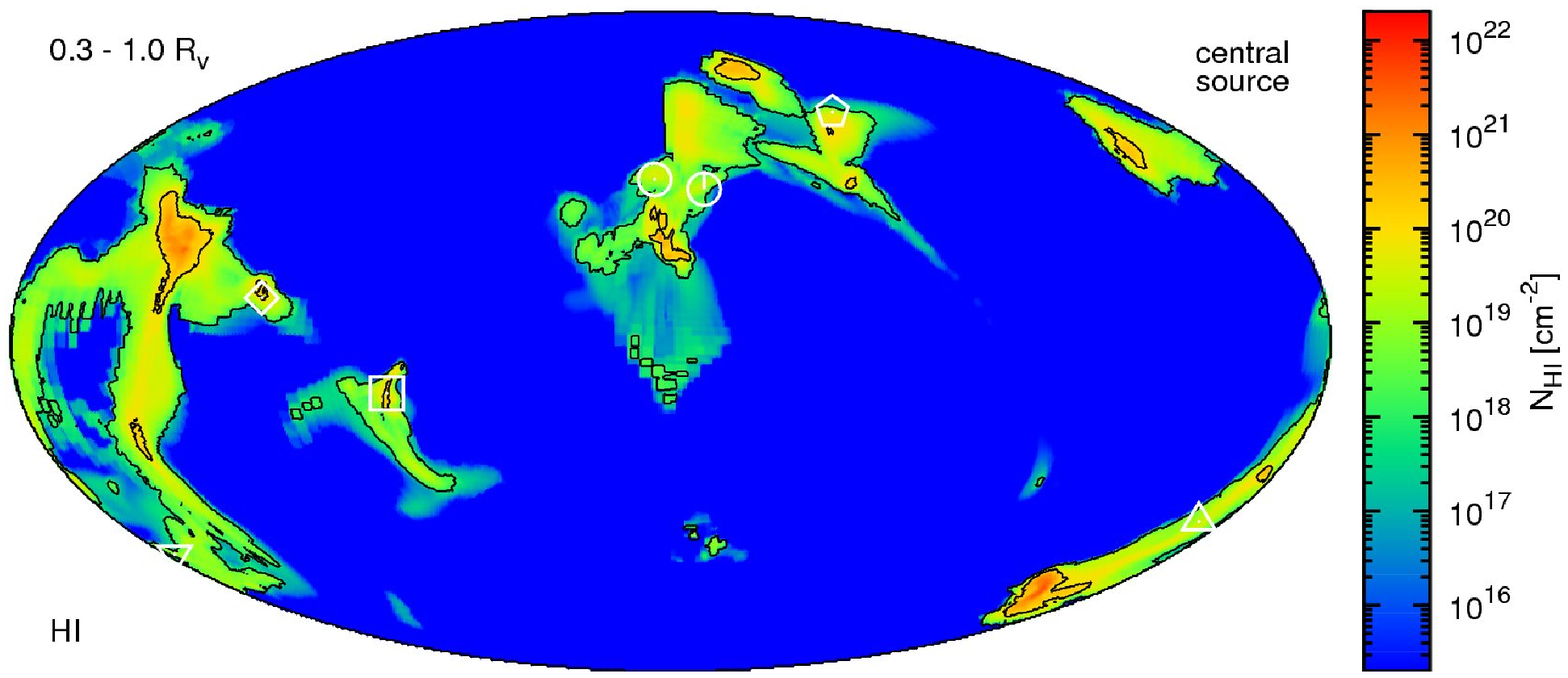}
\includegraphics[width=17.73cm]{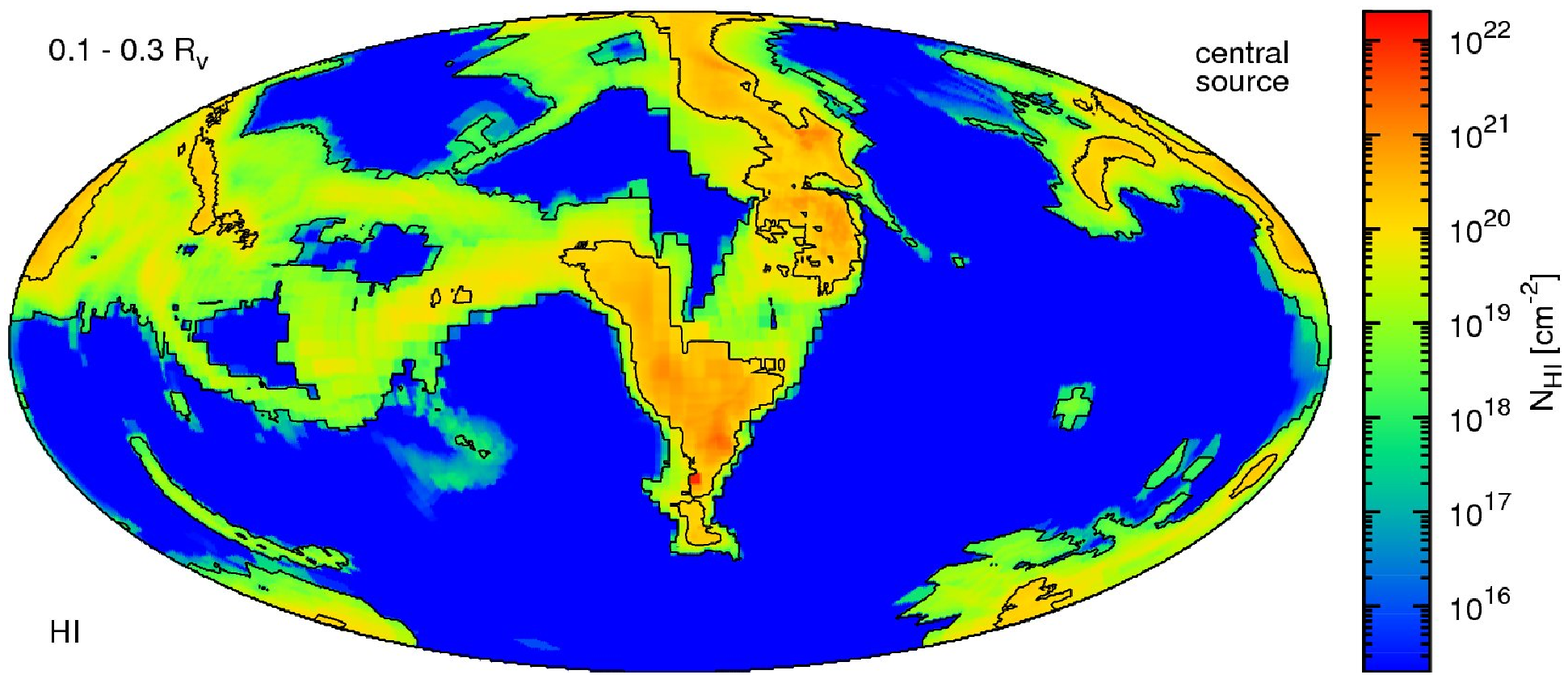}
\end{center}
\caption{Hammer equal area projections of {\HI} column densities in our
fiducial simulated galaxy (resolution $70\pc$, $z=2.3$, $\Mv = 3.5 \times
10^{11}\msun$). The upper panel is integrated from 0.3 to 1.0 $\Rv$, the lower
panel from 0.1 to 0.3 $\Rv$. The contour lines indicate $10^{20}$ and
$10^{18}$\,cm$^{-1}$. The narrow major streams with column densities up to $\sim
10^{22}$ cm$^{-2}$ can be seen, as well as the vast majority of the
“stream-free” sky (as seen from the central galaxy) having only column
densities of $< 10^{15}$ cm$^{-2}$. The white open boxes in the upper panel
identify the main features such as the streams. The same type of points is used
to identify these features in figure \ref{fig:denmap}. The reader should be
aware of the fact that these plots are theoretical and not observed in nature
since one always looks down to the very centre at $r = 0$.}
\label{fig:hammer}
\end{figure*}

As in the observations of {\s10}, we consider separately absorption features 
of radiation emitted from {\it background sources\,} and from {\it central
sources}, namely along lines of sight with finite and with vanishing impact
parameter with respect to the galaxy at the centre of the halo that hosts the
absorbing gas. A theoretical study of absorption in the central-source geometry
has not been attempted so far. The same is true for a statistical study of
absorption signal in individual galaxies without stacking. We focus on
predicting absorption line profiles that correspond to the measurements of
{\s10}. We refer to the same ten different lines and mimic their observational
procedure by averaging data for several simulated galaxies and many lines of
sight. As in {\f11}, we employ high-resolution AMR simulations that are
described in \citet{cd}. The spatial resolution in these simulations is 20
times better than in the Horizon MareNostrum simulation \citep{ocvirk} used by
\citet[][they also used higher resolution simulations for verification
purposes]{kimm}. Unlike in {\f11}, we do not employ a full radiative transport
analysis, and instead account for self-shielding by a simple density criterion
that approximates the radiative-transport results. The computations include
explicitly highly ionised species (e.g. {\CIV}) created in the warm highly
ionised intergalactic medium (as opposed to {\CII} and others that are created
in the cold low-ionised neutral medium).

Our paper is organised as follows. In section \ref{sec:sim} we introduce the
simulations used for this analysis. In section \ref{sec:linestrength} we
explain the computation of the absorption features. In section \ref{sec:cg} we
show the results for the central geometry. In section \ref{sec:bg} we show the
results for the background geometry. Finally, in section \ref{sec:con} we draw
our conclusions.

\section{The Simulations}
\label{sec:sim}
We use three simulated galaxies from a suite of simulations, employing Eulerian
adaptive mesh refinement (AMR) hydrodynamics in a cosmological setting. These
are zoom-in simulations in dark-matter haloes with masses $\sim 5\times 10^{11}
\msun$ at $z = 2.3$, with a maximum resolution of $35-70\pc$ in physical
coordinates \citep[][hereafter \CD]{cd}. \Fig{denmap} shows a density map of
one of the three galaxies from the simulation, which serves as our fiducial
galaxy. It demonstrates the dominance of typically three co-planar narrow cold
streams \citep[see][]{danovich}, originating outside the virial radius along
the dark-matter filaments of the cosmic web, penetrating into the discs at the
halo centres. The streams consist of a smooth component and clumps with a
spectrum of sizes. The typical densities in the streams are in the range
$n= 0.003-0.1\cmmc$, and they reach $n = 1 \cmmc$ near the central disk and at
the clump centres. Some of those clumps are satellites with dark matter haloes.

The {\CD} simulations were run with the \textsc{art} \citep[adaptive refinement
tree;][]{kkk,andrey} code. It incorporates relevant physical processes for
galaxy formation, including gas cooling, photo-ionisation, heating, star
formation, metal enrichment and stellar feedback \citep{cak}. Cooling rates
were computed for the given gas density, temperature, metallicity and UV
background \citep[based on \textsc{cloudy},][]{ferland}. Cooling is assumed at
the centre of a cloud of uniform density with thickness 1 kpc \citep{ceverino,
rk}. Metallicity dependent, metal-line cooling is included, assuming a relative
abundance of elements equal to the solar composition. The code implements a
``constant" feedback model, in which the combined energy from stellar winds and
supernova explosions is released as a constant heating rate over 40 Myr. This
is the typical age of the lightest star that explodes as a type-II supernova.
Photo-heating is also taken into account self-consistently with radiative
cooling. A uniform UV background based on the \citet{haardt} model is assumed.
Local sources are ignored.

This code has a unique feature for the purpose of simulating the detailed
structure of the streams and the gravitational instability in the disk. It
allows gas cooling to well below $10^4$K. This enables high densities in
pressure equilibrium with the hotter and more dilute medium. A non-thermal
pressure floor has been implemented to ensure that the Jeans length is resolved
by at least seven resolution elements preventing artificial fragmentation on
the smallest grid scale \citep{truelove,rk,cd}. The pressure floor is effective
in the dense $(n > 10$ cm$^{-3})$ and cold $(T<10^4 K)$ regions inside galactic
disks. Most of the absorbing gas in question, which is far outside the disk, is
not affected by this pressure floor.

The equation of state remains unchanged at all densities. Stars form in cells
where the gas temperature is below $10^4$K and its density is above the
threshold $n = 1\cmmc$ according to a stochastic model that is roughly
consistent with the \citet{kennicutt} law. The ISM is enriched by metals from
supernovae type II and type Ia. Metals are released from each stellar particle
by SNII at a constant rate for 40 Myr since its birth. A \citet{miller} IMF is
assumed. This procedure matches the results of \citet{woosley}. The metal
ejection by SNIa assumes an exponentially declining SNIa rate from a maximum at
1 Gyr. The code treats the advection of metals self-consistently and it
distinguishes between SNII and SNIa ejecta \citep{ceverino}.

\begin{figure}
\begin{center}
\psfrag{fc(>N)}[B][B][1][0] {$f_{\rm c}(> N_{\rm HI})$}
\psfrag{Ncm2}[B][B][1][0] {$N_{\rm HI}$ [cm$^{-2}$]}
\psfrag{HI}[B][B][1][0] {\HI}
\psfrag{cg}[Bl][Bl][1][0] {central source}
\psfrag{visibility limit}[Bl][Bl][1][0] {visibility limit}
\psfrag{ze138me138e12}[Br][Br][1][0] {\textcolor{red}{$z = 1.4 \, \Mv = 1.4
\times 10^{12}\,$M$_\odot$}}
\psfrag{ze233me35e11}[Br][Br][1][0] {\textcolor{blue}{$z = 2.3 \, \Mv = 3.5
\times 10^{11}\,$M$_\odot$}}
\includegraphics[width=8.45cm]{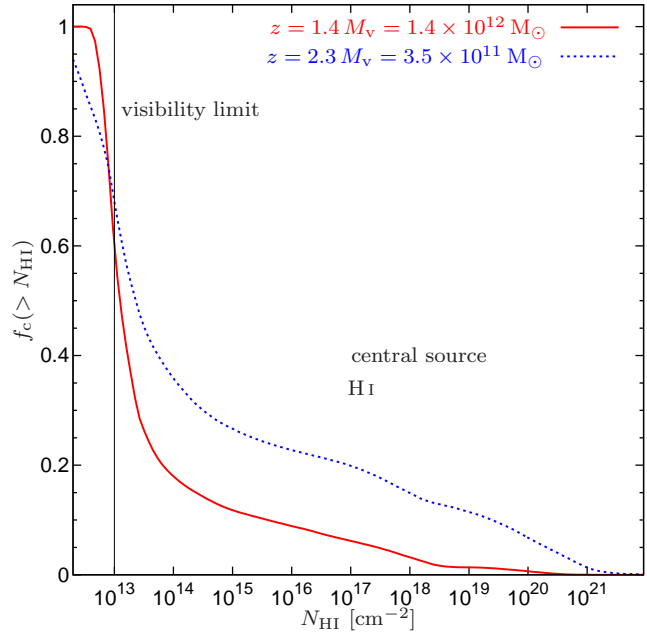}
\end{center}
\caption{Cumulative {\la} sky covering fractions $f_{\rm c}$ higher than column
density $N_{\rm HI}$. It is integrated from 0.3 to 1.0 $\Rv$ for all gas
inflowing and outflowing. For the blue dashed line we use our three simulated
galaxies (resolution $70\pc$) at $z=2.3$ with $\Mv = 3.5 \times 10^{11}\msun$
and $\Rv = 74\,$kpc, for the red solid line we use two simulated galaxies at
$z=1.38$ with $\Mv = 1.38 \times 10^{12}\msun$ and $\Rv = 150\,$kpc. The
observational visibility limit at 10$^{13}\,$cm$^{-2}$ is indicated by the
vertical line. We see very low $(<0.3)$ covering fraction over almost the whole
range of column densities and a a sharp jump of the covering fraction at the
visibility limit. The higher redshift galaxy has a considerably higher covering
fraction over the whole range of column densities.}
\label{fig:laskycover}
\end{figure}

The dark matter particle mass is $5.5 \times 10^5 \msun$. The minimum star
particle mass is $10^4 \msun$. The initial conditions for the {\CD} simulations
were created using a low-resolution cosmological $N$-body simulation in a
co-moving box of side 29 Mpc. Its cosmological parameters were motivated by
WMAP5 \citep{WMAP5}. The values are: $\Omega_{\rm m} = 0.27$, $\Omega_\Lambda =
0.73$, $\Omega_{\rm b} = 0.045$, $h = 0.7$ and $\sigma_8 = 0.82$. At $z=1$,
three haloes of $\Mv \simeq 10^{12}\msun$ each have been selected, avoiding
haloes that were subject to a major merger at that time. The three halo masses
at $z=2.3$ are $3.5, 4, 6\times 10^{11} \msun$, their virial radii are around
74 kpc and they end up as $(3-4)\times 10^{12}$ M$_\odot$ haloes today. Around
each halo, a concentric sphere of radius twice the virial radius was selected
for resimulation with high resolution. Gas was added to the box following the
dark matter distribution with a fraction $f_{\rm b} = 0.15$. The whole box was
then re-simulated, with refined resolution only in the
selected volume about the respective galaxy.

\begin{figure*}
\begin{center}
\psfrag{fc(>N)}[B][B][1][0] {$f_{\rm c}(> N_{\rm Ai})$}
\psfrag{Ncm2}[B][B][1][0] {$N_{\rm Ai}$ [cm$^{-2}$]}
\psfrag{CII}[B][B][1][0] {\textcolor{red}{\CII}}
\psfrag{CIV}[B][B][1][0] {\textcolor{green}{\CIV}}
\psfrag{OI}[B][B][1][0] {\textcolor{blue}{\OI}}
\psfrag{SiII}[B][B][1][0] {\textcolor{magenta}{\SiII}}
\psfrag{SiIV}[B][B][1][0] {\textcolor{cyan}{\SiIV}}
\psfrag{MgII}[B][B][1][0] {\textcolor{yellow}{\MgII}}
\psfrag{FeII}[B][B][1][0] {\textcolor{black}{\FeII}}
\psfrag{OII}[B][B][1][0]  {\textcolor{Orange}{\OII}}
\psfrag{OIII}[B][B][1][0] {\textcolor{Gray}{\OIII}}
\psfrag{cg}[Bl][Bl][1][0] {central source}
\psfrag{visibility limit}[Br][Br][1][0] {visibility limit}
\psfrag{ze138}[Bl][Bl][1][0] {$z = 1.4$}
\psfrag{ze233}[Bl][Bl][1][0] {$z = 2.3$}
\psfrag{me1.38e12}[Bl][Bl][1][0] {$\Mv = 1.4 \times 10^{12}$ M$_\odot$}
\psfrag{me3.5e11}[Bl][Bl][1][0]  {$\Mv = 3.5 \times 10^{11}$ M$_\odot$}
\includegraphics[width=9.24cm]{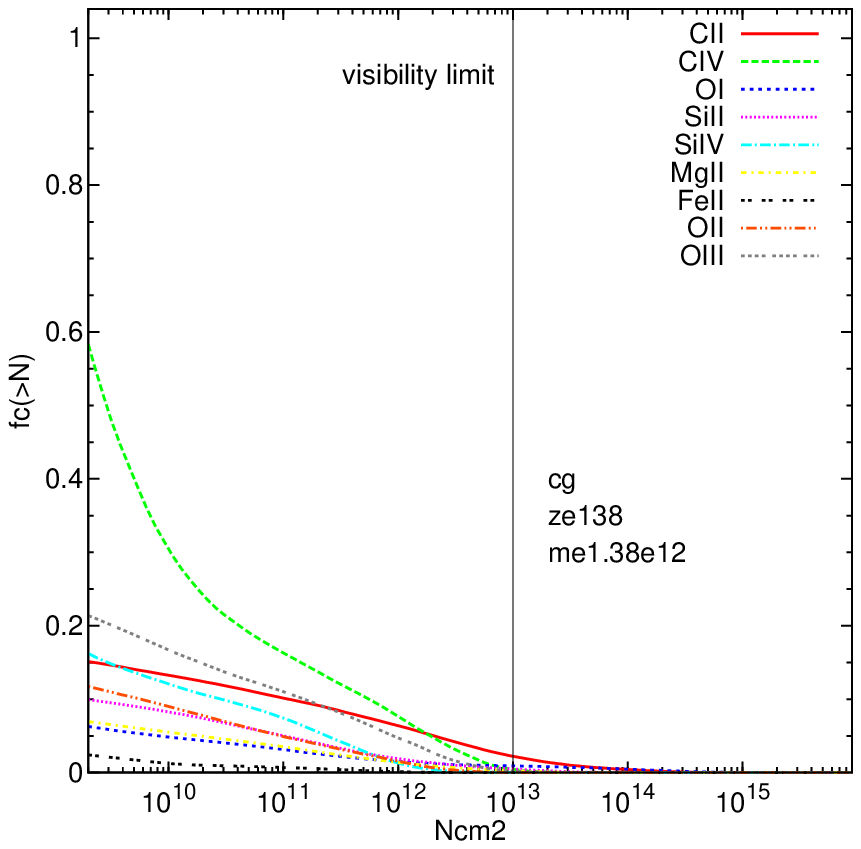}
\includegraphics[width=8.38cm]{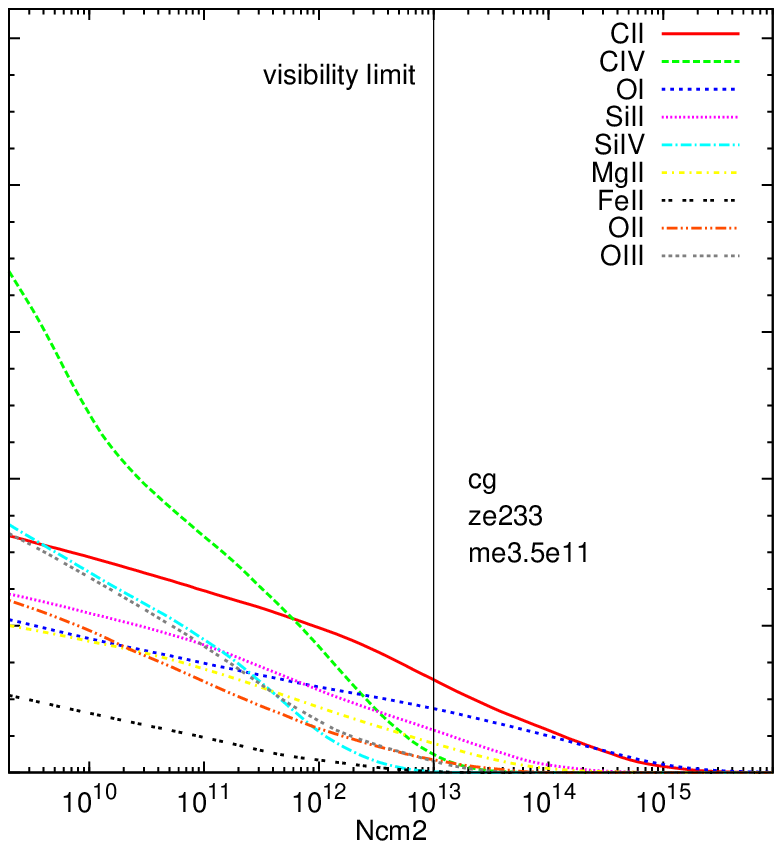}
\end{center}
\caption{Same cumulative sky covering fractions as in figure
\ref{fig:laskycover} this time for metal lines. In the right panel we use our
three simulated galaxies (resolution $70\pc$, $z=2.3$, $\Mv = 3.5 \times
10^{11}\msun$ and $\Rv = 74\,$kpc), in the left panel we use two simulated
galaxies at $z=1.38$ with $\Mv = 1.38 \times 10^{12}\msun$ and $\Rv = 150\,$kpc.
The observational visibility limit at 10$^{13}\,$cm$^{-2}$ is indicated by the
vertical line. Together with figure \ref{fig:laskycover} this plot shows that a
full sky map of column densities exhibits a few very high peaks corresponding
to the streams extending only over a fairly tiny solid angle and a majority of
very low column density solid angle corresponding to the space without streams.
This effect is stronger for the low redshift panels.}
\label{fig:skycover}
\end{figure*}

\section{Computing the ionisation states}
\label{sec:linestrength}
A reliable estimate of the ionisation state of the gas is essential for any
study of absorption line systems. Ideally, the determination of the ionisation
states of the gas should be coupled to the hydrodynamic calculations, but due
to the numerical complexity of the problem this type of calculation remains
computationally expensive for high resolution simulations especially at lower
redshifts. In order to compute them via post-processing we take the densities,
temperatures and metallicities from the simulation. We assume a primordial
helium mass fraction $Y=0.24$, corresponding to a helium particle abundance of
$1/12$ relative to hydrogen. For the heavy elements we assume the
\citet{asplund} solar photosphere pattern.

We determine whether a given simulation cell is ``self-shielded", i.e.
optically thick to Lyman continuum radiation, via a simple density criterion.
Cells with total hydrogen exceeding $n_{\rm shield}\equiv 0.01$~cm$^{-3}$ are
assumed to be self-shielded. For lower densities the cells are assumed to be
optically thin. We find that setting $n_{\rm shield}=0.01$~cm$^{-3}$ gives the best
agreement with a radiative transfer calculation as we demonstrate in figure
\ref{fig:danieltest} (see discussion below). For a further discussion of the
density criterion, see the discussion by \citet{cox} or the appendix A2 of
{\f11}. We calculate the atomic and ionic fractions $x_{\rm Ai}$ using
\textsc{cloudy} \citep{ferland}, where $x_{\rm Ai}$ is defined as the fractions
of element $A$ in ionisation state $i$. The specific species we are interested
in are {\HI}, {\OI}, {\CII}, {\CIV}, {\SiII} and {\SiIV}. These are the most
important ions that are visible and detected in optical spectra at $z = 2 - 3$
and discussed in great detail by {\s10}. We also consider {\MgII} and {\FeII}
since they are assumed to be very suitable for future observational programmes.
Table \ref{tab:lines} lists the wavelengths and oscillator strengths of the
various absorption lines we consider.

\begin{table}
\begin{center}
\setlength{\arrayrulewidth}{0.5mm}
\begin{tabular}{lllc}
\hline
ion & $\lambda_0$ [\AA] & $f_\lambda$ & $\gamma_\lambda$ [s$^{-1}$] \\
\hline
{\la}   & 1215.6701 & 0.4164  & $6.265 \times 10^{8}$ \\
{\CII}  & 1334.5323 & 0.1278  & $2.870 \times 10^{8}$ \\
{\SiII} & 1260.4221 & 1.115   & $2.533 \times 10^{9}$ \\
{\OI}   & 1302.1685 & 0.04887 & $5.750 \times 10^{8}$ \\
{\SiII} & 1526.7066 & 0.1155  & $1.960 \times 10^{9}$ \\
{\SiII} & 1304.3702 & 0.09345 & $1.720 \times 10^{9}$ \\
{\CIV}  & 1548.195  & 0.1908  & $2.654 \times 10^{8}$ \\
{\SiIV} & 1393.755  & 0.514   & $8.825 \times 10^{8}$ \\
{\CIV}  & 1550.770  & 0.09522 & $2.641 \times 10^{8}$ \\
{\SiIV} & 1402.770  & 0.2553  & $8.656 \times 10^{8}$ \\
{\MgII} & 2796.352  & 0.6123  & $2.612 \times 10^{8}$ \\
{\FeII} & 2382.765  & 0.3006  & $3.100 \times 10^{8}$ \\
\hline
\end{tabular}
\end{center}
\caption{Basic parameters of the various absorption lines, we quote wavelength
$\lambda_0$, oscillator strength $f_\lambda$ and the damping width
$\gamma_\lambda$ which is the sum over the spontaneous emission coefficients.
Data is taken from \citet{morton}, except for the $f_\lambda$-values for the
three {\SiII} lines, which are from \citet{dufton}.}
\label{tab:lines}
\end{table}

\begin{figure*}
\begin{center}
\psfrag{relative intensity}[B][B][1][0] {relative intensity}
\psfrag{vks}[B][B][1][0] {$\Delta w$ [km s$^{-1}$]}
\psfrag{cg}[Bl][Bl][1][0] {central source}
\psfrag{corrfig}[Bl][Bl][1][0] {see figure \ref{fig:ewhammer}}
\psfrag{lya}[Bl][Bl][1][0] {\la}
\psfrag{A}[B][B][1][0] {A}
\psfrag{B}[B][B][1][0] {B}
\psfrag{C}[B][B][1][0] {C}
\psfrag{D}[B][B][1][0] {D}
\psfrag{E}[B][B][1][0] {E}
\psfrag{F}[B][B][1][0] {F}
\psfrag{G}[B][B][1][0] {G}
\psfrag{H}[B][B][1][0] {H}
\psfrag{I}[B][B][1][0] {I}
\psfrag{J}[B][B][1][0] {J}
\psfrag{K}[B][B][1][0] {K}
\psfrag{L}[B][B][1][0] {L}
\psfrag{NHI=3.7e19}[Bl][Bl][1][0] {$N_{\rm HI} = 3.3 \times 10^{20}$ cm$^{-2}$}
\psfrag{NHI=3.2e12}[Bl][Bl][1][0] {$N_{\rm HI} = 2.5 \times 10^{13}$ cm$^{-2}$}
\psfrag{NHI=4.7e11}[Bl][Bl][1][0] {$N_{\rm HI} = 3.7 \times 10^{12}$ cm$^{-2}$}
\psfrag{NHI=7.6e19}[Bl][Bl][1][0] {$N_{\rm HI} = 3.4 \times 10^{20}$ cm$^{-2}$}
\psfrag{NHI=9.1e12}[Bl][Bl][1][0] {$N_{\rm HI} = 3.1 \times 10^{20}$ cm$^{-2}$}
\psfrag{7.9e16}[Bl][Bl][1][0] {$1.9 \times 10^{19}$ cm$^{-2}$}
\psfrag{NHI=}[Bl][Bl][1][0] {$N_{\rm HI} =$}
\psfrag{8.2e16}[Bl][Bl][1][0] {$8.9 \times 10^{19}$ cm$^{-2}$}
\psfrag{1.5e17}[Bl][Bl][1][0] {$7.4 \times 10^{19}$ cm$^{-2}$}
\psfrag{5.7e14}[Bl][Bl][1][0] {$4.1 \times 10^{17}$ cm$^{-2}$}
\psfrag{1.3e13}[Bl][Bl][1][0] {$3.1 \times 10^{19}$ cm$^{-2}$}
\psfrag{3.7e13}[Bl][Bl][1][0] {$2.7 \times 10^{14}$ cm$^{-2}$}
\psfrag{1.1e19}[Bl][Bl][1][0] {$4.7 \times 10^{19}$ cm$^{-2}$}
\psfrag{W0=1.6A}[Bl][Bl][1][0] {$W_0 = 1.6$ \AA}
\psfrag{W0=0.45A}[Bl][Bl][1][0] {$W_0 = 0.45$ \AA}
\psfrag{W0=2.8A}[Bl][Bl][1][0] {$W_0 = 2.8$ \AA}
\psfrag{W0=5.2A}[Bl][Bl][1][0] {$W_0 = 5.2$ \AA}
\psfrag{W0=4.3A}[Bl][Bl][1][0] {$W_0 = 4.3$ \AA}
\psfrag{W0=1.7A}[Bl][Bl][1][0] {$W_0 = 1.7$ \AA}
\psfrag{W0=0.082A}[Bl][Bl][1][0] {$W_0 = 0.082$ \AA}
\psfrag{W0=0.012A}[Bl][Bl][1][0] {$W_0 = 0.012$ \AA}
\psfrag{W0=1.5A}[Bl][Bl][1][0] {$W_0 = 1.5$ \AA}
\psfrag{W0=0.44A}[Bl][Bl][1][0] {$W_0 = 0.44$ \AA}
\psfrag{W0=3.3A}[Bl][Bl][1][0] {$W_0 = 3.3$ \AA}
\psfrag{W0=1.7A}[Bl][Bl][1][0] {$W_0 = 1.7$ \AA}
\includegraphics[width=5.66cm]{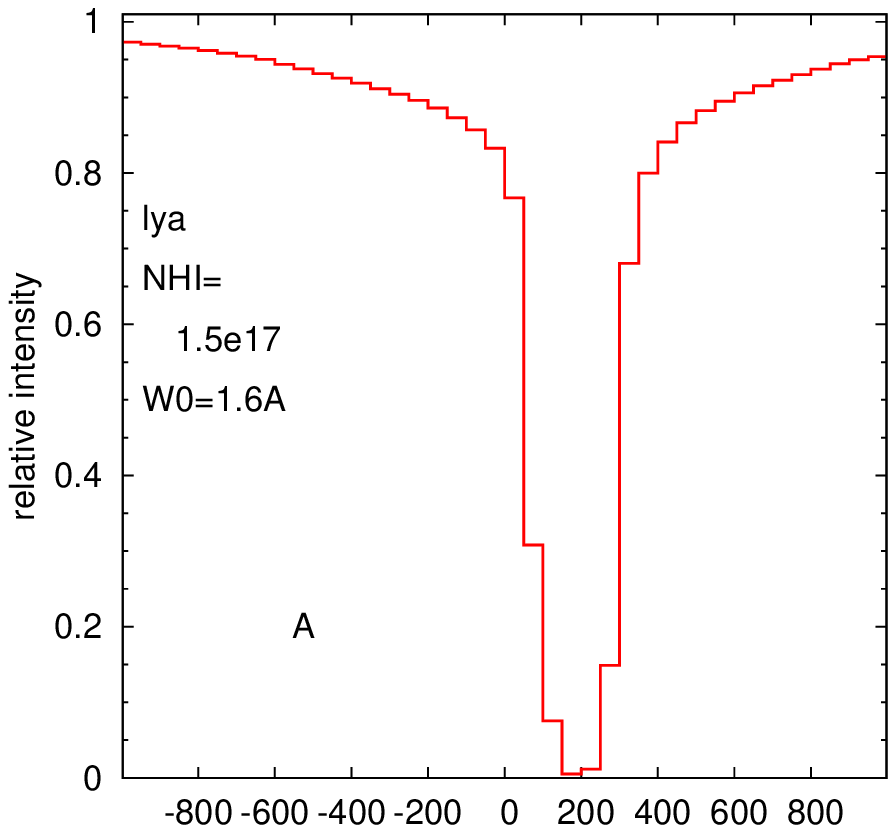}
\includegraphics[width=4.91cm]{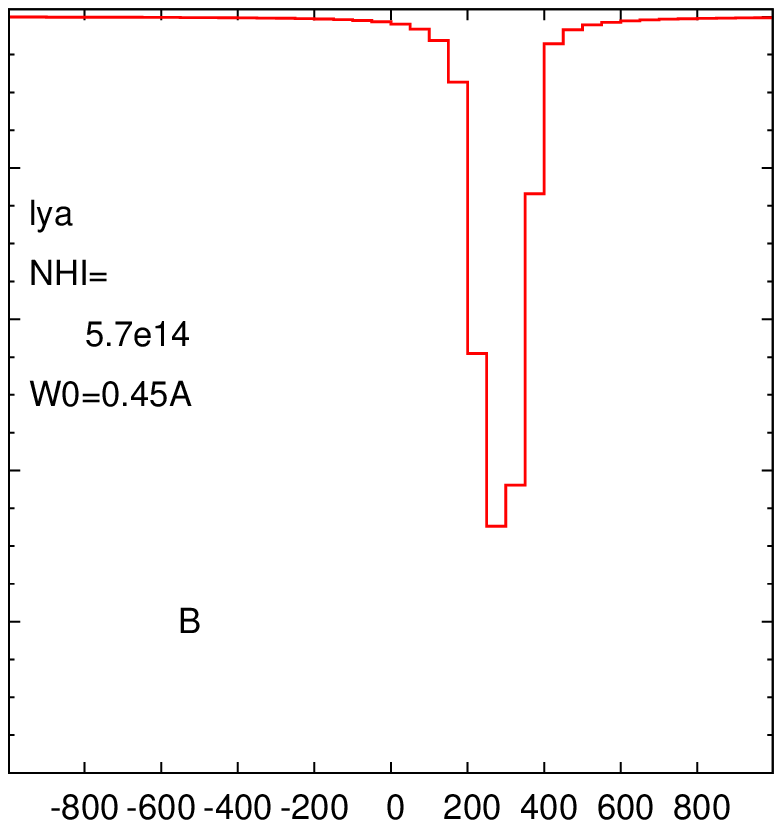}
\includegraphics[width=4.91cm]{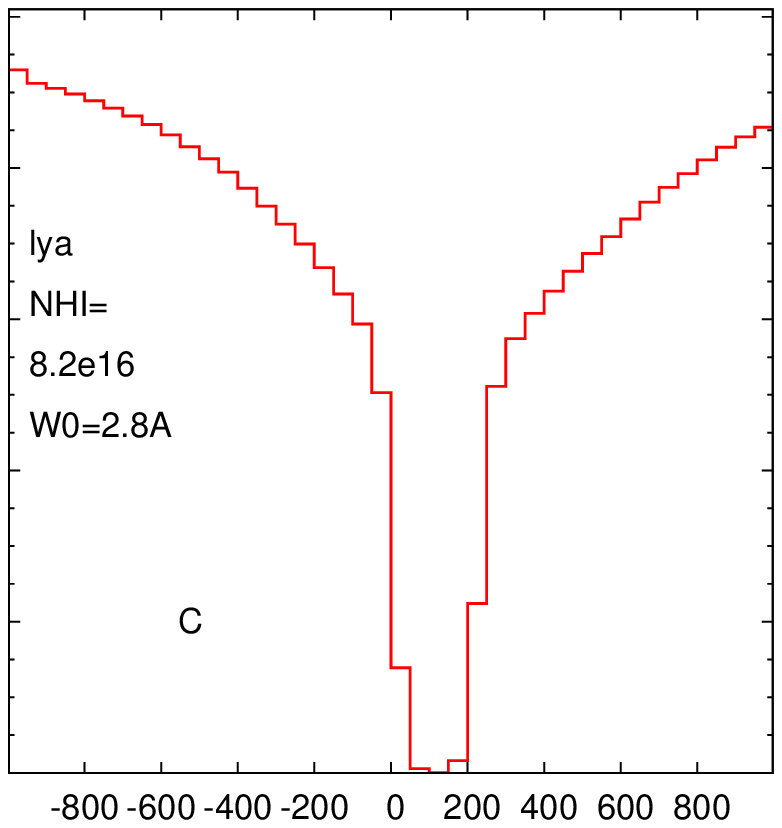}
\includegraphics[width=5.66cm]{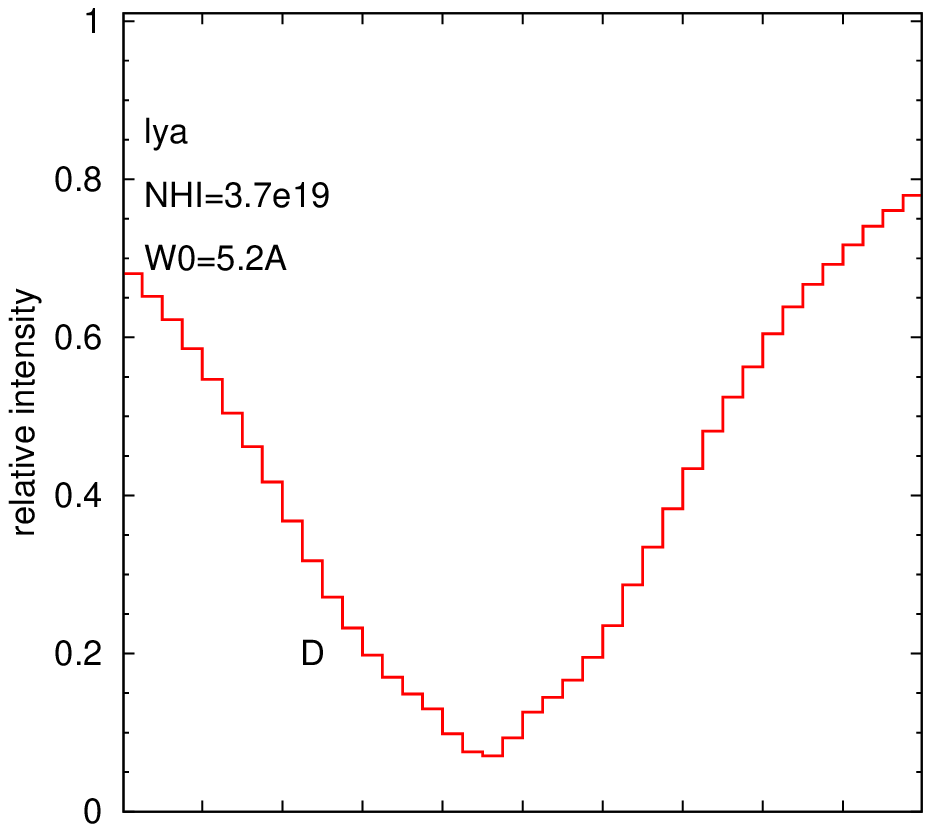}
\includegraphics[width=4.91cm]{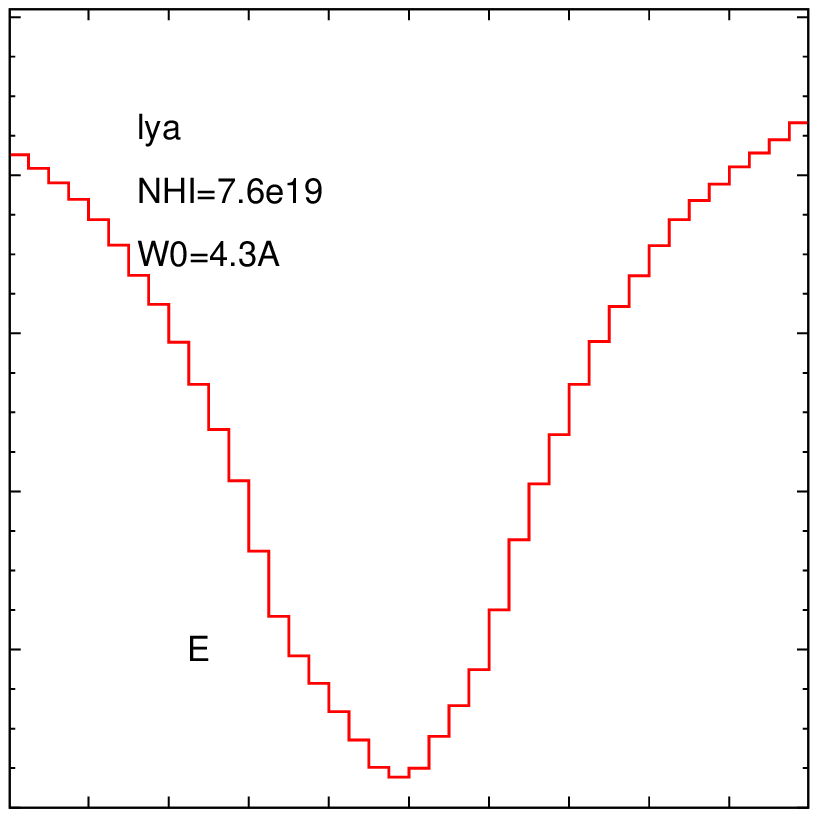}
\includegraphics[width=4.91cm]{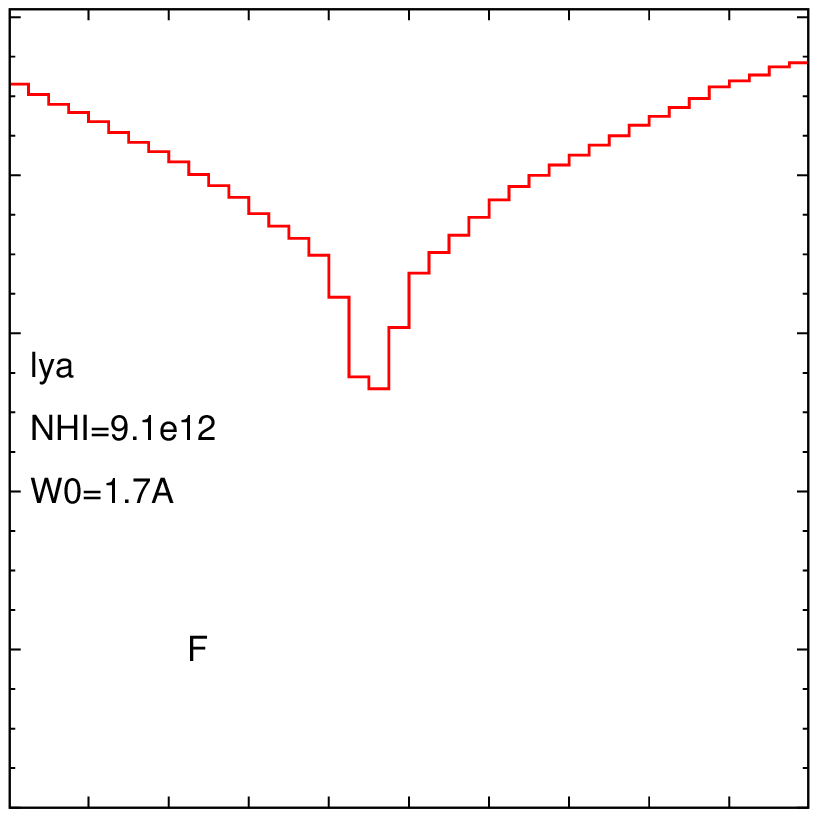}
\includegraphics[width=5.66cm]{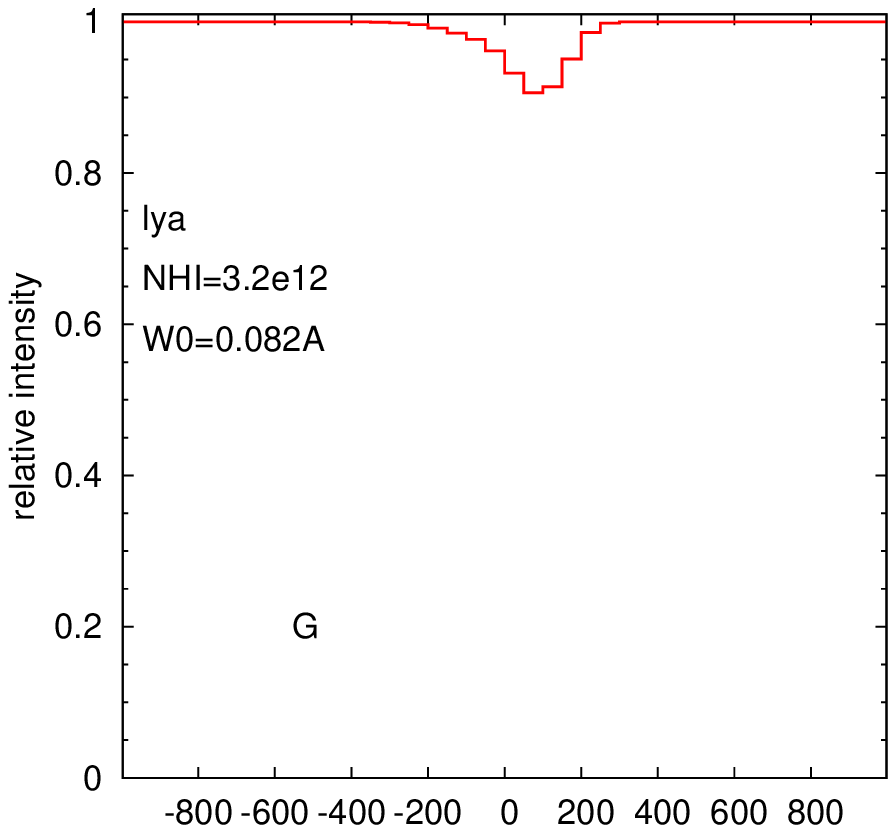}
\includegraphics[width=4.91cm]{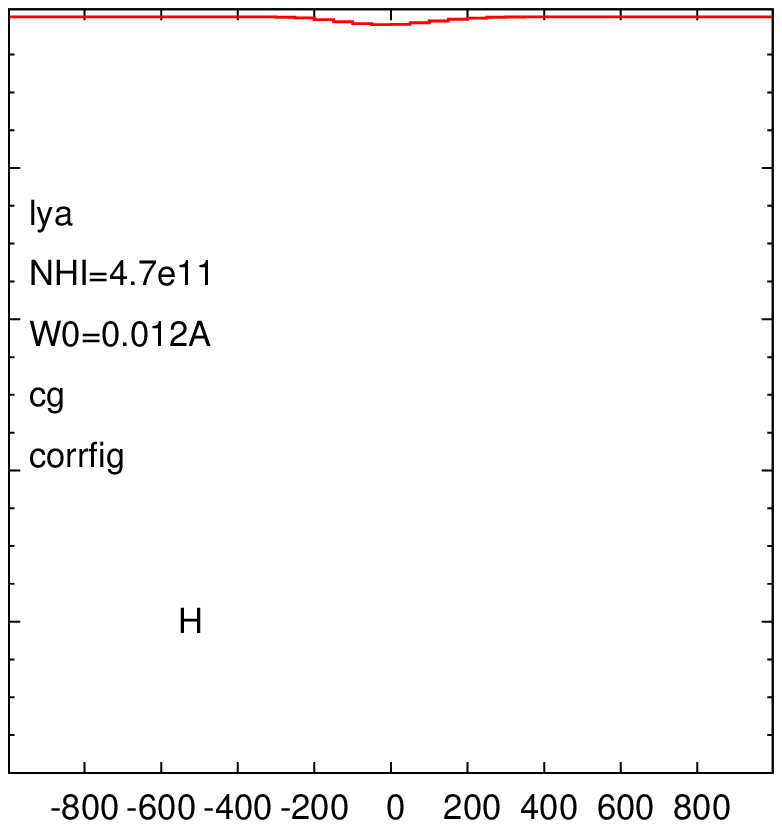}
\includegraphics[width=4.91cm]{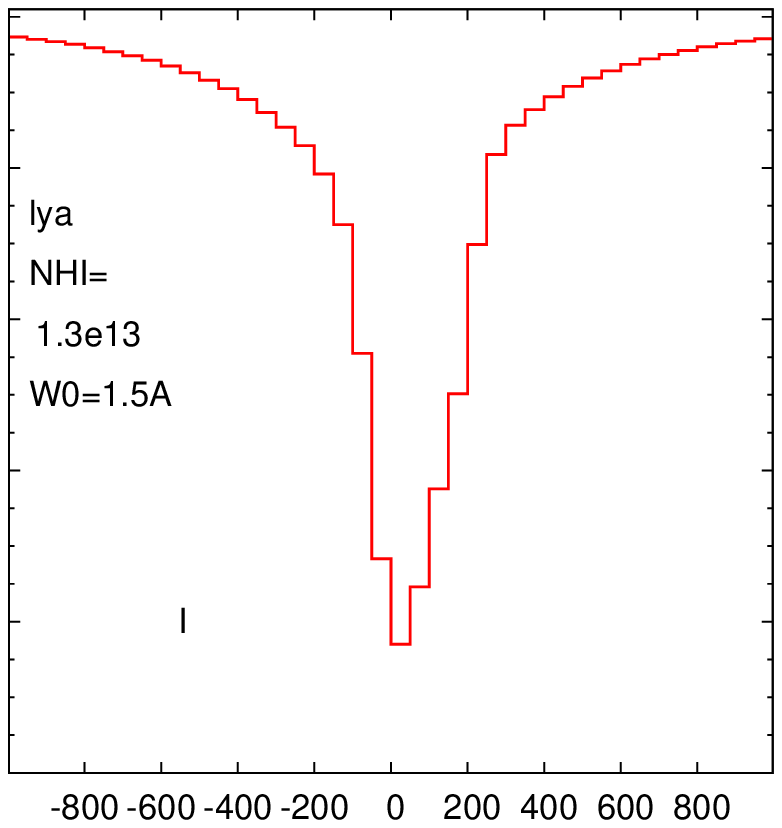}
\includegraphics[width=5.66cm]{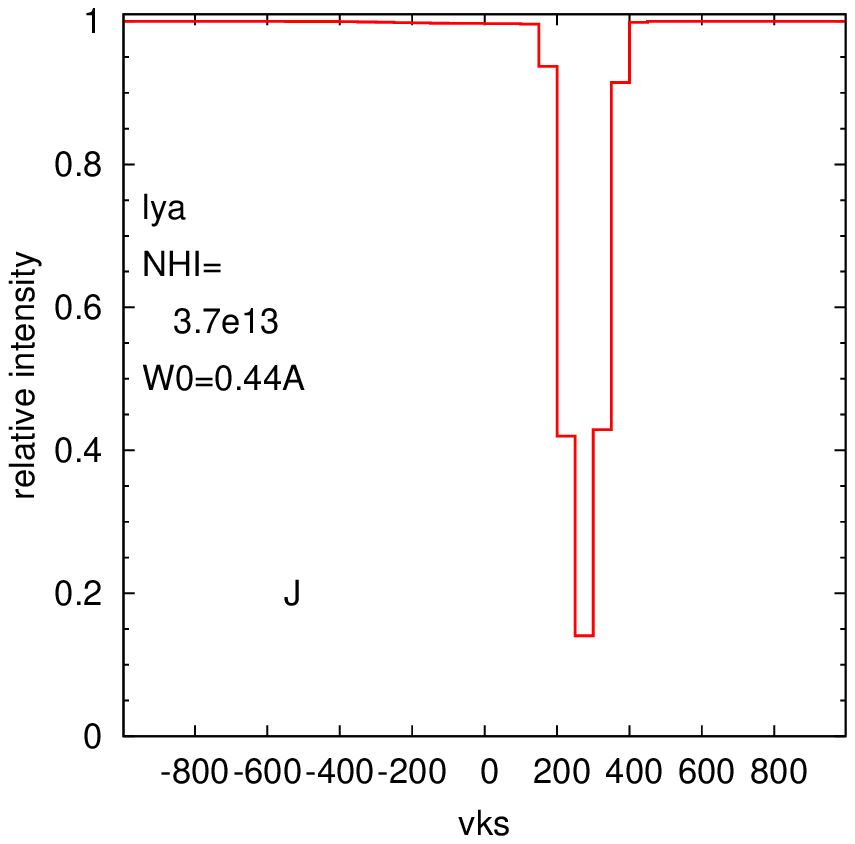}
\includegraphics[width=4.91cm]{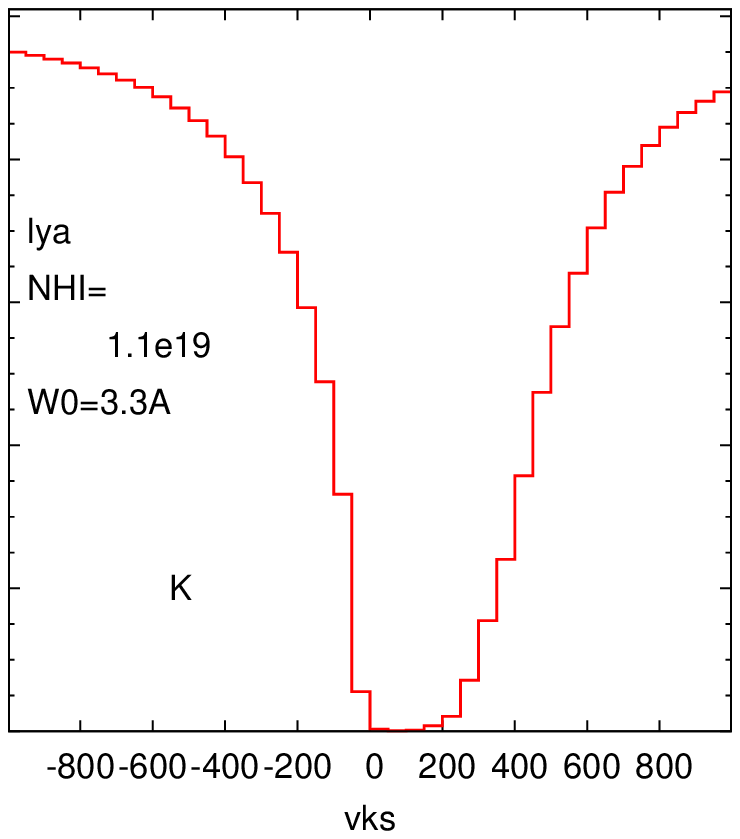}
\includegraphics[width=4.91cm]{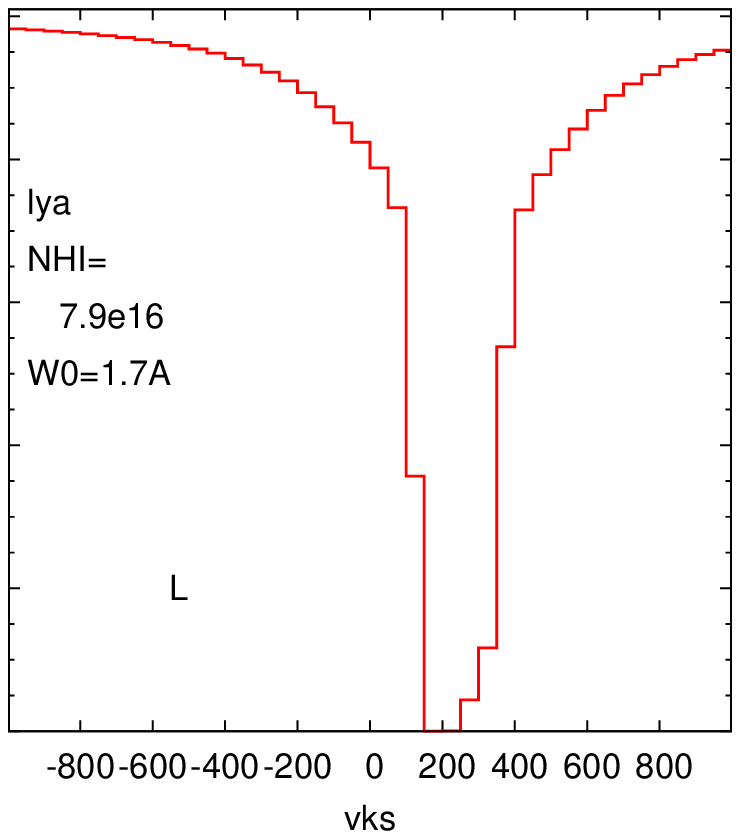}
\end{center}
\caption{Example {\la} absorption line profiles seen from a single direction in
the central source geometry integrated from 0.3 to 1.0 $\Rv$. Positive
velocities are inflowing into the galaxy and negative velocities are out of the
galaxy. The letters correspond to the indicated position of the Hammer
projection from the upper left panel of figure \ref{fig:ewhammer}. Values for
neutral hydrogen column density and {\EW} are quoted in each panel. Uppermost
panel: various modes of inflowing streams, intermediate upper panel: different
examples of outflowing material, intermediate lower panel: low absorption
profiles and lowermost panel: extraordinary cases which are all showing clear
signatures of inflows. Some of the lines are saturated like A, C, K and L
whereas others are not.}
\label{fig:viewang}
\end{figure*}

\begin{figure*}
\begin{center}
\psfrag{relative intensity}[B][B][1][0] {relative intensity}
\psfrag{cg}[Bl][Bl][1][0] {central source}
\psfrag{corrfig}[Bl][Bl][1][0] {see figure \ref{fig:ewhammer}}
\psfrag{vks}[B][B][1][0] {$\Delta w$ [km s$^{-1}$]}
\psfrag{D}[B][B][1][0]{D}
\psfrag{W0=0.064A}[Bl][Bl][1][0] {$W_0 = 0.064$ \AA}
\psfrag{W0=0.055A}[Bl][Bl][1][0] {$W_0 = 0.055$ \AA}
\psfrag{W0=0.10A}[Bl][Bl][1][0] {$W_0 = 0.10$ \AA}
\psfrag{W0=0.056A}[Bl][Bl][1][0] {$W_0 = 0.056$ \AA}
\psfrag{NCII=1.2e14}[Bl][Bl][1][0] {$N_{\rm CII} = 1.2 \times 10^{14}$ cm$^{-2}$}
\psfrag{NSiII=1.5e13}[Bl][Bl][1][0] {$N_{\rm SiII} = 1.5 \times 10^{13}$ cm$^{-2}$}
\psfrag{NMgII=1.0e13}[Bl][Bl][1][0] {$N_{\rm MgII} = 1.0 \times 10^{13}$ cm$^{-2}$}
\psfrag{NFeII=1.2e13}[Bl][Bl][1][0] {$N_{\rm FeII} = 1.2 \times 10^{13}$ cm$^{-2}$}
\psfrag{CII}[Bl][Bl][1][0] {{\CII} (1334 \AA)}
\psfrag{SiIIb}[Bl][Bl][1][0] {{\SiII} (1260 \AA)}
\psfrag{MgIIb}[Bl][Bl][1][0] {{\MgII} (2796 \AA)}
\psfrag{FeIIb}[Bl][Bl][1][0] {{\FeII} (2383 \AA)}
\includegraphics[width=4.94cm]{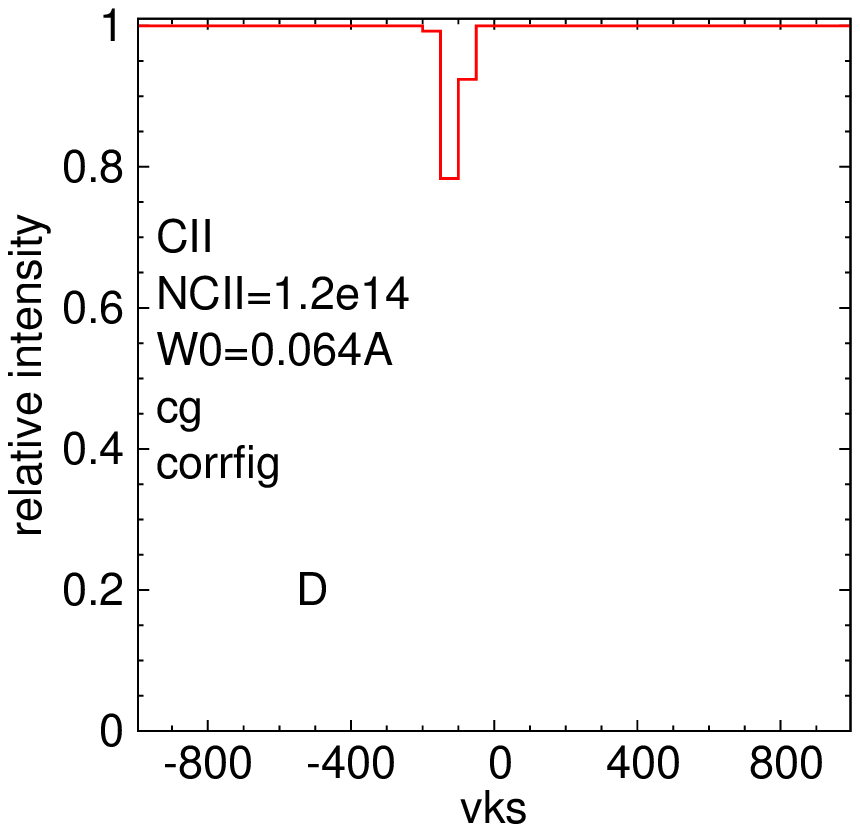}
\includegraphics[width=4.17cm]{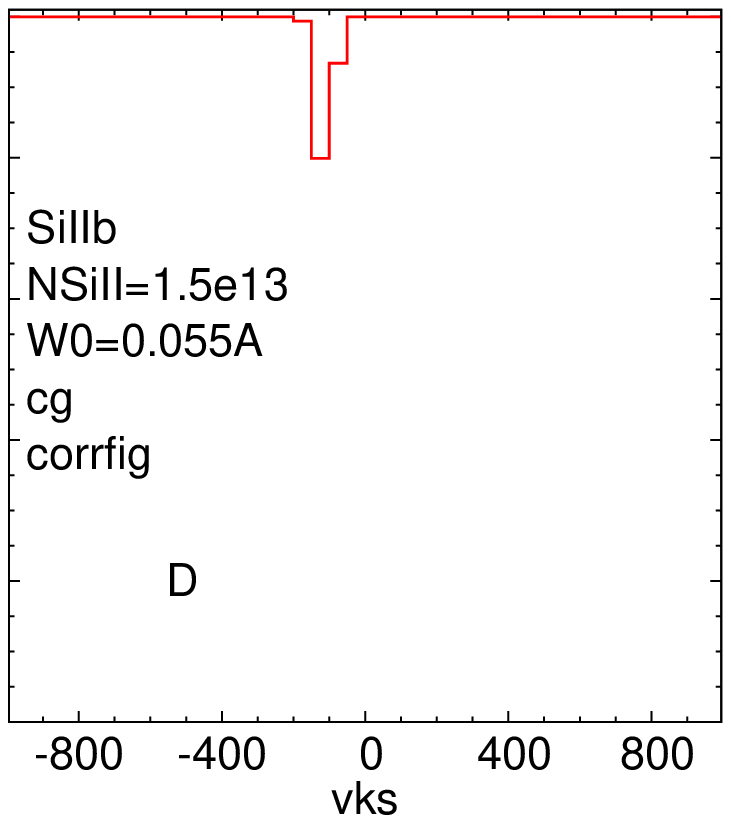}
\includegraphics[width=4.17cm]{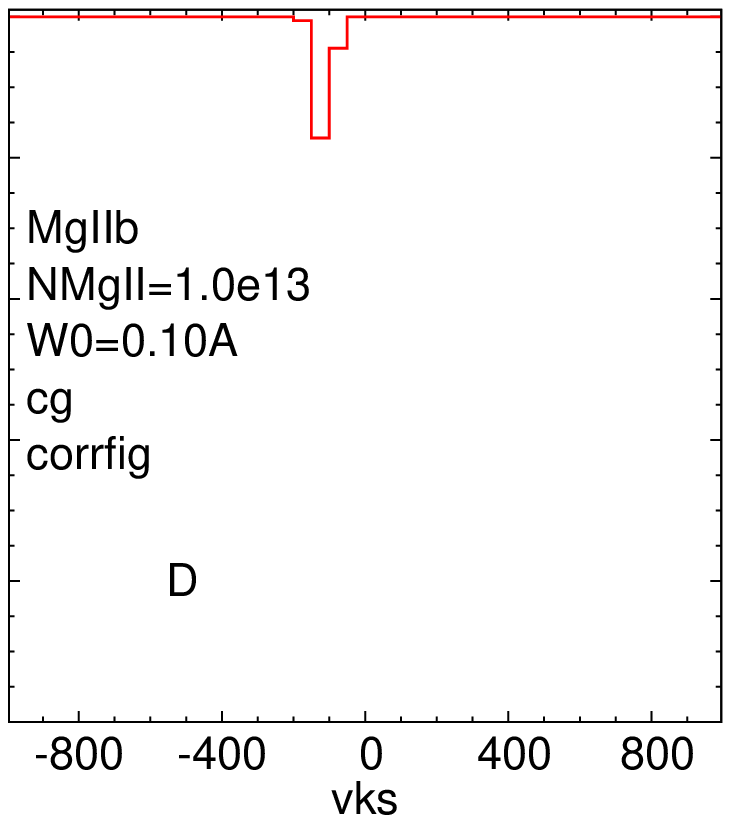}
\includegraphics[width=4.17cm]{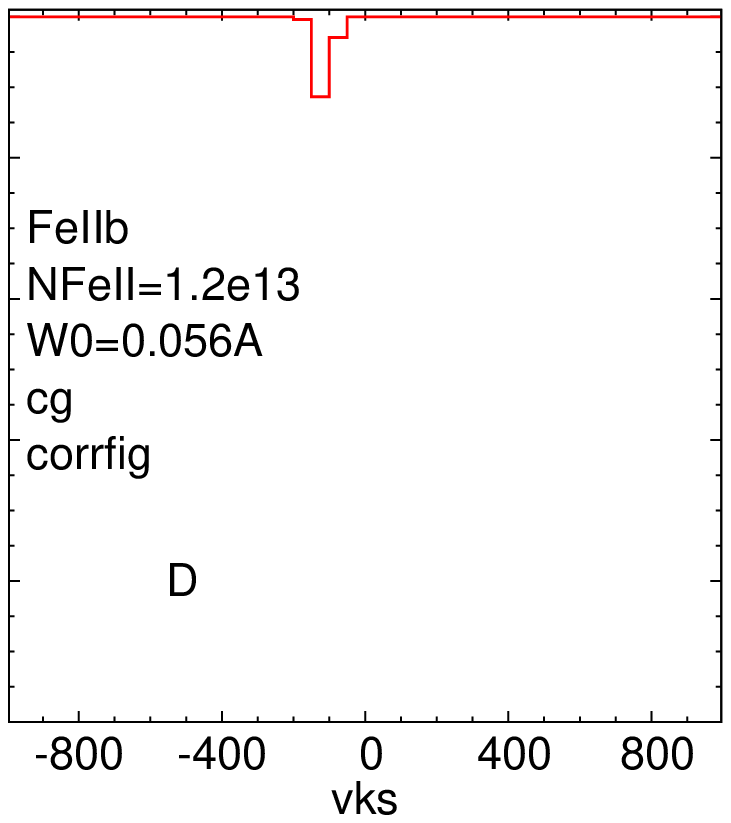}
\end{center}
\caption{Same figure as figure \ref{fig:viewang}, panel D (the panel with the
deepest signal) this time for selected metal lines providing also the deepest
signal, namely: {\CII} (1334 \AA), {\SiII} (1260 \AA), {\MgII} (2796 \AA) and
{\FeII} (2383 \AA). Panel D is fully saturated in {\la} but still shows peak
absorption line depths of 0.2 and {\EW} of $\sim 0.1$ {\AA} for these metal
lines. The {\CII} panel of this plot disagrees with the results of
\citet[][their figure 2, lower panel]{kimm} due to different line computing
algorithms.}
\label{fig:metalviewang}
\end{figure*}

In our \textsc{cloudy} computations we distinguish between self-shielded and
optically thin cells as follows. For optically thin cells, we assume that the
gas is exposed to the full \citet{haardt} UV to X-ray (UVX) metagalactic
background field at the appropriate redshift. We do not include any radiation
from ``local sources". The ionisation state is then controlled by the combined
effects of electron-impact collisional ionisation and photoionisation for the
full spectral range of the background field. For shielded cells, we truncate the
radiation field and include only photons with energies below the Lyman limit.
Hydrogen and other species with ionisation potentials (IP) greater than 13.6 eV
are then produced by collisional ionisation only\footnote{The reader should
note that all our species listed in Table \ref{tab:lines} have IPs greater than
13.6 eV.}. For the shielded cells we neglect photoionisation by penetrating
X-rays. So for pure collisional ionisation in shielded cells the hydrogen
becomes ionised for $T \gtrsim 2 \times 10^4$~K \citep{orly}. However,
``low-ions" with IPs below 13.6 eV continue to be produced by photoionisation
as well. Thus, for a given cell, the ionisation state is determined by its
total hydrogen density $n_{\rm H}$ and temperature $T$, and the adopted
radiation field, which is either the ``full" or ``truncated" \citet{haardt}
field, depending on whether the cell is characterised as shielded or not.

To verify our procedure for distinguishing between shielded and optically thin
cells, we compare our results for the hydrogen ionisation states with the more
detailed radiative transfer computations presented by {\f11}. In figure
\ref{fig:danieltest} we plot the distributions by volume of the neutral
hydrogen fractions $x_{\rm HI}$ and total hydrogen gas densities $n_{\rm H}$ in
the simulation cells, for three different computations of the ionisation
structure, all for the same simulated galaxy. The left-hand panel shows the
neutral fraction versus density distributions for the detailed radiative
transfer computation of {\f11}. This includes the combined effects of the
external UVX and internal sources of photoionising radiation. The middle panel
shows the {\f11} results for UVX only. Cells with densities higher than 0.1
cm$^{-3}$, which usually are close to the disc or the satellites, are not
included in figure \ref{fig:danieltest}. The right-hand panel shows our results
for $x_{\rm HI}$ versus $n_{\rm H}$ using our simple criterion of assuming the
cells are fully shielded for  $n_{\rm H} > n_{{\rm H, shield}} = 0.01$~cm$^{-3}$.
Again, in the post-processing we assume that the hydrogen is collisionally
ionised in the shielded cells, and collisionally plus photoionised in the
optically thin cells. It is apparent that this simplified model provides a
useful crude approximation for the results of the UVX radiative transfer
(middle panel). The only significant disparity occurs near the transition
density of 0.01~cm$^{-3}$ and neutral fraction of $x_{\rm HI}\approx 0.1$, where
the ionisation states appear to be sensitive to the details of the radiative
transfer computations. At the end of section \ref{sec:bg} we will compare our
``final product'' (figures \ref{fig:lalinebg} and \ref{fig:metallinesbg}) to
similar results already published in the literature computed with full
radiative transfer including ionising radiation from local sources (figure 13
of {\f11}). The differences are small, justifying our simplifying model.

\section{Central source}
\label{sec:cg}
In this section we consider the \lya and metal line absorption that occurs as
UV light emitted by the central galaxy is absorbed by gas in the
circum-galactic environment. As defined by {\s10}, the circumgalactic medium is
situated in the spherical zone from just outside the galactic disk to around
the virial radius $\Rv$. Observations of absorption against the central galaxy
itself have the advantage of being able to discriminate between inflows and
outflows because the absorptions may be assumed to occur in foreground material
only. Radiation emitted or scattered from behind the galaxy is blocked by the
galaxy itself. However, such absorptions do not provide spatial information
about the distance from the galaxy centre as they are all by definition at an
impact parameter $b = 0$ from the galaxy centre. {\s10} employed this technique
of observing absorptions against the central galaxy by stacking a sample of 89
galaxies with $z = 2.3 \pm 0.3$ using both rest-frame far-UV and H$\alpha$
spectra, to investigate the kinematics of the gas flows in the circumgalactic
regions. Here we predict the spectral line profiles and equivalent widths (=\EW)
produced in absorption against the central galaxy, for a direct comparison to
the {\s10} observations (as presented in their section 4).

\begin{figure*}
\begin{center}
\includegraphics[width=8.81cm]{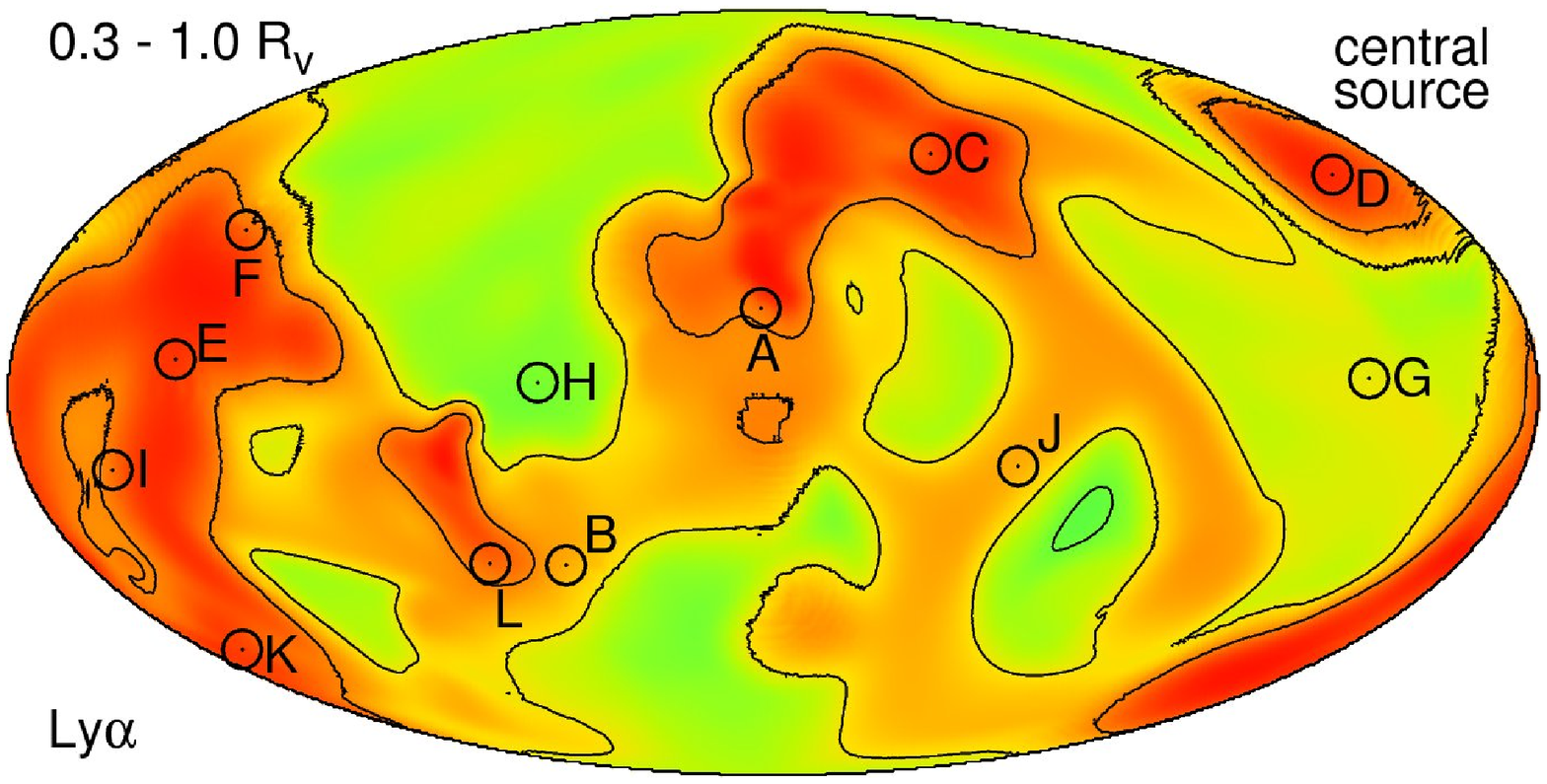}
\includegraphics[width=8.81cm]{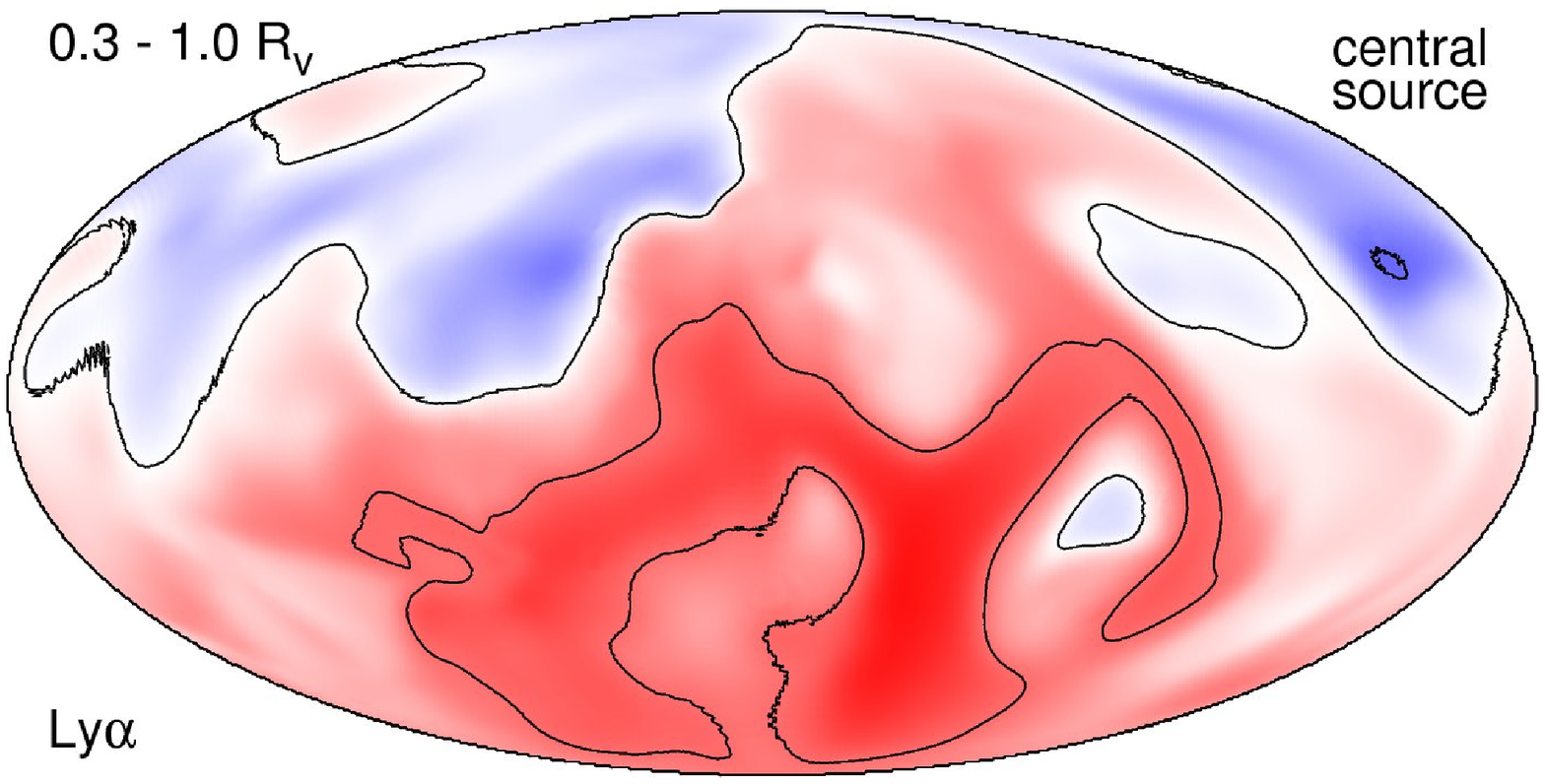}
\includegraphics[width=8.81cm]{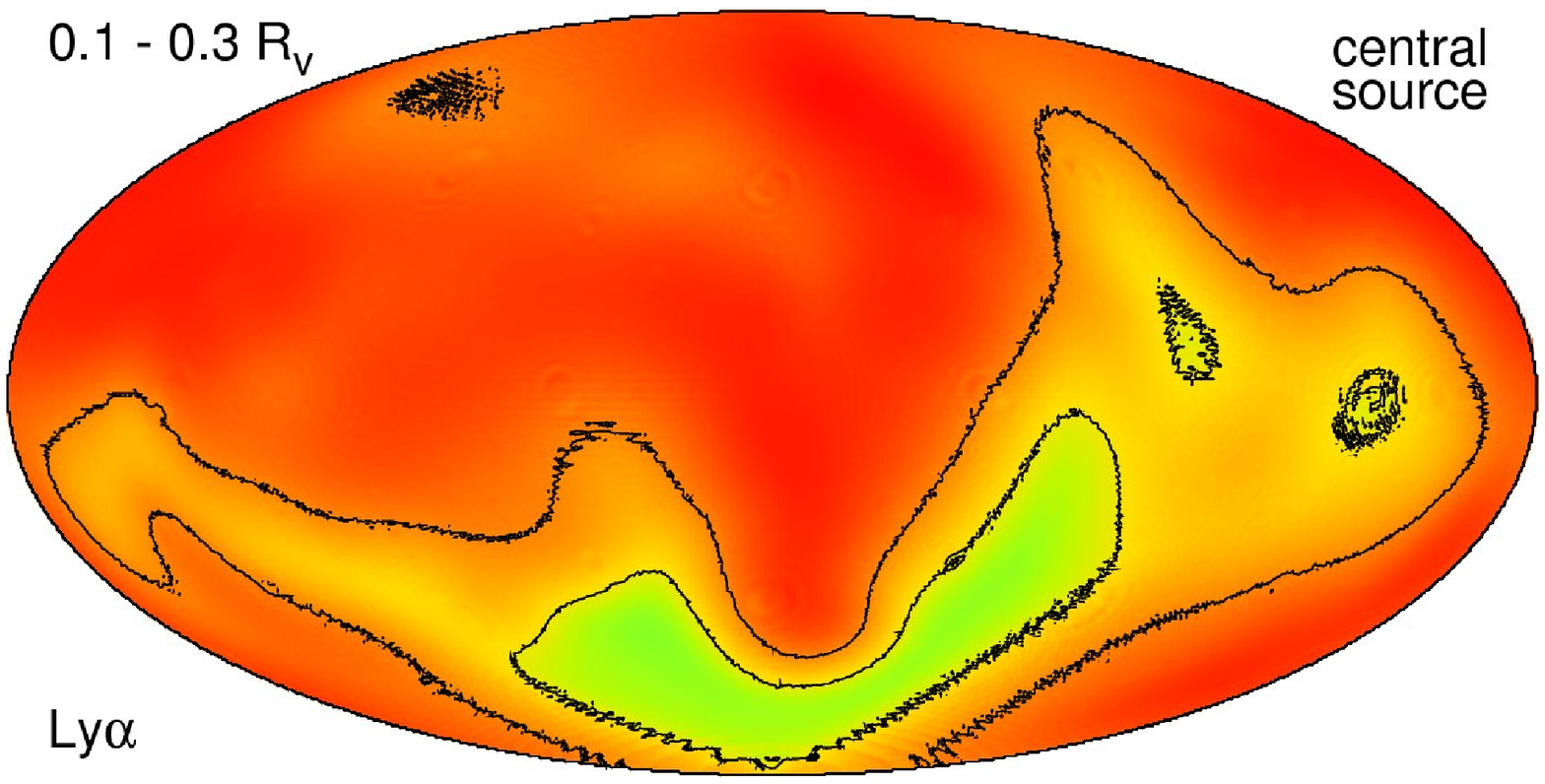}
\includegraphics[width=8.81cm]{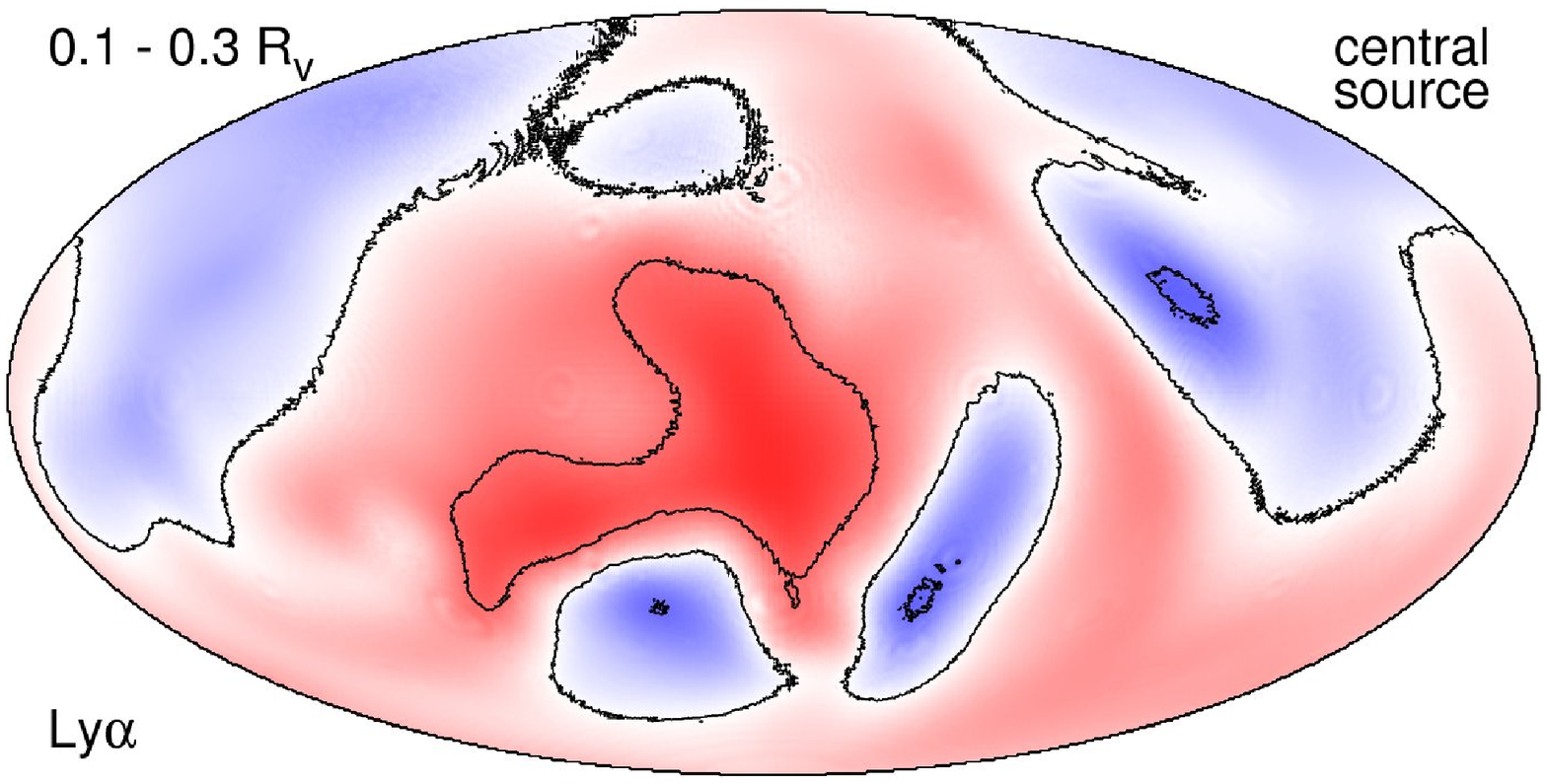}
\includegraphics[width=8.81cm]{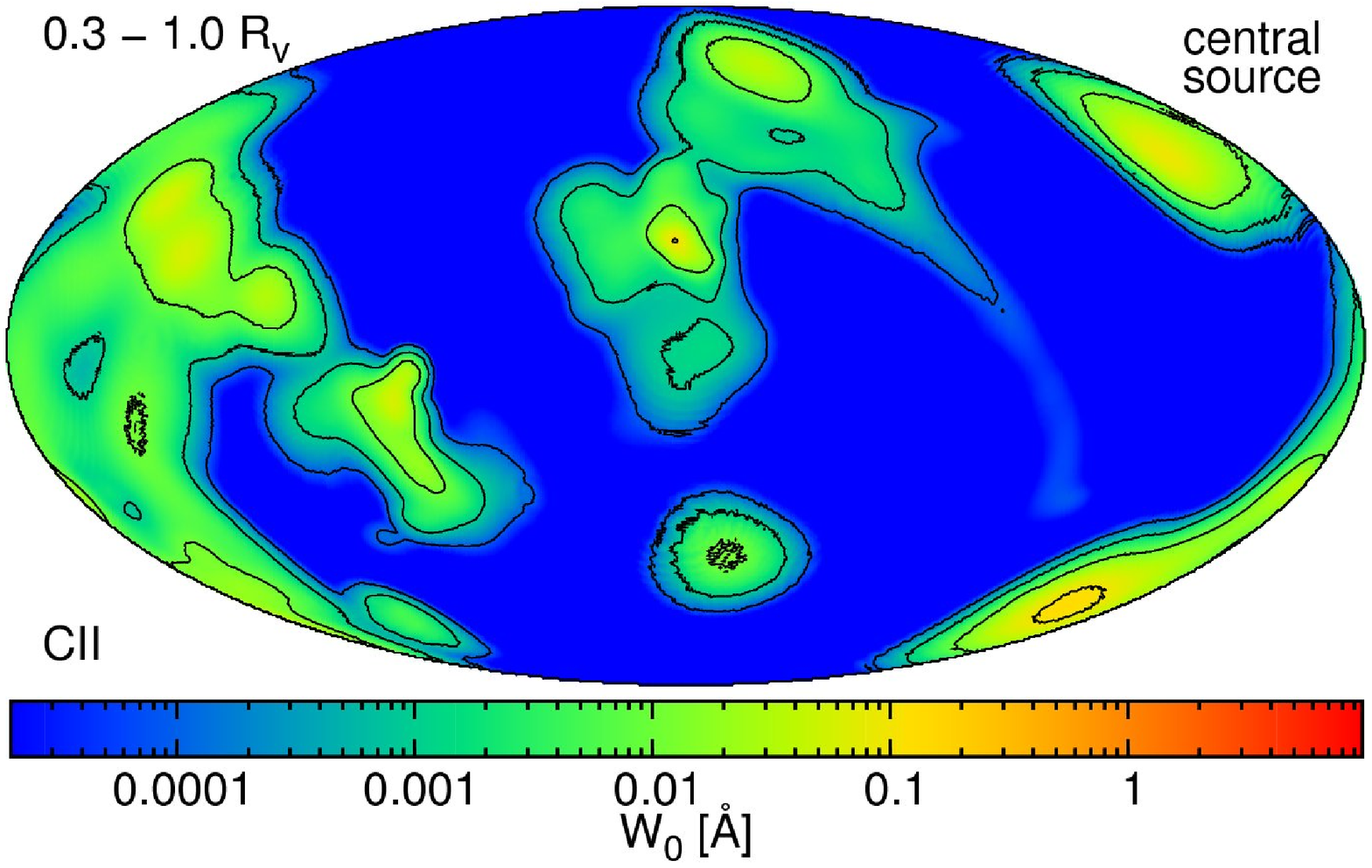}
\includegraphics[width=8.81cm]{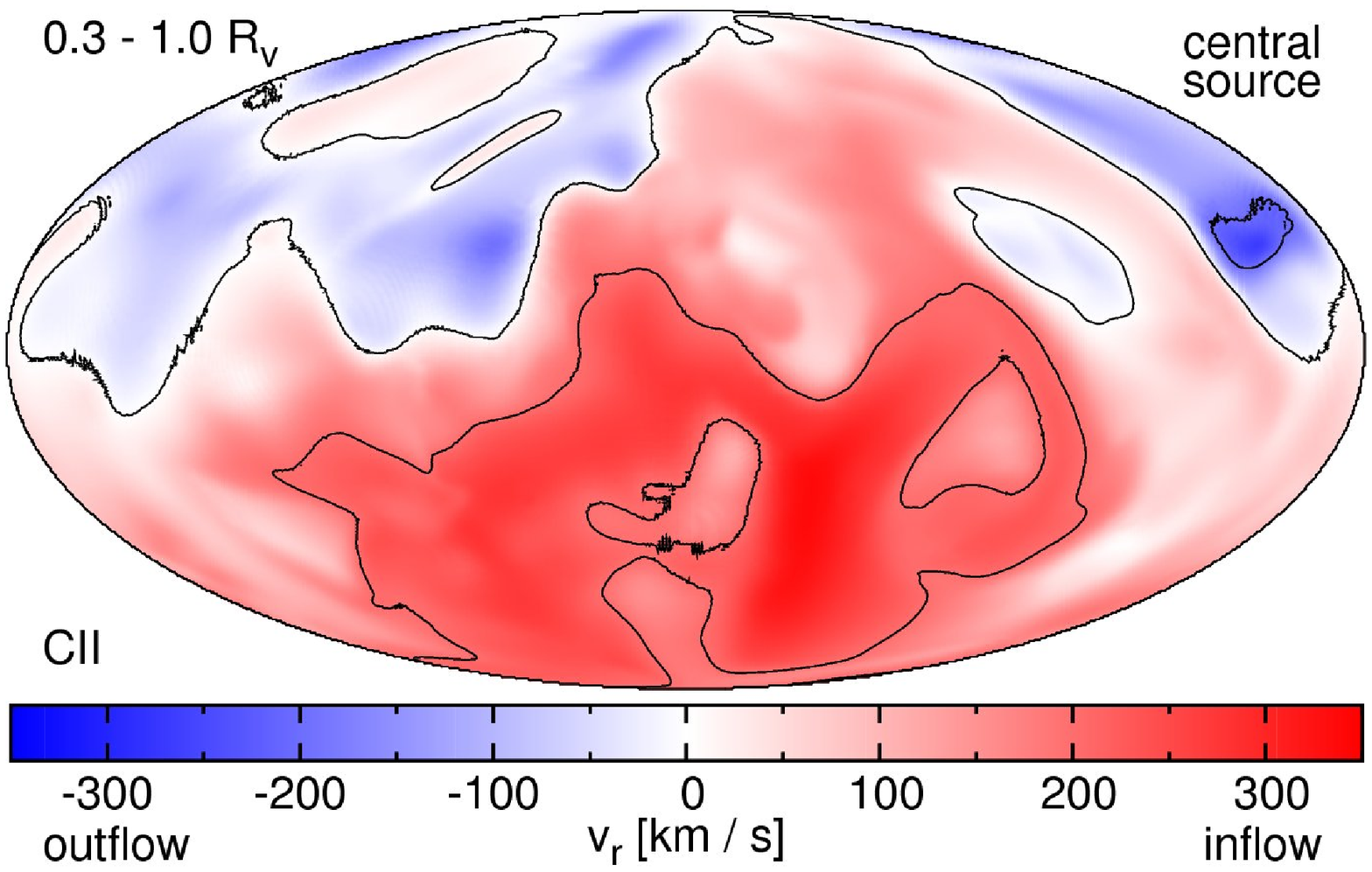}
\end{center}
\caption{Hammer equal area projections of {\EW} (left panels) and of
{\EW}-weighted velocity (right panels) in our fiducial galaxy. The upper four
panels show {\la} whereas the lowermost panels show {\CII}. In the uppermost as
well as in the lowermost panels the data were computed between 0.3 and 1.0
$\Rv$, whereas in the middle panel the data were computed between 0.1 and 0.3
$\Rv$. Positive velocities are inflowing into the galaxy (shown in red) and
negative velocities are out of the galaxy (shown in blue). The contour lines
indicate {\EW}s of 1.0, 0.1, 0.01, 0.001 and 0.0001\,{\AA} or velocities of
+200, 0 and -200 km\,s$^{-1}$, respectively. In the upper left panel the
positions of the example line profiles from figure \ref{fig:viewang} are
indicated by the circles having letters attached. Comparing right and left
panels gives a comprehensive picture of inflows and outflows. The velocity
pattern of {\la} and {\CII} is the same even if their {\EW}s are very
different. In the panels spanning the outer radius range the narrow major
streams are clearly seen as well as the vast majority of the ``stream-free''
sky (as seen from the central galaxy). In the middle panel representing a
narrow radius range very close to the central galaxy this bi-modality is not as
obvious as in the other two cases. The radial velocities show different
behaviour in the two different radius regimes, the {\la} and {\CII} panels of
the same radius regime on the other hand are very similar.}
\label{fig:ewhammer}
\end{figure*}

The absorption-line profiles depend on the column densities along the
line-of-sight. We compute the column densities along radial rays from an
innermost radius $r_{\rm i}$ (usually around the galactic disk) to the virial
radius $\Rv$. The column $N_{\rm Ai}(\phi, \theta)$ of element $A$ in ionisation
state $i$, at angular position $(\phi, \theta)$ is calculated by
\begin{equation}
\label{eq:colden}
N_{\rm Ai}(\phi, \theta) = \int^{\rm \Rv}_{\rm r_i} x_{\rm Ai}(\vec{r}) \
n_{\rm A}(\vec{r}) \ dr,
\end{equation}
where $n_{\rm A}(\vec{r})$ is the total gas density of element $A$ at position
$\vec{r} = (\phi, \theta, r)$ and $x_{\rm Ai}$ is the ionisation fraction. In
figure \ref{fig:hammer} we show two Hammer equal area projections
\citep{snyder} of {\HI} column densities for our fiducial galaxy ($z = 2.3$,
$\Mv = 3.5 \times 10^{11} \msun$, $\Rv = 74$\,kpc). In the upper panel the
column densities are integrated from 0.3 to 1.0 $\Rv$ whereas in the lower
panel they are integrated from 0.1 to 0.3 $\Rv$. Excluding the very inner
sphere is somewhat artificial since in practise absorption will occur from
close to the very centre at $r = 0$. However, these plots are useful in giving
an impression of the distribution of column densities along different
sight-lines. In the upper panel the main stream features are identified and
tagged to the same features shown in figure \ref{fig:denmap} by white open
symbols. The same symbols types in figures \ref{fig:denmap} and
\ref{fig:hammer} indicate identical features in the galaxy but for different
geometries. In figure \ref{fig:hammer} the narrow major streams with column
densities up to $\sim 10^{22}$ cm$^{-2}$ can be seen, as well as the vast
majority of the ``stream-free'' sky (as seen from the central galaxy) having
only column densities of $< 10^{15}$ cm$^{-2}$.

To quantify the column density distribution, we compute the cumulative fraction
$f_{\rm c}$ of the sky (as viewed from the central galaxy) that is covered with
column densities greater than $N_{\rm Ai}$. The covering fractions for hydrogen
and the various metal ions are plotted in figures \ref{fig:laskycover} and
\ref{fig:skycover}. These plots include inflowing and outflowing gas. The data
in the blue dashed line of figure \ref{fig:laskycover} as well as in the right
panel of figure \ref{fig:skycover} is averaged over our three simulated
galaxies at $z=2.3$ with $\Mv = 3.5 \times 10^{11}\msun$. For better
comparability with future observations we also use a second set of data
consisting of two galaxies at $z=1.38$ with $\Mv = 1.4 \times 10^{12}\msun$
(red, solid line). The plot shows that the whole sky, as seen from the central
galaxy, has a minimum $N_{\rm HI}$ of $10^{12}$ cm$^{-2}$ and above this value
$f_{\rm c}$ decreases strongly from 1.0 to 0.3 at $10^{14}$ cm$^{-2}$. The {\HI}
covering fraction then decreases more slowly down to $\sim 1\%$ at $10^{21}$
cm$^{-2}$. The $f_{\rm c}$ for the metals are of course much lower. The highest
metal columns are usually around $10^{15}$ cm$^{-2}$ and at $10^9$ cm$^{-2}$ the
covering fractions vary between 0.10 (\FeII) and 0.75 (\CIV). This verifies our
initial qualitative impression that a full sky map of column densities exhibits
a few very high peaks corresponding to the streams extending only over a fairly
tiny solid angle and a majority of very low column density solid angle
corresponding to the space without streams. This effect is even more impressive
for the low redshift panels. Also there is a decrease of the covering fraction
with decreasing redshift.

Given the computed column densities we can construct simulated absorption line
profiles. We start with a single line of sight (i.e. a single pixel in figure
\ref{fig:hammer}). Along a given sight-line the gas has a varying density,
temperature and radial velocity as function of radial position $r$ from the
central galaxy. The convolutions of the different densities and radial
velocities in the gas along the line of sight are the main ingredients to
compute an absorption line profile. This is done as follows: the radial
velocity offset $\Delta w$ relative to central source at rest of all the gas is
measured. We assume Voigt profiles with a thermal Doppler broadening parameter
\begin{equation}
b = \sqrt{2\,k \ T \over m_{\rm A}},
\label{eq:doppler}
\end{equation}
where k is the Boltzmann constant, $T$ is the temperature of the gas and
$m_{\rm A}$ is the mass of the element $A$. So for angular position $(\phi,
\theta)$ we can compute the optical depth $\tau_\nu (\phi, \theta, \Delta w)$
at the velocity offset $\Delta w$ as \citep[cf. equation 10.24 of][]{erika}
\begin{eqnarray}
\tau_\nu(\phi, \theta, \Delta w) & = & {\sqrt{\pi} \ e^2 \ f_\lambda \ \lambda_0
\over m_{\rm e} \ c} \int_{\rm r_i}^{\Rv} {n_{\rm A}(\vec{r}) \ x_{\rm Ai}(\vec{r})
\over b(\vec{r})} \nonumber \\ & \times & H \left[{\gamma_\lambda \ \lambda_0
\over 4\, \pi \ b(\vec{r})}\ , {\Delta w - v(\vec{r}) \over b(\vec{r})}\right]
\ dr,
\label{eq:freoff}
\end{eqnarray}
where $e$ is the electron charge, $m_{\rm e}$ is the electron mass, $c$ is the
speed of light, $\lambda_0$ is the transition wavelength, $n_{\rm A}(\vec{r})$
is the gas density of element $A$ at position $\vec{r} = (\phi, \theta,r)$,
$x_{Ai}$ is the ionisation fraction of element $A$ in state $i$, $v$ is the
velocity of the gas, $f_\lambda$ is the oscillator strength of the absorption
line and $\gamma_\lambda$ is the sum over the spontaneous emission coefficients
or the damping width. We took these $f_\lambda$ and $\gamma_\lambda$ values from
\citet{morton}. The Voigt profile is 
\begin{equation}
H(a,u) = {a \over \pi} \int^\infty_{-\infty} {\exp (-y^2)\over(u - y)^2 + a^2} \
dy,
\label{eq:voigt}
\end{equation}
where $a$ is the ratio of the damping width to the Doppler width and $u$ is the 
offset from line centre in units of Doppler widths. Knowing $\tau_\nu(\Delta
w)$ it is now possible to compute an absorption line profile $I(\Delta w)$ for
a given direction which is simply the function
\begin{equation}
I(\Delta w) = {\rm exp}[-\tau_\nu(\Delta w)].
\label{eq:ablipro}
\end{equation}
For a fair comparison to the observations done by {\s10}, we mimic a Gaussian
point spread function. This will be done from now on up to the end of this
section. It is in contrast to figures \ref{fig:hammer}, \ref{fig:laskycover} or
\ref{fig:skycover} where no point spread function was applied. It has a
beam-size ($=$ FWHM $\eta$ of the Gaussian) of 4\,kpc. It is done by splitting
up a cylinder with a radius of three times the beam-size into as many parallel
fibres as the resolution permits, determining the absorption line profile for
every individual fibre and then computing a Gaussian weighted average
absorption line profile from all fibres. Also from now on up to the end of this
section (except for figures \ref{fig:statlaline} and \ref{fig:statmetallines})
we degrade our velocity space resolution down to 50 km/s, to match the velocity
resolution used in the corresponding figures of \s10.

\begin{figure}
\begin{center}
\psfrag{sky covering fraction}[B][B][1][0] {$f_{\rm c}(>W_0)$}
\psfrag{EW}[B][B][1][0] {$W_0$ [\AA]}
\psfrag{cg}[Bl][Bl][1][0] {central source}
\psfrag{La}[Bl][Bl][1][0] {\la}
\psfrag{z23m35e11}[Br][Br][1][0] {\textcolor{red}{$z = 2.3$, $\Mv = 3.5 \times
10^{11}$ M$_\odot$}}
\psfrag{z14m14e12}[Br][Br][1][0] {\textcolor{blue}{$z = 1.4$, $\Mv = 1.4 \times 
10^{12}$ M$_\odot$}}
\includegraphics[width=8.45cm]{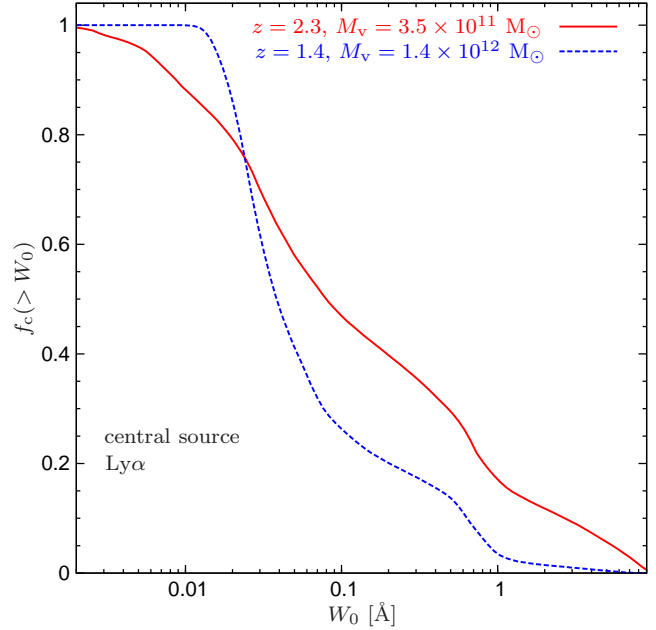}
\end{center}
\caption{Cumulative {\EW} sky covering fractions $f_{\rm c}$ higher than $W_0$
for a central source integrated from 0.3 to 1.0 $\Rv$ for {\la}. Please note
the different x-axis scaling with respect to figure \ref{fig:EWcovfrac}. In
solid red we show the average of our three simulated galaxies (resolution
$70\pc$, $z=2.3$, $\Mv = 3.5 \times 10^{11}\msun$, $\Rv = 74\,$kpc). Whereas in
dashed blue we show the average of two simulated galaxies at $z=1.38$ ($\Mv =
1.38 \times 10^{12}\msun$, $\Rv = 150\,$kpc). At higher redshift there is a
higher covering fraction of high {\EW} visible, implying that there are more
pronounced streams present.}
\label{fig:laEWcovfrac}
\end{figure}

A selection of the large variety of different possible absorption line profiles
depending on the respective viewing angles are shown in figures
\ref{fig:viewang} (\la) and \ref{fig:metalviewang} (selected metal lines). One
can see various modes of inflowing streams, different examples of outflowing
material as well as very shallow absorption line profiles corresponding to a
line of sight with very low column density. In {\la} some of the lines are
saturated like A, C, K and L, whereas the others are not. Panel D which shows
the strongest lines is fully saturated in {\la} but still shows peak absorption
line depths of 0.2 and {\EW} of $\sim 0.1$ {\AA} for selected metal lines. The
reader should note that some of the weaker lines (panels G or H for example)
might be completely erased by noise and the ISM component at $v = 0$. 

\begin{figure*}
\begin{center}
\psfrag{sky covering fraction}[B][B][1][0] {$f_{\rm c}(>W_0)$}
\psfrag{EW}[B][B][1][0] {$W_0$ [\AA]}
\psfrag{cg}[Bl][Bl][1][0] {central source}
\psfrag{z23}[Bl][Bl][1][0] {$z = 2.3$}
\psfrag{z14}[Bl][Bl][1][0] {$z = 1.4$}
\psfrag{m35e11}[Bl][Bl][1][0] {$\Mv= 3.5 \times 10^{11}$ M$_\odot$}
\psfrag{m14e12}[Bl][Bl][1][0] {$\Mv= 1.4 \times 10^{12}$ M$_\odot$}
\psfrag{CII}[Br][Br][1][0] {\textcolor{red}{{\CII} (1334 \AA)}}
\psfrag{CIVa}[Br][Br][1][0] {\textcolor{cyan}{{\CIV} (1548 \AA)}}
\psfrag{OI}[Br][Br][1][0] {\textcolor{green}{{\OI} (1302 \AA)}}
\psfrag{SiIIa}[Br][Br][1][0] {\textcolor{blue}{{\SiII} (1526 \AA)}}
\psfrag{SiIVa}[Br][Br][1][0] {\textcolor{black}{{\SiIV} (1393 \AA)}}
\psfrag{CIVb}[Br][Br][1][0] {\textcolor{red}{{\CIV} (1550 \AA)}}
\psfrag{SiIIb}[Br][Br][1][0] {\textcolor{magenta}{{\SiII} (1260 \AA)}}
\psfrag{SiIIc}[Br][Br][1][0] {\textcolor{cyan}{{\SiII} (1304 \AA)}}
\psfrag{SiIVb}[Br][Br][1][0] {\textcolor{green}{{\SiIV} (1402 \AA)}}
\psfrag{MgIIa}[Br][Br][1][0] {\textcolor{red}{{\MgII} (2803 \AA)}}
\psfrag{MgIIb}[Br][Br][1][0] {\textcolor{green}{{\MgII} (2796 \AA)}}
\psfrag{FeIIa}[Br][Br][1][0] {\textcolor{blue}{{\FeII} (2600 \AA)}}
\psfrag{FeIIb}[Br][Br][1][0] {\textcolor{magenta}{{\FeII} (2383 \AA)}}
\psfrag{FeIIc}[Br][Br][1][0] {\textcolor{cyan}{{\FeII} (2344 \AA)}}
\psfrag{FeIId}[Br][Br][1][0] {\textcolor{black}{{\FeII} (2587 \AA)}}
\psfrag{FeIIe}[Br][Br][1][0] {\textcolor{Orange}{{\FeII} (2374 \AA)}}
\includegraphics[width=6.43cm]{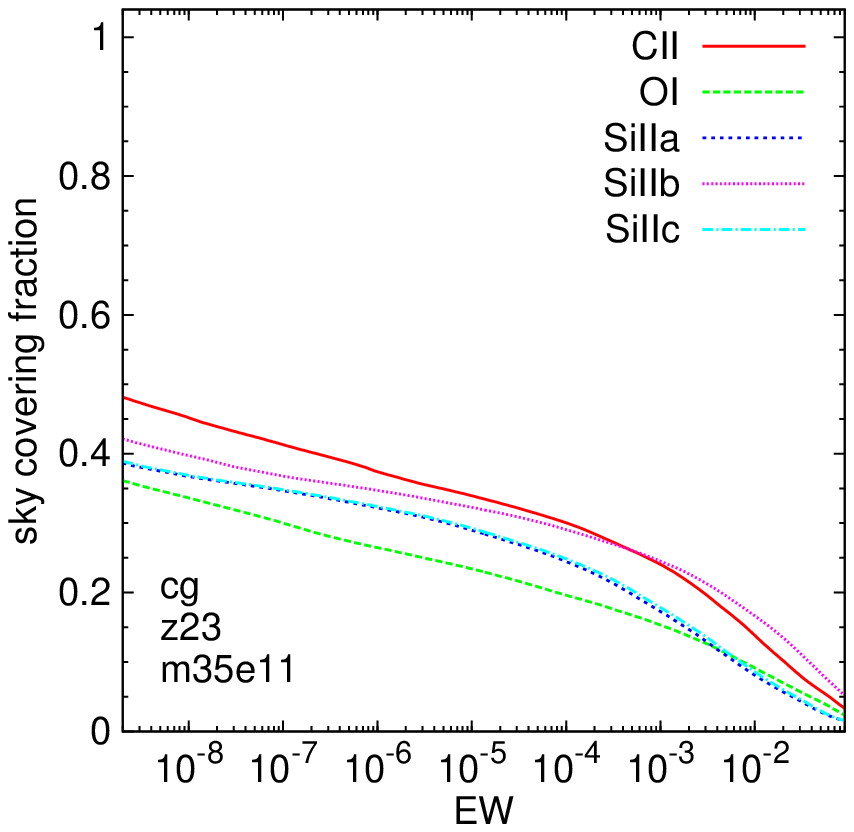}
\includegraphics[width=5.55cm]{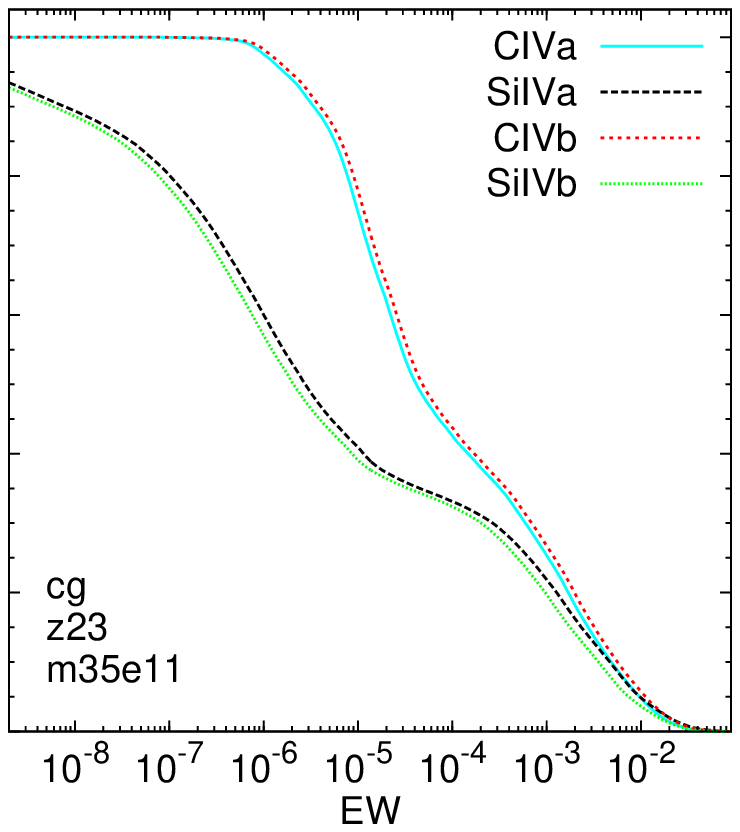}
\includegraphics[width=5.55cm]{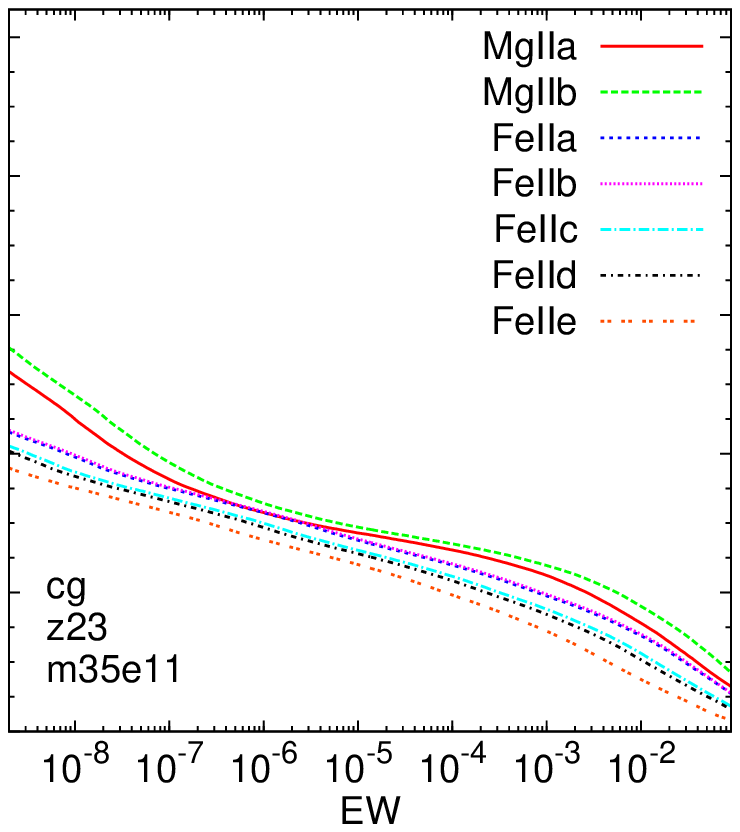}
\includegraphics[width=6.43cm]{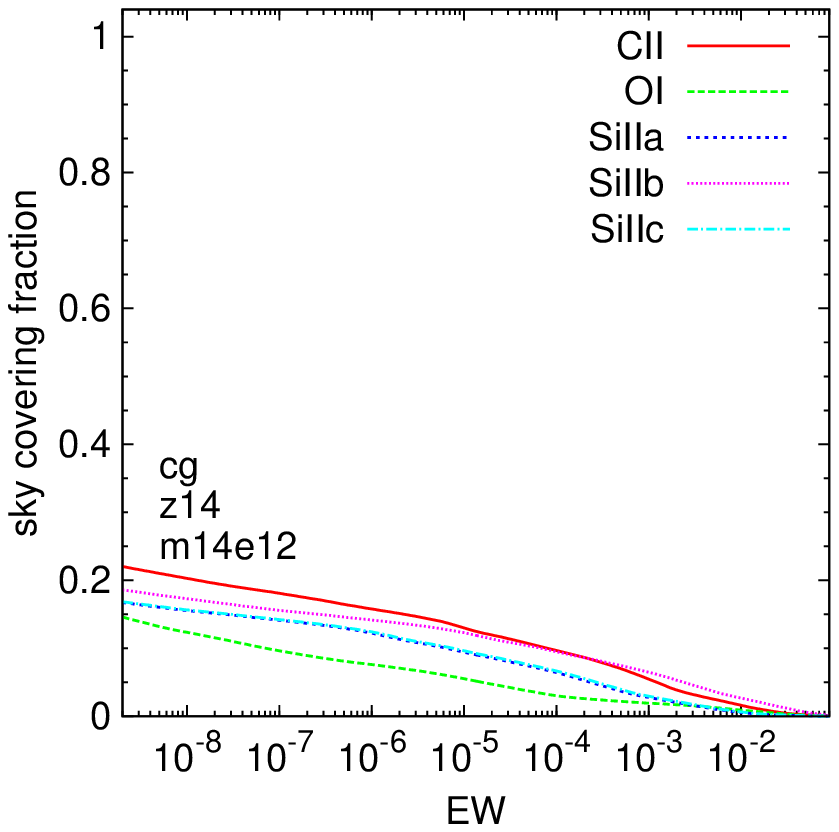}
\includegraphics[width=5.55cm]{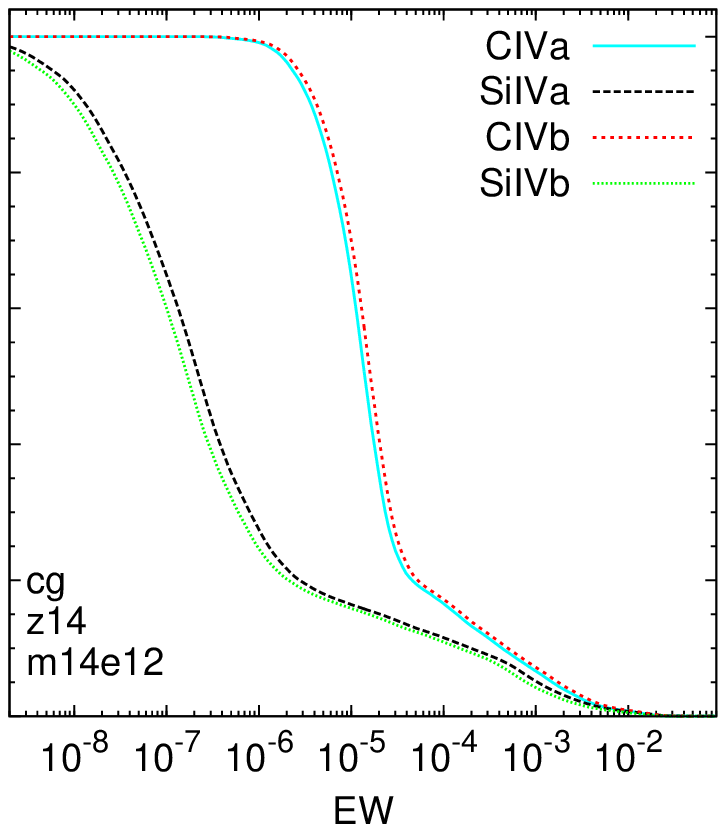}
\includegraphics[width=5.55cm]{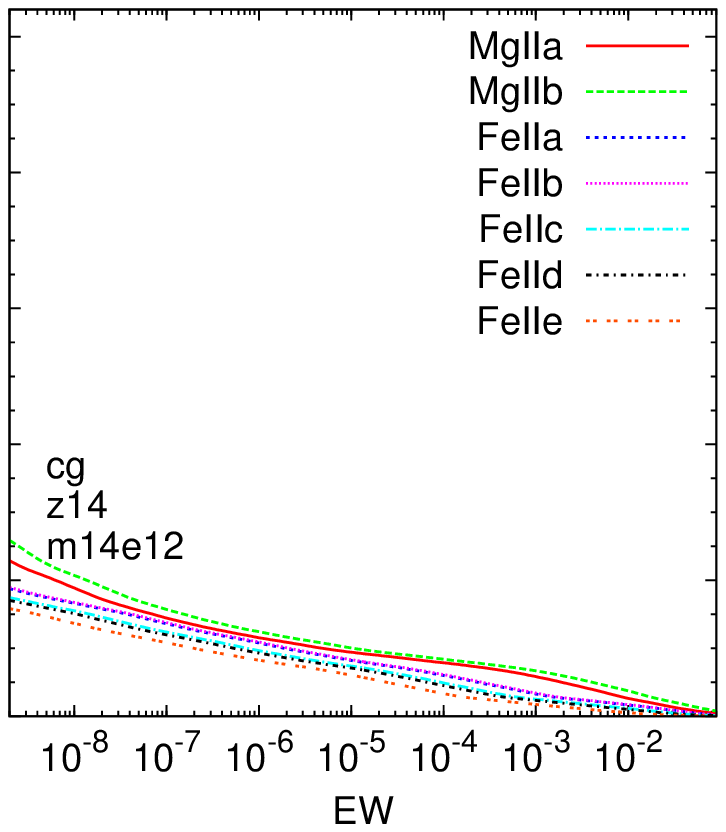}
\end{center}
\caption{Same cumulative {\EW} sky covering fractions $f_{\rm c}$ as in figure
\ref{fig:laEWcovfrac} this time for the metal lines. Note the different x-axis
scaling with respect to figure \ref{fig:laEWcovfrac}. In the upper panels we
show our three simulated galaxies (resolution $70\pc$, $z=2.3$, $\Mv =
3.5 \times 10^{11}\msun$, $\Rv = 74\,$kpc). Whereas in the lower panels we
use two simulated galaxies at $z=1.38$ with $\Mv = 1.38 \times 10^{12}\msun$
and $\Rv = 150\,$kpc. The {\CIV} and {\SiIV} lines in the middle panels clearly
stand out. Those lines can be produced in hot medium whereas all the others can
only be produced in cold gas only. So {\CIV} and {\SiIV} reach covering
fractions as high as unity. The low ionised metal lines (left and right panel)
have higher peak values of {\EW} by up to one order of magnitude.}
\label{fig:EWcovfrac}
\end{figure*}

The leftmost panel of figure \ref{fig:metalviewang} is comparable to the lower
panel of figure 3 in \citet{kimm}. They get for {\CII} a maximum line depth
(MLD) of $\sim 0.9$ with a FWHM $\eta$ of 650 km s$^{-1}$. The differences stem
from the fact that they used a lower resolution simulation, the Horizon
MareNostrum simulation \citep{ocvirk}. Further on they used higher mass haloes
($\Mv > 10^{12}$ M$_\odot$) and a very different prescription for the Gaussian
velocity distribution. They obtain their dispersion from the neighbouring 26
cells instead of the algorithm from equations (\ref{eq:doppler}) to
(\ref{eq:voigt}). Some statistical properties of the ensemble of all possible
example line profiles will be shown later.

Each of these absorption line profiles has a certain total absorption which can
be expressed as a single number, the "rest equivalent width" (\EW). This is
usually given in wavelength units and it is defined as:
\begin{eqnarray}
W_0 & = & {\lambda_0 \over c} \int_{-\infty}^\infty \left\{1 -
{\rm exp}[-\tau_\nu (w)]\right\} \,dw \nonumber \\ & = &
{\lambda_0 \over c} \int_{-\infty}^\infty \left[1 - I(w)\right] \, dw .
\label{eq:EW}
\end{eqnarray}
An {\EW} can now be calculated for every angular position on the sky. With these
numbers we can plot a whole full sky map of {\EW}. This is done in the left
panels of figure \ref{fig:ewhammer}. There we show the resulting Hammer
projections of {\EW} for our fiducial galaxy. The upper two panels show {\la}
whereas the lowermost panel shows {\CII}. In the uppermost and lowermost panels
the {\EW}s were computed for the gas in the radius range between 0.3 and 1.0
$\Rv$, whereas in the middle panels the {\EW}s were determined for gas in a
somewhat narrower and innermore radius range between 0.1 and 0.3 $\Rv$. In the
upper left panel the positions of the example line profiles from figure
\ref{fig:viewang} are indicated by the circles with the letters attached to
them. In the panels spanning the outer radius range one can again clearly see
the narrow major streams having {\EW}s of $> 1${\AA} (\la) or $> 0.01${\AA}
(\CII). One can also see the vast majority of the ``stream-free'' sky (as seen
from the central galaxy) with {\EW}s of $\sim 0.01${\AA} (\la) or $<
10^{-5}$\,{\AA} (\CII). In the middle panel representing a narrow radius range
very close to the central galaxy $(0.3 \Rv > r > 0.1 \Rv)$ this bi-modality is
by far not as obvious as in the other two cases. This is due to the fact that
at this distance from the galaxy centre the inflowing streams slowly dissipate
into a region with more complex gas motions (``messy region''). Since the
{\EW}s are significantly lower for the {\CII} line and even lower for the other
lines, as we will see later, we do not show Hammer projections of {\EW} of
other metal lines than {\CII}. We do show Hammer projections of {\EW}-weighted
velocity in the right panels of this figure. Positive velocities are inflowing
into the galaxy (shown in red) and negative velocities are out of the galaxy
(shown in blue). The contour lines indicate velocities of +200, 0 and -200
km\,s$^{-1}$. Regions with very high velocity outflows have a low {\EW}, while
velocities close to systemic velocity are at very high {\EW}. This agrees very
well with the high {\EW} seen by {\s10} at $v_{\rm r} = 0\,$km\,s$^{-1}$. The
velocity pattern of {\la} and {\CII} is the same even if their {\EW}s are so
different. Comparing the upper two panels one can see that the distribution of
velocities in the two different radius regimes varies. The term ``outflow''
can be misleading here. In our simulations some of the material that is
receding from the centre of the galaxy is not necessarily diffuse gas driven by
some kind of feedback processes. It could also be a clump or satellite galaxy
which is merging with the central galaxy. It just had its first flyby with the
central object and is therefore now ``outflowing''.

\begin{figure}
\begin{center}
\includegraphics[width=8.45cm]{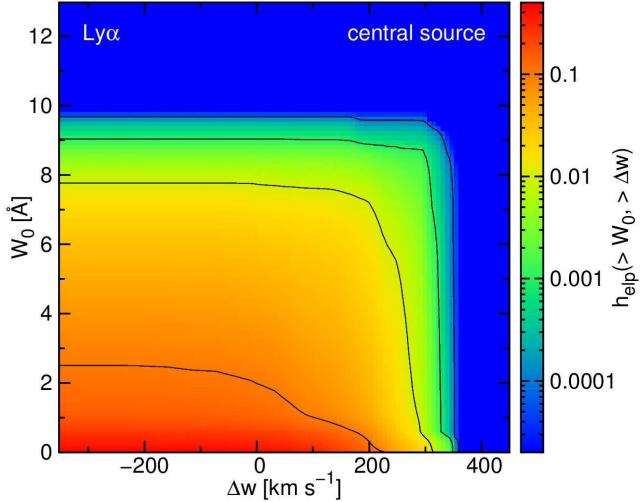}
\end{center}
\caption{Fraction $h_{\rm elp}$ of all possible example line profiles in the
central geometry whose {\EW} is higher than $W_0$ and whose line centre is at a
velocity offset indicating an inflow at least as fast as $\Delta w$. Here for
the {\la} line. The example line profiles are integrated from $1.0\,\Rv$ down
to $0.3\,\Rv$. Positive velocities are inflowing into the galaxy and negative
velocities are out of the galaxy. Contour lines are at 0.1, 0.01, 0.001 and
0.0001 respectively. Note the different y-axis scalings with respect to figure
\ref{fig:statmetallines}. This figure shows that the probability of detecting
an inflow which is flowing in with at least 150\,km\,s$^{-1}$ with a signal of
at least 3.9\,{\AA} in a single observation of a single galaxy without
averaging is $\sim 4\,\%$.}
\label{fig:statlaline}
\end{figure}

To present the statistics from these maps we determine the sky covering
fractions. In figures \ref{fig:laEWcovfrac} and \ref{fig:EWcovfrac} we show the
fraction $f_{\rm c}$ of the sky as seen from the central galaxy that is covered
with absorption line profiles which have {\EW} higher than $W_0$. Here we
integrate in the central source geometry from 0.3 to 1.0 $\Rv$. In figure
\ref{fig:laEWcovfrac} this is done for {\la} whereas in figure
\ref{fig:EWcovfrac} this is done for the metal lines (note the different x-axis
scaling in both figures). We use our three simulated galaxies (resolution
$70\pc$) at $z = 2.3$ with $\Mv = 3.5 \times 10^{11}\msun$ and $\Rv = 74\,$kpc.
These are shown in solid red in figure \ref{fig:laEWcovfrac} and on the upper
panels of figure \ref{fig:EWcovfrac}. Again for better comparability with
future observations we also use a second set of data consisting of two galaxies
at $z=1.38$ with $\Mv = 1.4 \times 10^{12}\msun$ and $\Rv = 150\,$kpc. These
are shown in dashed blue in figure \ref{fig:laEWcovfrac} and on the upper
panels of figure \ref{fig:EWcovfrac}. We concentrate on the lines studied in
great detail in {\s10} which are: {\la} (1216 \AA), {\CII} (1334\,\AA), {\OI}
(1302 \AA), {\SiII} (1260 {\AA}, 1304\,{\AA} and 1526 {\AA}), {\CIV}
(1548\,{\AA} and 1550 \AA) as well as {\SiIV} (1393 {\AA} and 1402 \AA). 
In addition to these ten lines observed by {\s10} we also use here seven new
ones: {\MgII} (2803 {\AA} and 2796 \AA) and {\FeII} (2600 {\AA}, 2383 {\AA},
2344 {\AA}, 2587 {\AA} and 2374 {\AA}). These lines have not been observed in
this geometry or redshift range yet, but may be very suitable for future
observational programmes at $z \sim 1$. For {\la} one can see that at higher
redshift there is a higher covering fraction of large {\EW}, implying that the
streams are much more pronounced. The {\CIV} and {\SiIV} lines in the middle
panels of figure \ref{fig:EWcovfrac} clearly stand out since they reach
covering fractions as high as unity but for low column densities only. This is
because these lines can be produced in the hot and diffuse medium whereas all
the others can only be produced in cold gas only. Another interesting feature
for the metal lines is that low ionised metal lines have higher peak values of
{\EW} by up to one order of magnitude, but low overall covering fraction. This
is because low-ionised ions are more abundant in high-density, cold regions
with low covering fraction, like streams.

As already mentioned earlier we want to show statistics of how many absorption
lines show signatures of inflow versus how many absorption lines show
signatures of outflow and how strong are those. This is done in figures
\ref{fig:statlaline} and \ref{fig:statmetallines}. These figures are intended
to give the likelihood of a detection of a cold stream while looking at a
single galaxy from a single direction without averaging. They show the fraction
$h_{\rm elp}$ of all possible example line profiles in the central geometry
whose {\EW} is higher than $W_0$ and whose line centre is at a velocity offset
indicating an inflow at least as fast as $\Delta w$. The example line profiles
are integrated from $1.0\,\Rv$ down to $0.3\,\Rv$. Positive velocities are
inflowing into the galaxy and negative velocities are out of the galaxy. In
these two figures no velocity degrading is used. The {\la} panel shows that the
probability of detecting an inflow which is flowing in with at least
150\,km\,s$^{-1}$ with a signal of at least 3.9\,{\AA} in a single observation
of a single galaxy without averaging is $\sim 4\,\%$. In {\CII} one sees an
inflow $> 150$\,km\,s$^{-1}$ with an {\EW} $>$ 0.17\,{\AA} in 0.4\,\% of all
observations. For {\MgII} one sees an inflow $ > 150$\,km\,s$^{-1}$ with an
{\EW} $>$ 0.2\,{\AA} in 1.3\,\% of all observations. These values should be
achievable by future observations. Since it is the likely scenario that the
inflowing streams and the wide-angle outflowing gas avoid each other to a great
extent the streams as measured in our simulations where the outflows are a
factor of 2-3 weaker than in the extreme observed cases can be regarded as a
sensible approximation.

\begin{figure*}
\begin{center}
\includegraphics[width=6.00cm]{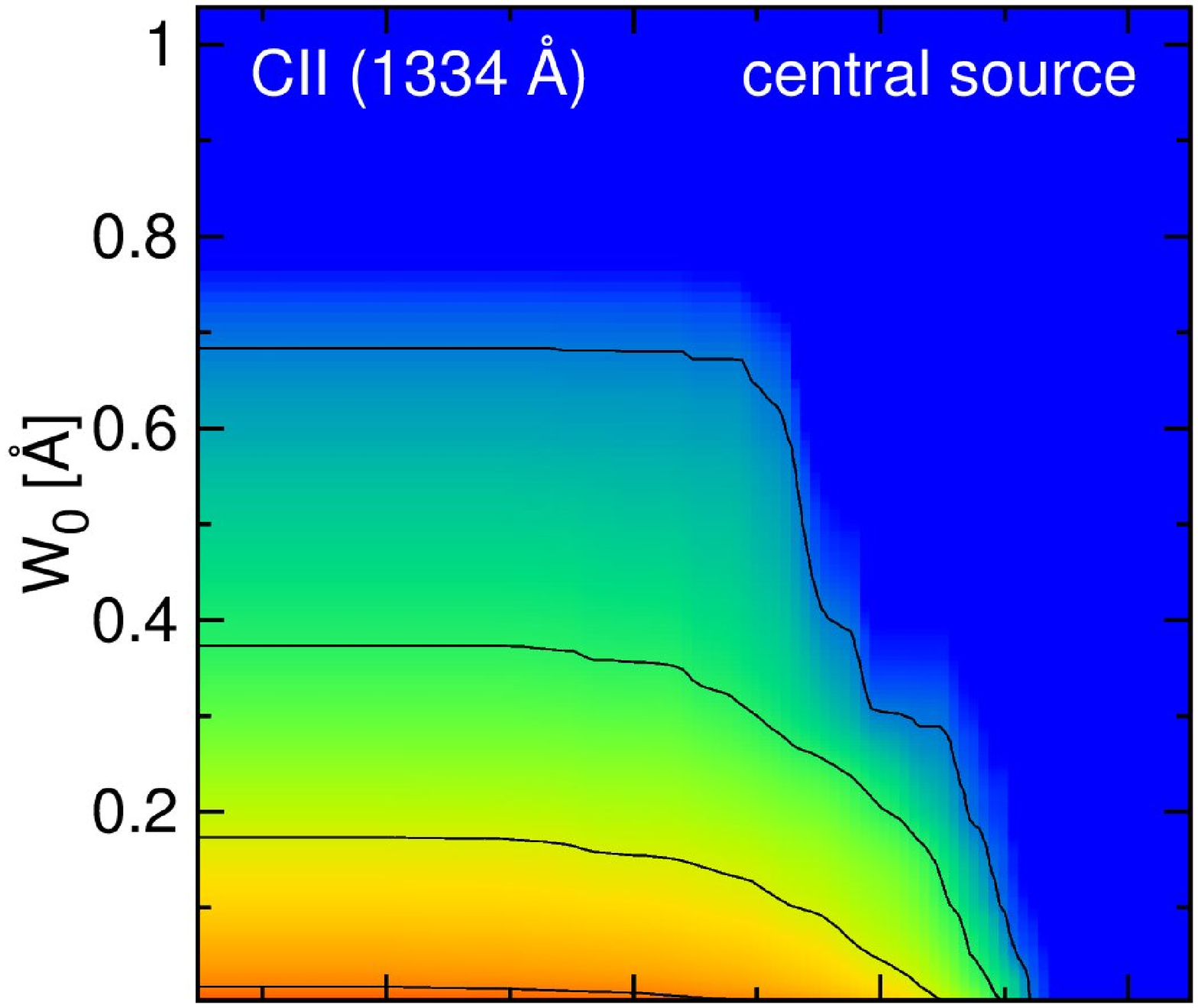}
\includegraphics[width=4.92cm]{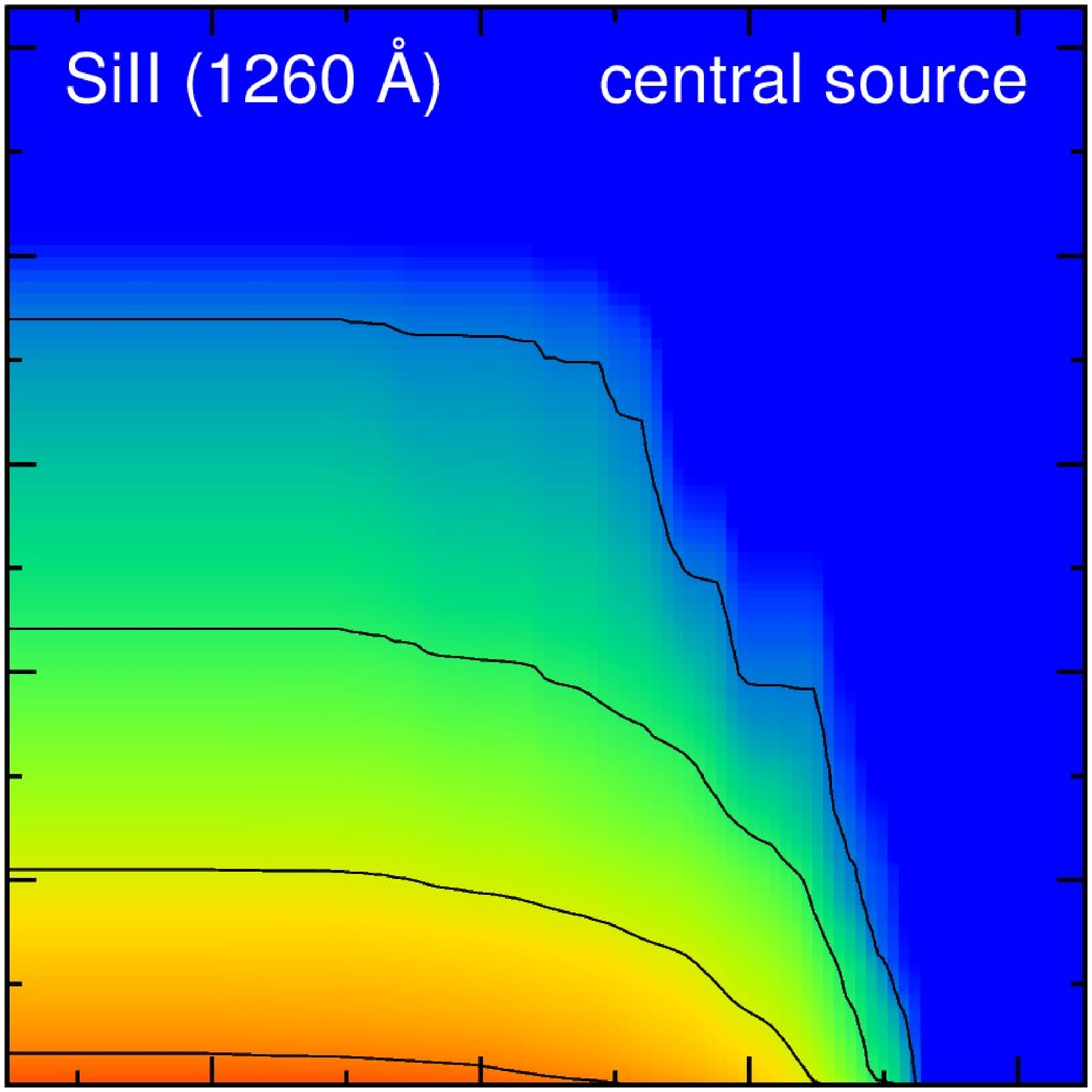}
\includegraphics[width=6.61cm]{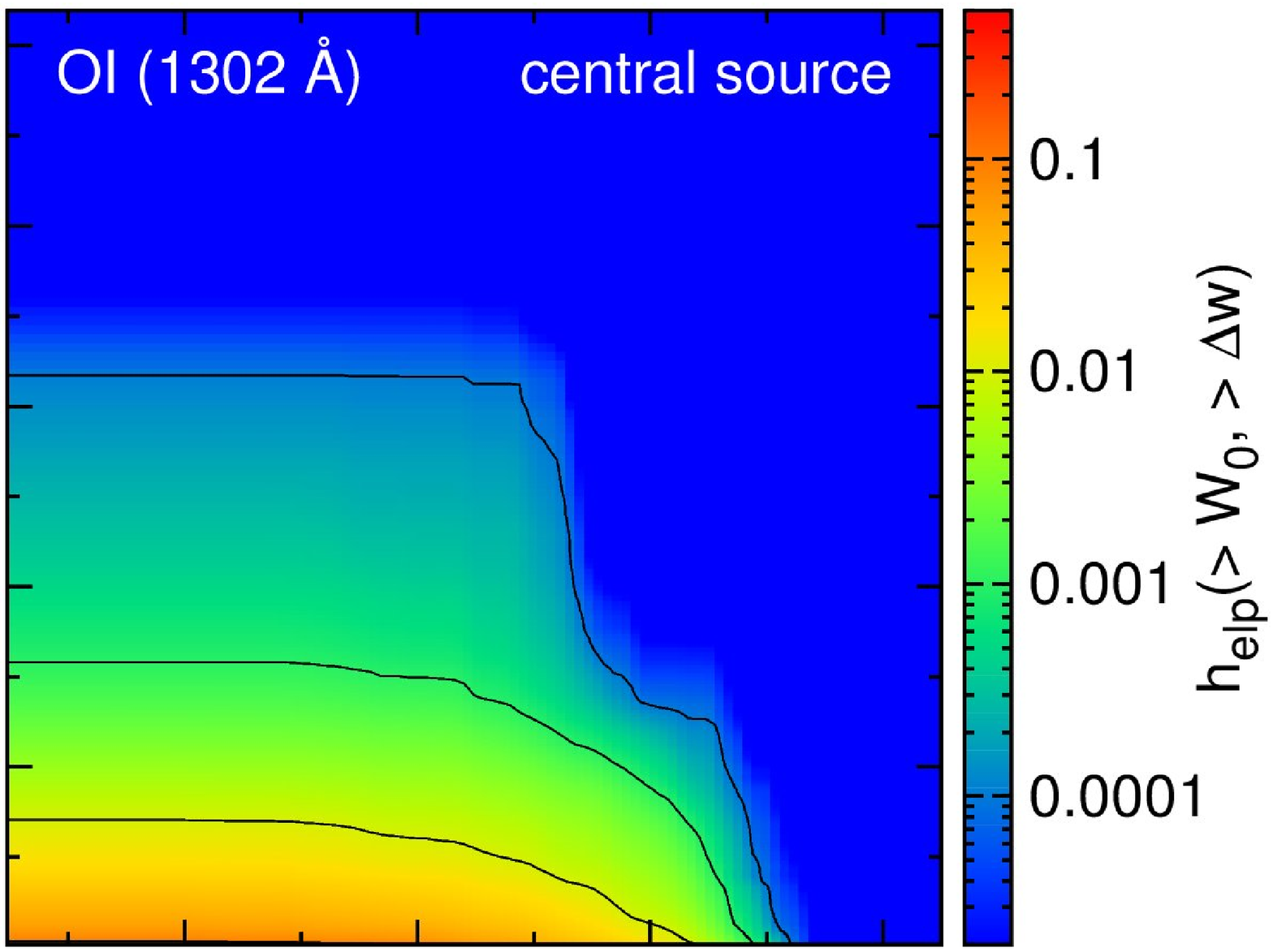}
\includegraphics[width=6.00cm]{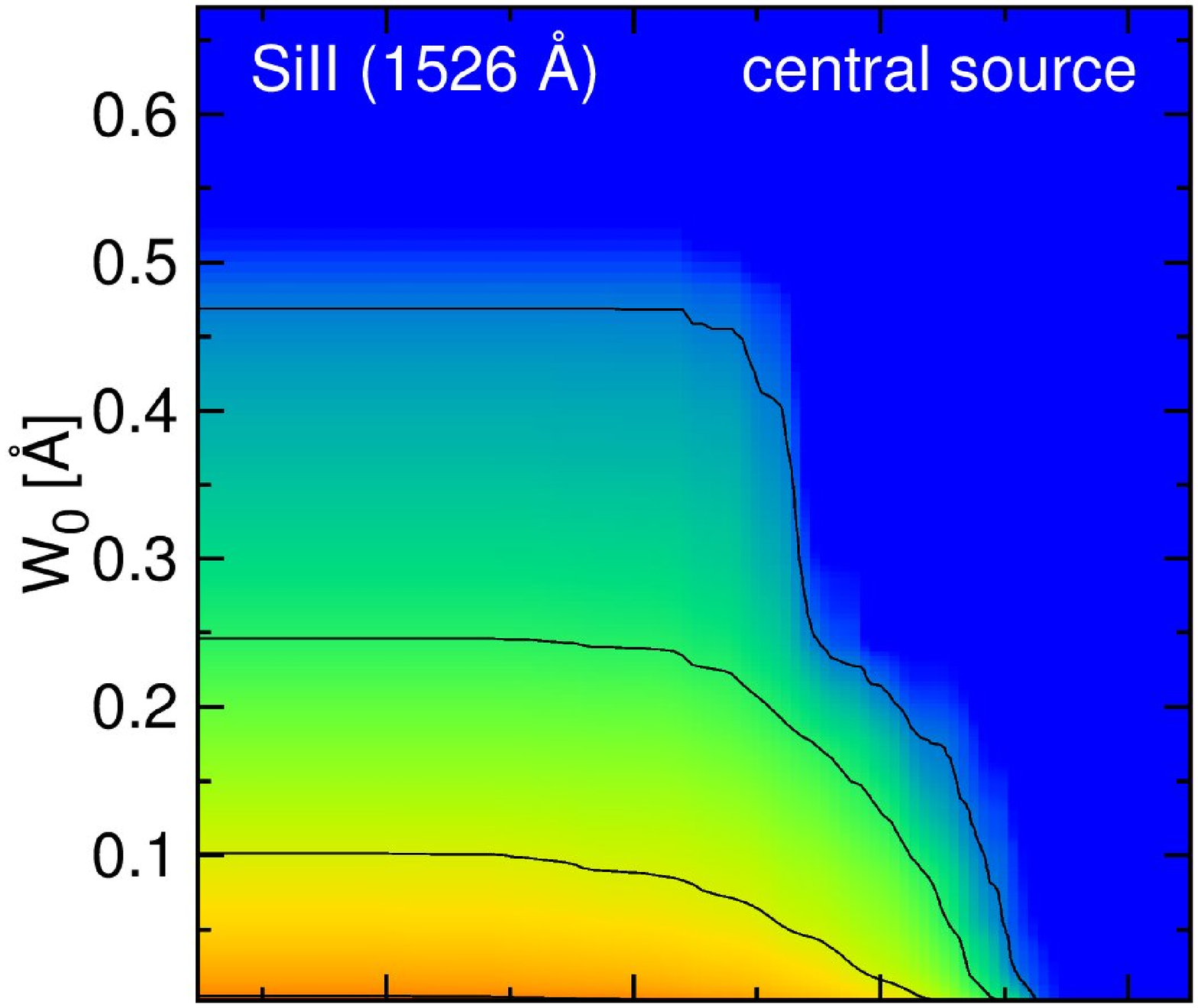}
\includegraphics[width=4.92cm]{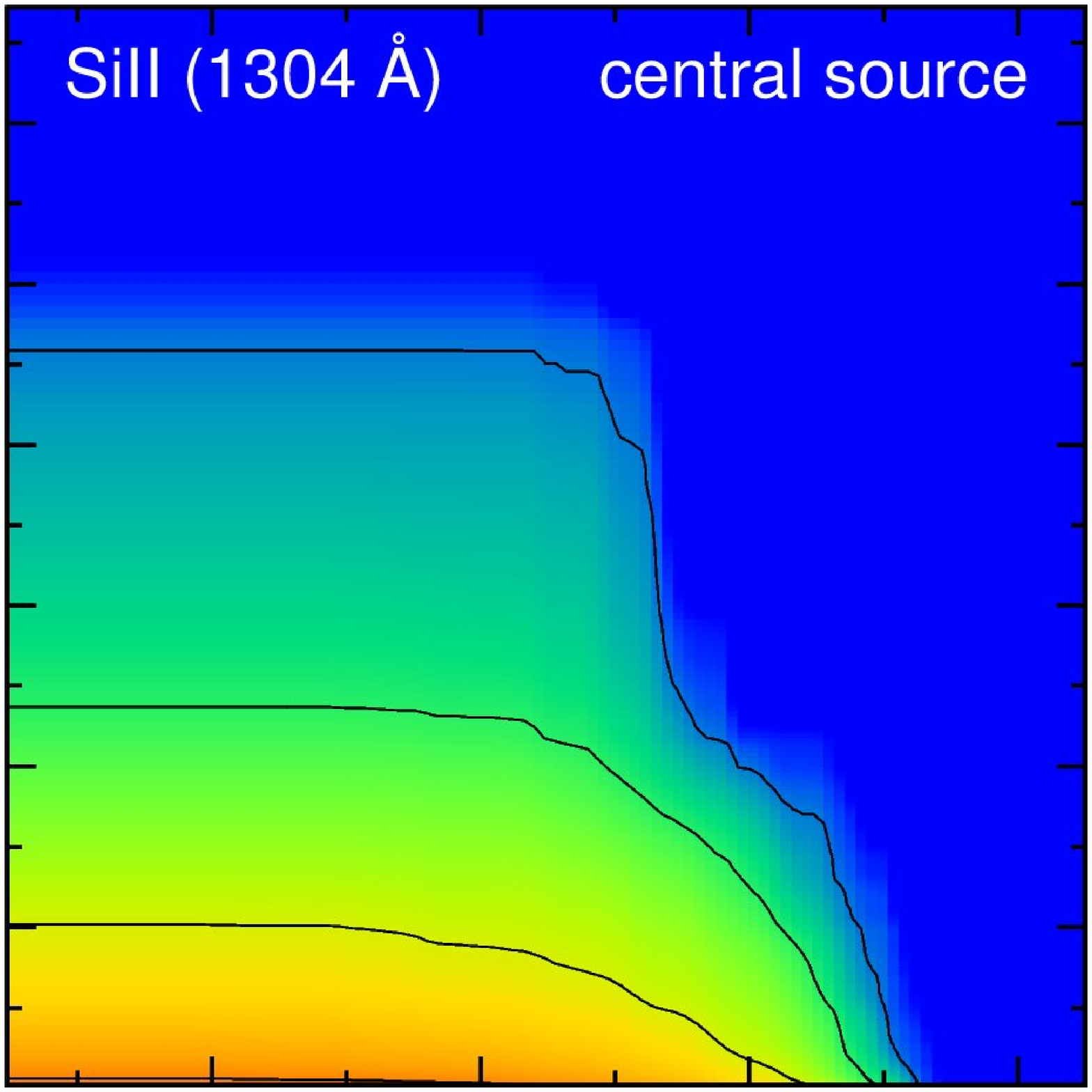}
\includegraphics[width=6.61cm]{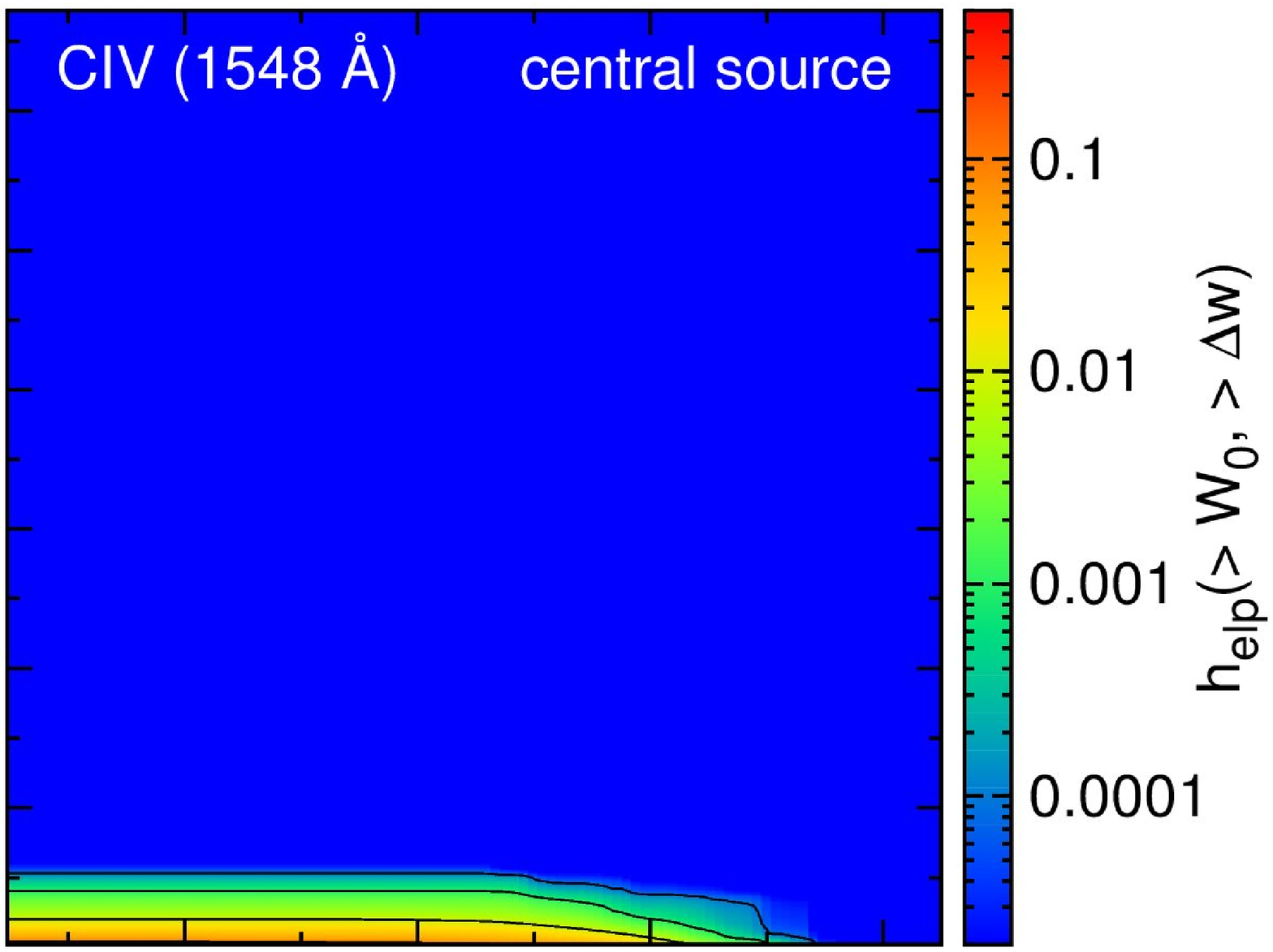}
\includegraphics[width=6.00cm]{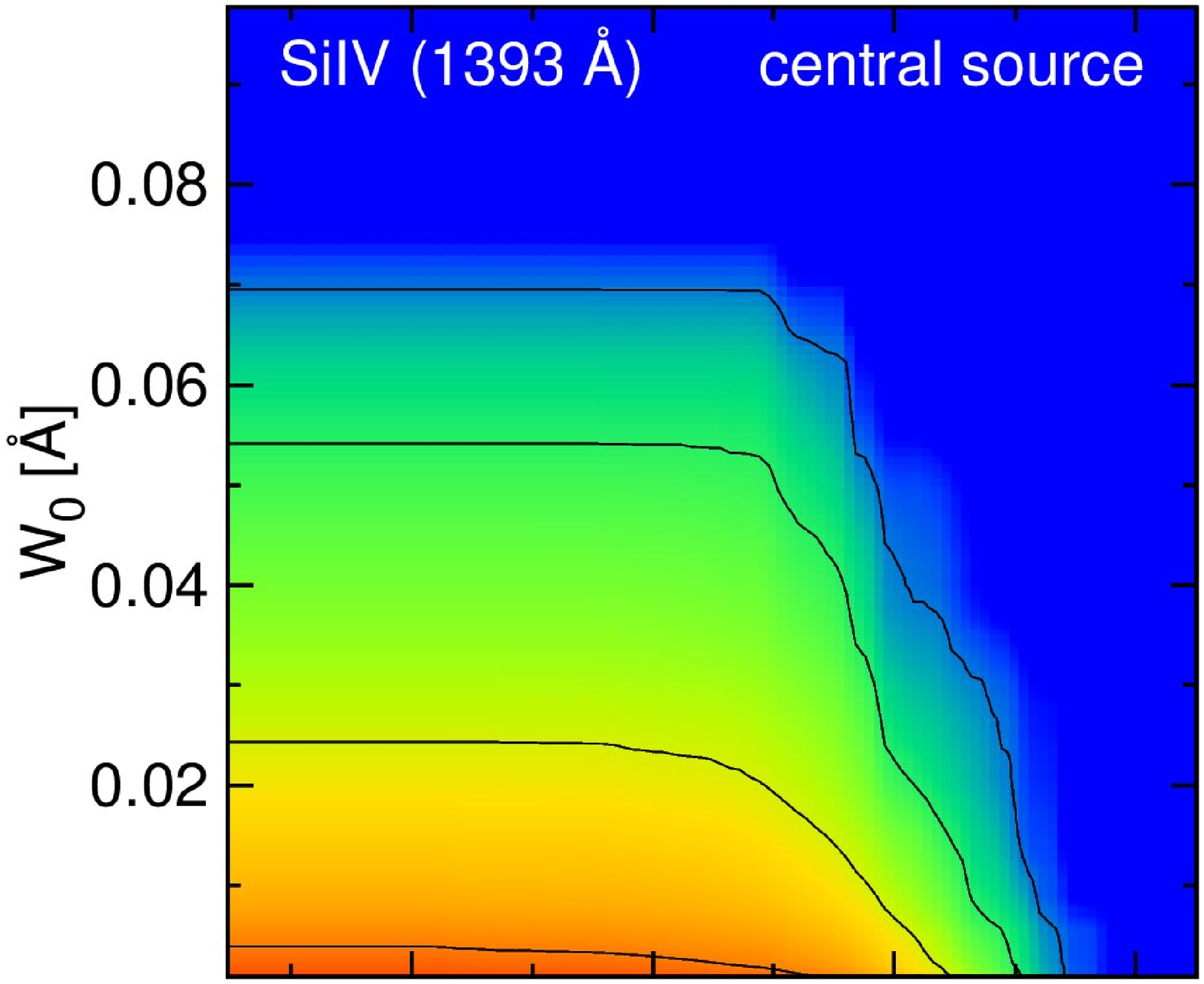}
\includegraphics[width=4.92cm]{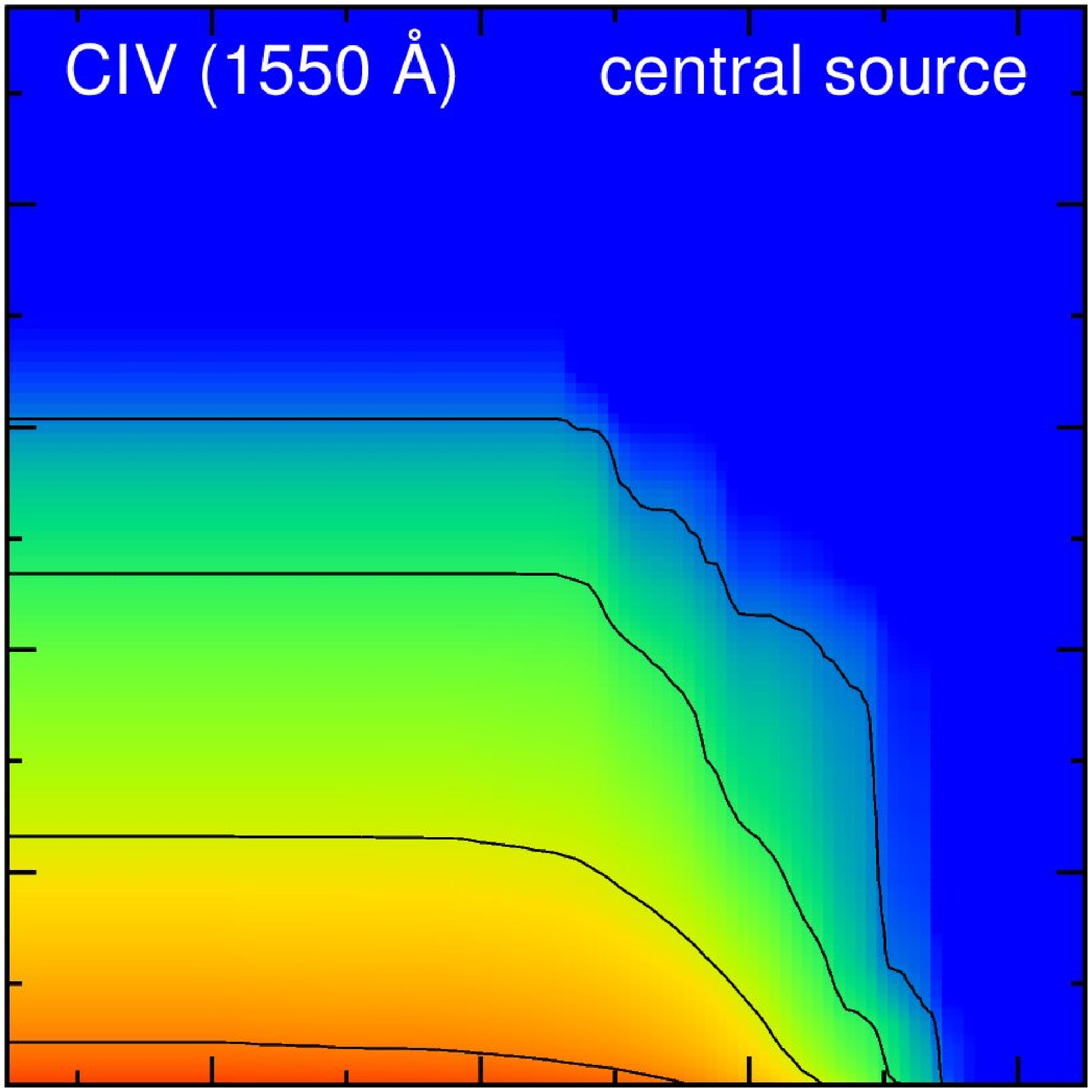}
\includegraphics[width=6.61cm]{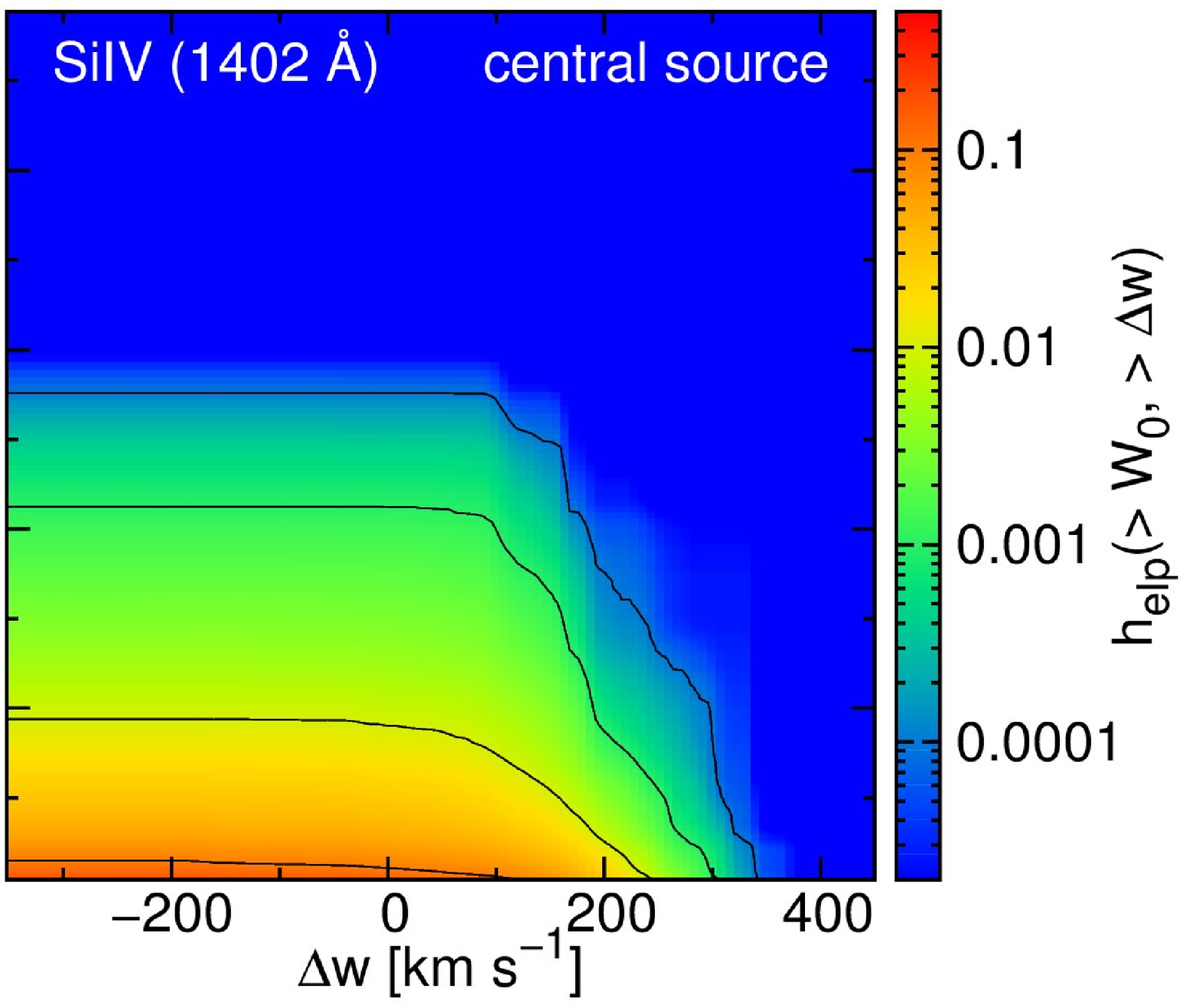}
\includegraphics[width=6.00cm]{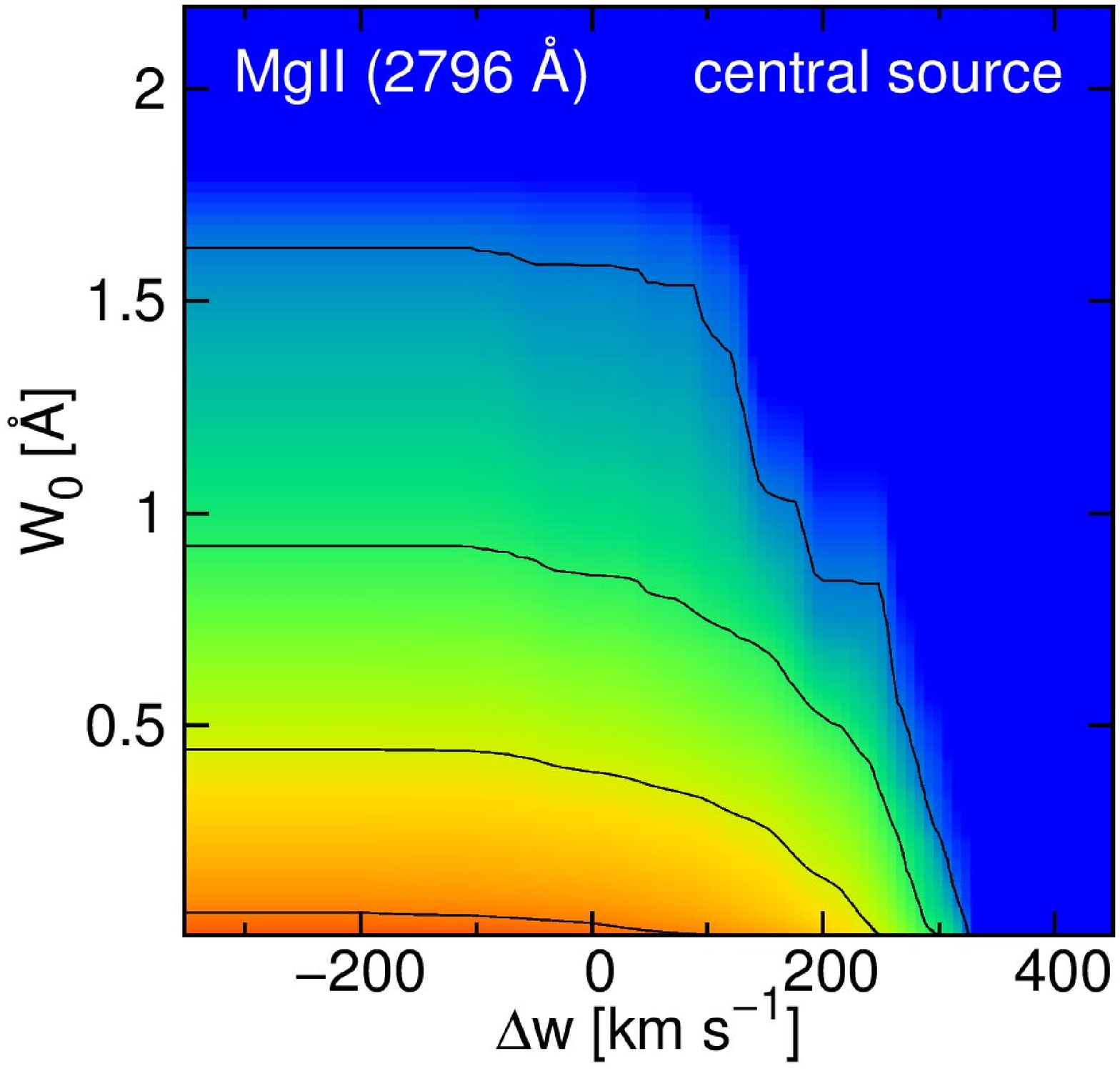}
\includegraphics[width=11.62cm]{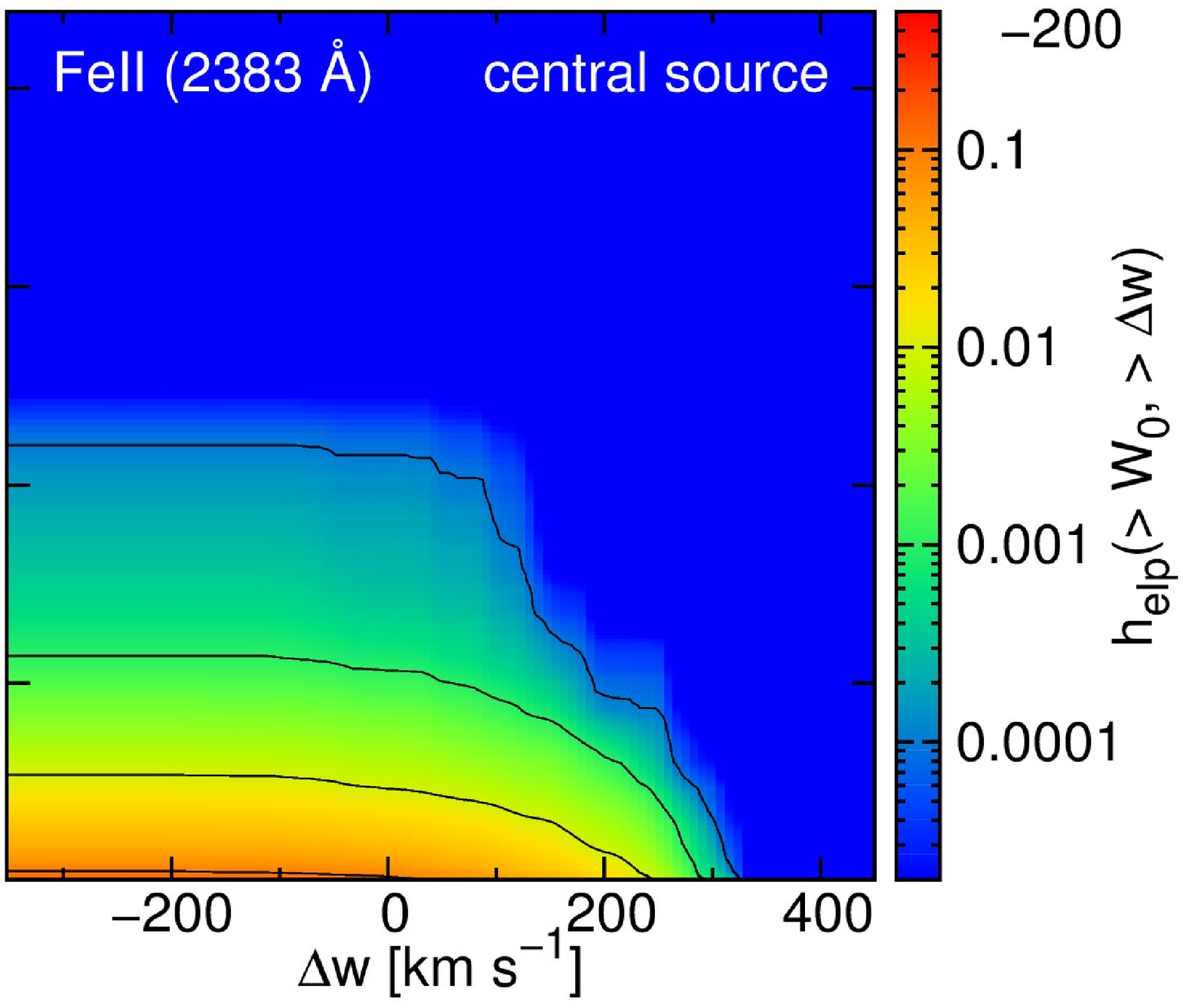}
\end{center}
\caption{Same cumulative fraction $h_{\rm elp}$ as in figure
\ref{fig:statlaline}. Here for the metal lines. Positive velocities are
inflowing into the galaxy and negative velocities are out of the galaxy.
Contour lines are at 0.1, 0.01, 0.001 and 0.0001 respectively. Note the
different y-axis scalings with respect to figure \ref{fig:statlaline} as well
as from row to row.  In {\CII} one sees an inflow $> 150$\,km\,s$^{-1}$ with an
{\EW} $>$ 0.17\,{\AA} in 0.4\,\% of all observations. For {\MgII} one sees an
inflow $> 150$\,km\,s$^{-1}$ with an {\EW} $>$ 0.2\,{\AA} in 1.3\,\% of all
observations. The line which has by far the strongest signal is the {\MgII}
line followed by the {\FeII} line. These values should be achievable by future
observations.}
\label{fig:statmetallines}
\end{figure*}

\begin{figure}
\begin{center}
\psfrag{relative intensity}[B][B][1][0] {relative intensity}
\psfrag{vks}[B][B][1][0] {$\Delta w$ [km s$^{-1}$]}
\psfrag{lya}[Br][Br][1][0] {\la}
\psfrag{cg}[Br][Br][1][0]  {central source}
\psfrag{r30}[Br][Br][1][0] {\textcolor{red}{$r_{\rm i} = 0.30 \ \Rv$}}
\psfrag{r20}[Br][Br][1][0] {\textcolor{green}{$r_{\rm i} = 0.20 \ \Rv$}}
\psfrag{r10}[Br][Br][1][0] {\textcolor{blue}{$r_{\rm i} = 0.10 \ \Rv$}}
\psfrag{r05}[Br][Br][1][0] {\textcolor{magenta}{$r_{\rm i} = 0.05 \ \Rv$}}
\psfrag{r02}[Br][Br][1][0] {\textcolor{cyan}{$r_{\rm i} = 0.02 \ \Rv$}}
\includegraphics[width=8.45cm]{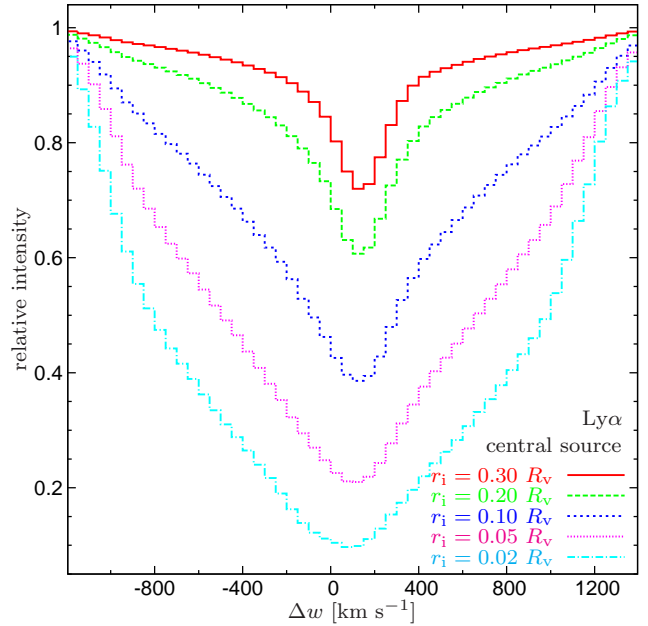}
\end{center}
\caption{{\la} absorption line profile for a central source geometry averaged
over different viewing angles and all three galaxies. We integrated from 1.0
$\Rv$ down to different inner radii $r_{\rm i}$. Positive velocities are
inflowing into the galaxy and negative velocities are out of the galaxy. Note
the different scaling of both axes in comparison to figure \ref{fig:metline}.
The profiles always peak in the positive, indicating inflow. Lines for high
$r_{\rm i}$ having a shallow total MLD show very extended wings at the edges of
the distribution whereas lines for small $r_{\rm i}$ having a deep total MLD the
edges of the profiles are much sharper since the centre of the line is much
deeper.}
\label{fig:laline}
\end{figure}

\begin{figure*}
\begin{center}
\psfrag{relative intensity}[B][B][1][0] {relative intensity}
\psfrag{vks}[B][B][1][0] {$\Delta w$ [km s$^{-1}$]}
\psfrag{cg}[Bl][Bl][1][0]   {central source}
\psfrag{r30}[Br][Br][1][0] {\textcolor{red}{$r_{\rm i} = 0.30 \ \Rv$}}
\psfrag{r20}[Br][Br][1][0] {\textcolor{green}{$r_{\rm i} = 0.20 \ \Rv$}}
\psfrag{r10}[Br][Br][1][0] {\textcolor{blue}{$r_{\rm i} = 0.10 \ \Rv$}}
\psfrag{r05}[Br][Br][1][0] {\textcolor{magenta}{$r_{\rm i} = 0.05 \ \Rv$}}
\psfrag{r02}[Br][Br][1][0] {\textcolor{cyan}{$r_{\rm i} = 0.02 \ \Rv$}}
\psfrag{CII}[Bl][Bl][1][0] {{\CII} (1334 \AA)}
\psfrag{OI}[Bl][Bl][1][0] {{\OI} (1302 \AA)}
\psfrag{SiIIa}[Bl][Bl][1][0] {{\SiII} (1526 \AA)}
\psfrag{SiIIb}[Bl][Bl][1][0] {{\SiII} (1260 \AA)}
\psfrag{CIVa}[Bl][Bl][1][0] {{\CIV} (1548 \AA)}
\psfrag{SiIVa}[Bl][Bl][1][0] {{\SiIV} (1393 \AA)}
\psfrag{CIVb}[Bl][Bl][1][0] {{\CIV} (1550 \AA)}
\psfrag{SiIVb}[Bl][Bl][1][0] {{\SiIV} (1402 \AA)}
\psfrag{SiIIc}[Bl][Bl][1][0] {{\SiII} (1304 \AA)}
\psfrag{MgIIb}[Bl][Bl][1][0] {{\MgII} (2796 \AA)}
\psfrag{FeIIb}[Bl][Bl][1][0] {{\FeII} (2383 \AA)}
\includegraphics[width=5.92cm]{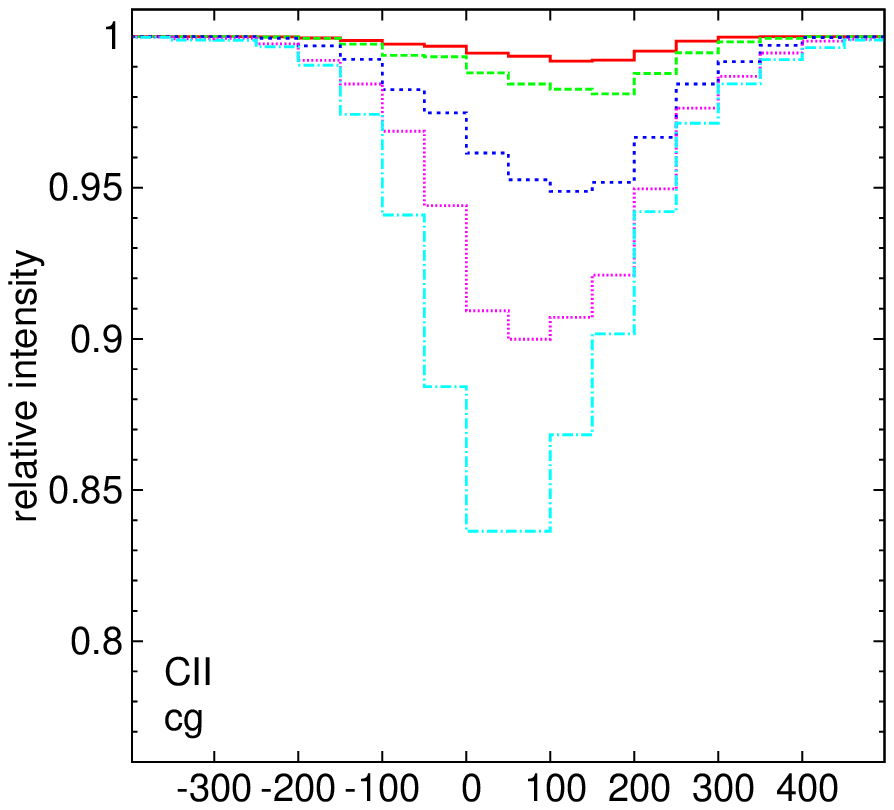}
\includegraphics[width=4.94cm]{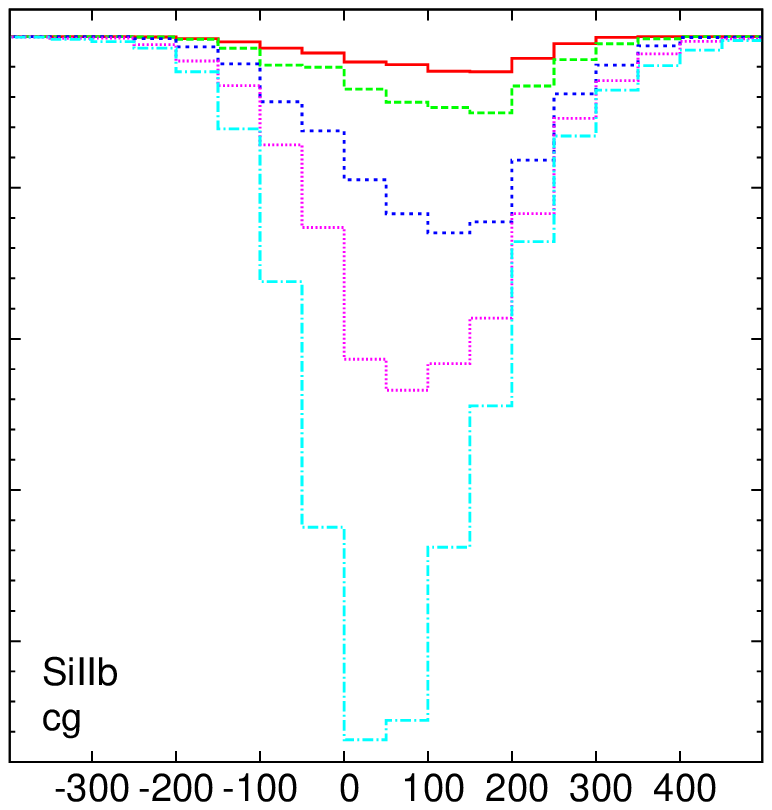}
\includegraphics[width=4.94cm]{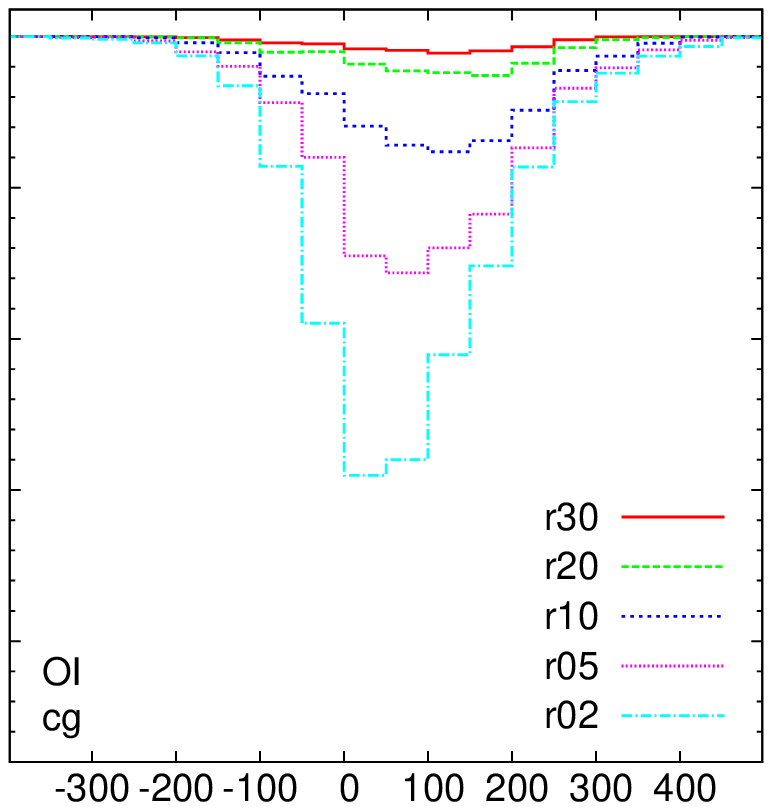}
\includegraphics[width=5.92cm]{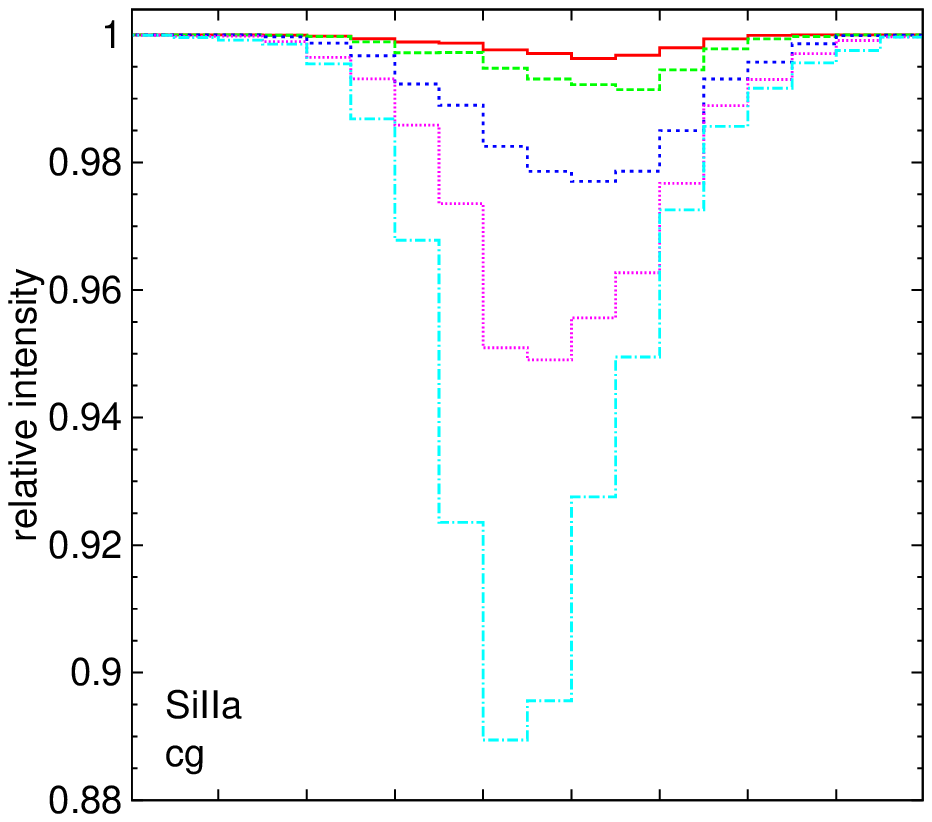}
\includegraphics[width=4.94cm]{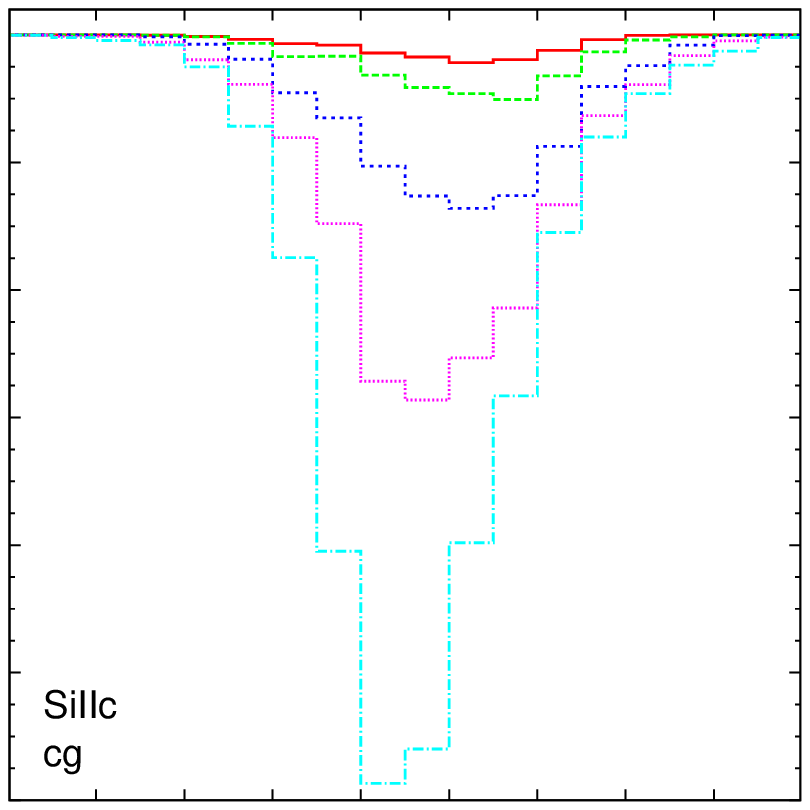}
\includegraphics[width=4.94cm]{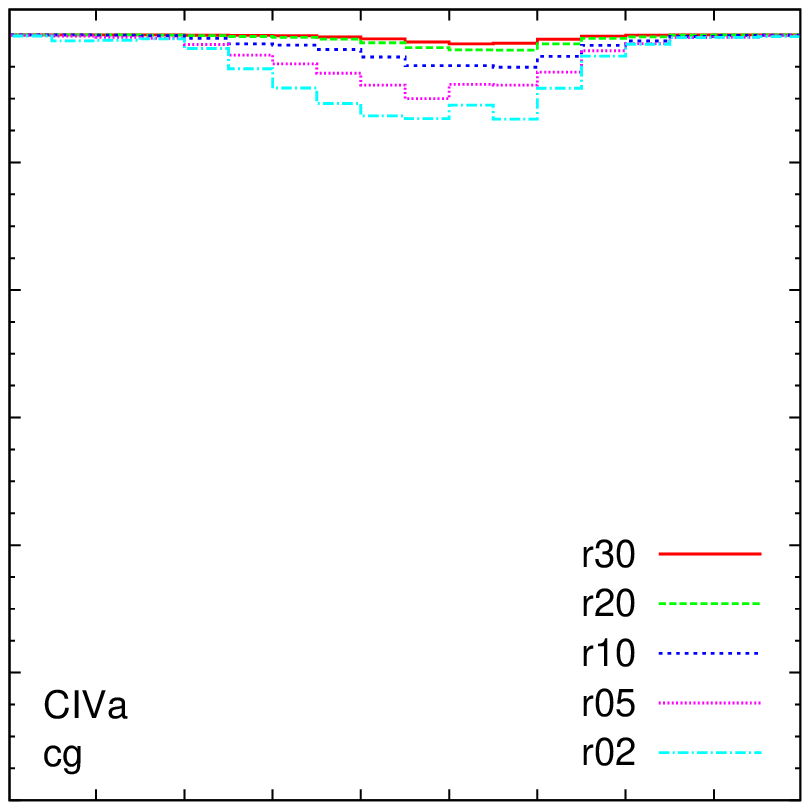}
\includegraphics[width=5.92cm]{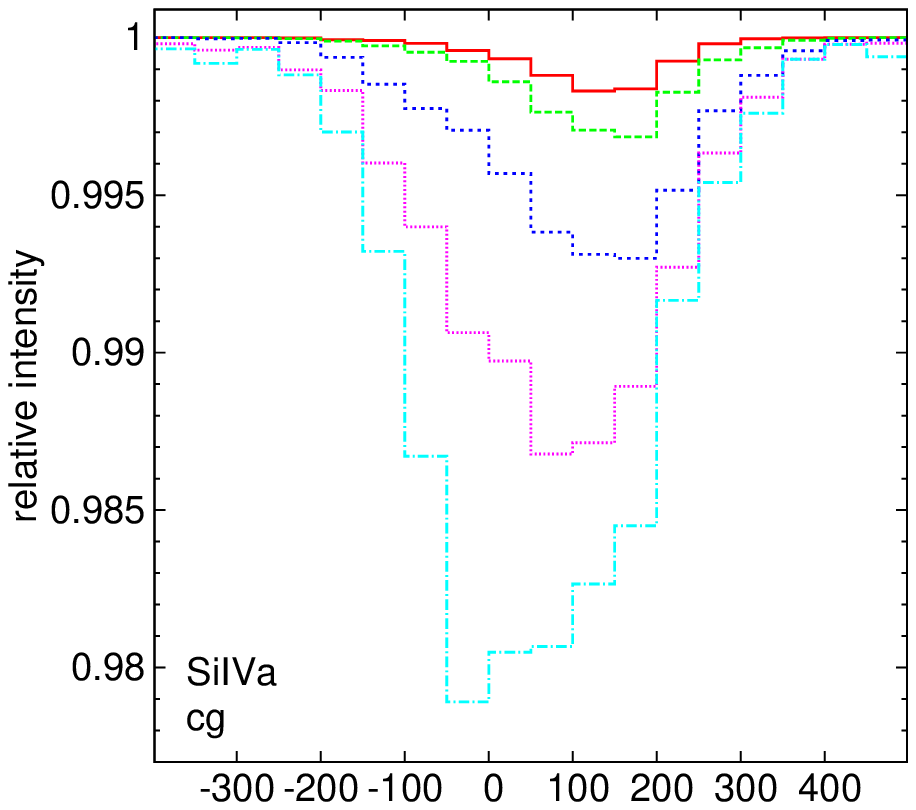}
\includegraphics[width=4.94cm]{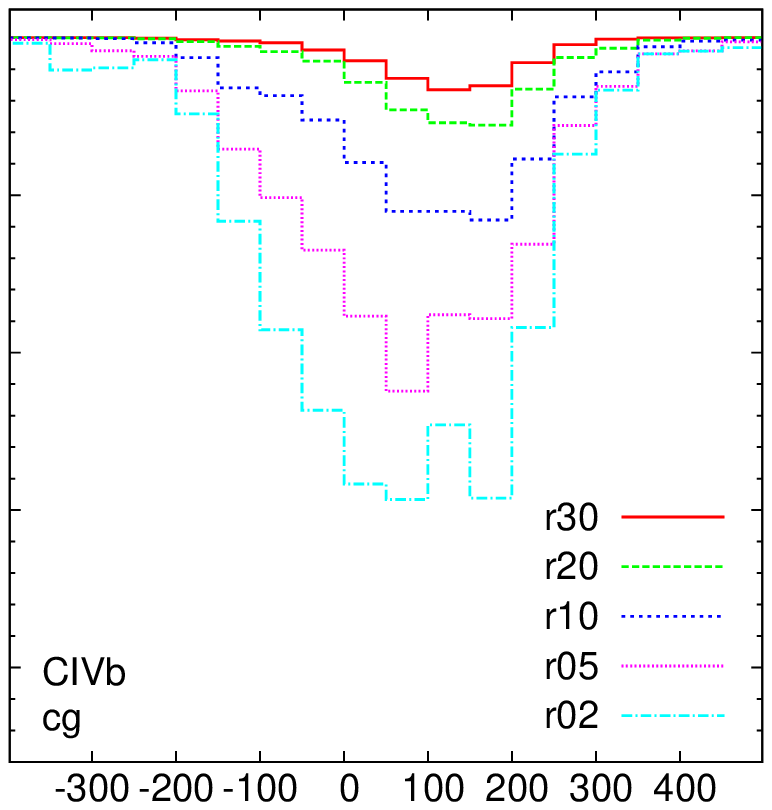}
\includegraphics[width=4.94cm]{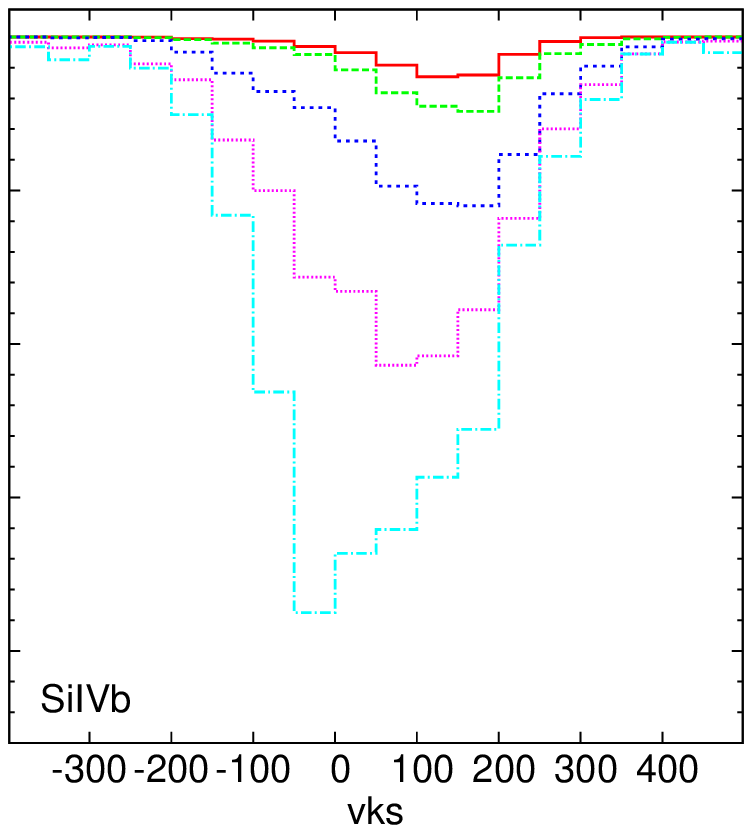}
\includegraphics[width=5.92cm]{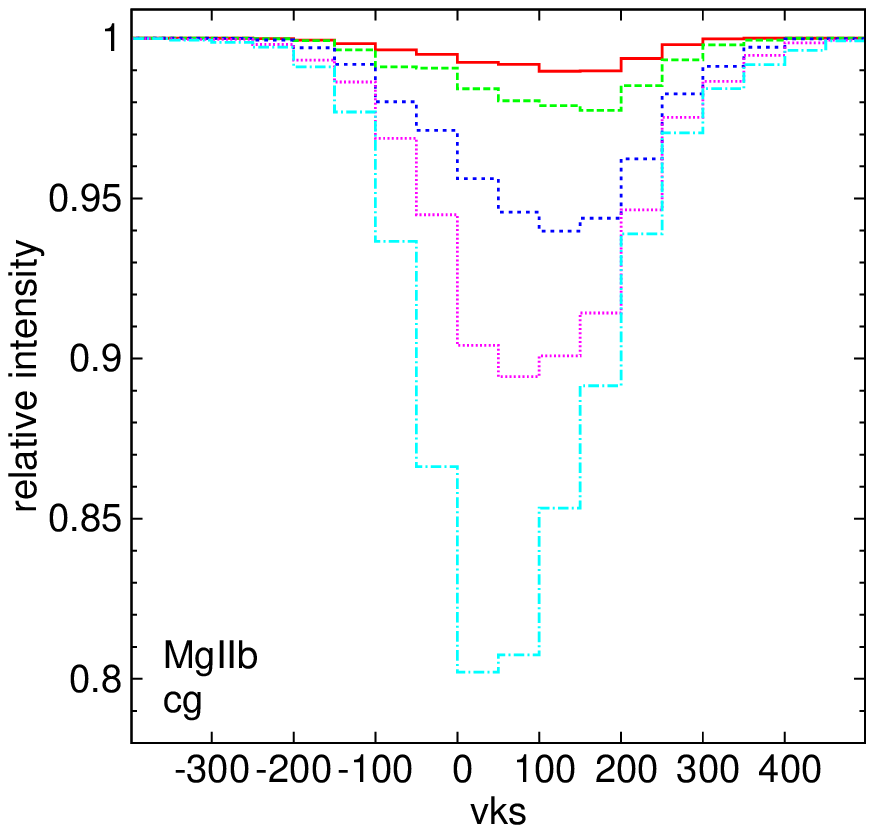}
\includegraphics[width=9.96cm]{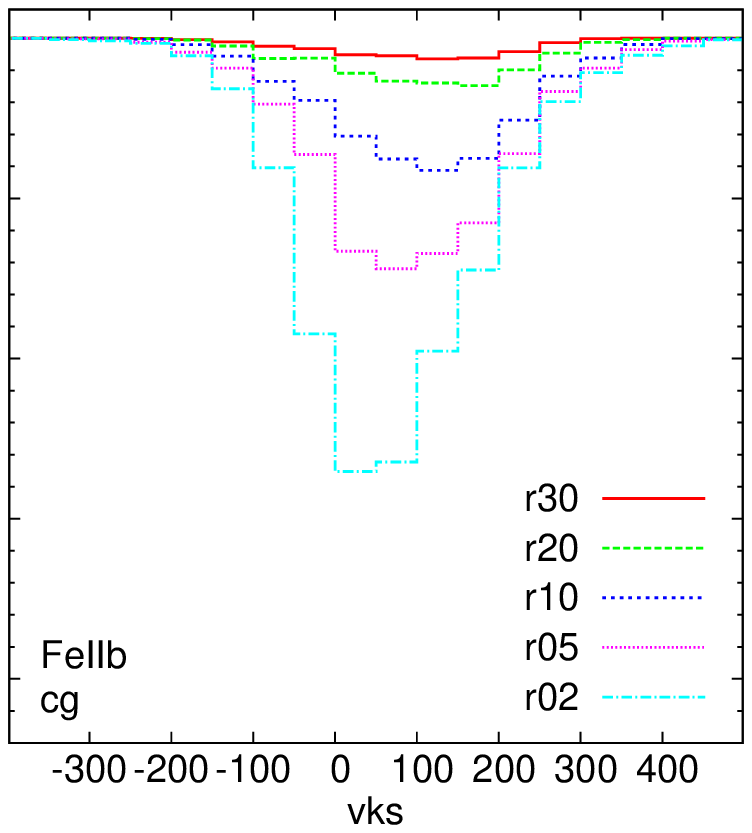}
\end{center}
\caption{Same absorption line profiles for a central source as in figure
\ref{fig:laline}, this time for the metal lines. Positive velocities are
inflowing into the galaxy and negative velocities are  out of the galaxy.
Note the different y-axis scaling from row to row and the different overall
axes scaling with respect to figure \ref{fig:laline}. The metal lines are much
weaker than the {\la} line, since the inflowing material is mainly unprocessed
primordial gas with very low metallicity. These metal lines also appear tiny
compared to the corresponding lines presented by {\s10}. The profiles always
peak in the positive, indicating inflow. The lines with the strongest signals
are {\SiII} and {\MgII}. A more detailed quantitative comparison to the
observations is presented in table \ref{tab:cg}.}
\label{fig:metline}
\end{figure*}

As a last application for this geometry we mimic the {\s10} stacking procedure
(see their figures 6 and 10) they used to increase the signal to noise ratio.
They stacked 89 galaxies with $z = 2.3 \pm 0.3$ to investigate the kinematics
of the galaxy-scale outflows. We produce stacked spectra using our simulations
by summing  up the line profiles of several thousand different directions for
each of the three galaxies and stack them together. We determine the absorption
line profile for a spherical shell between an outer radius and an inner radius
$r_{\rm i}$. The outer radius is always kept constant at 1.0 $\Rv$ which
corresponds roughly to 74\,kpc. The inner radius $r_{\rm i}$ however is varied
between 0.3 $\Rv$ and 0.02 $\Rv$. This is done for the following reason: As one
can see from figure \ref{fig:denmap}, there are very clean, unperturbed streams
into the galaxy down to as far as $\sim 0.25 \Rv$. In the immediate vicinity of
the galaxy there is the ``messy region'' without well defined streams and with
no well defined gas disk either. The galactic disk itself has no sharp edge and
therefore its extent is not determined easily either. In order to account for
these uncertainties we show several different absorption line profiles for
different inner radii for every wavelength in question. All are averaged in the
aforementioned way. In figures \ref{fig:laline} and \ref{fig:metline} the
resulting profiles are shown for all the lines. These figures can be directly
compared to figures 6 or 10 in {\s10}. The plots reveal that we have for {\la}
a FWHM $\eta$ of up to $\sim 1900$ km s$^{-1}$ with a line depth of up to 0.90
for an inner radius $r_{\rm i}$ going as deep as 0.02 $\Rv$ (1.5\,kpc). The
{\la} lines for high $r_{\rm i}$ having a shallow total MLD show very extended
wings at the edges of the line profile whereas lines for small $r_{\rm i}$
having a deep total MLD the edges of the profiles are much sharper since the
centre of the line is much deeper. All profiles of all lines always peak in the
positive, indicating inflow. Our strongest metal line is {\SiII}. It has a
maximum line depth of $\sim 0.23$ with a $\eta \sim 250\,$km\,s$^{-1}$ for
$r_{\rm i} = 0.02\, \Rv$. This confirms our initial guess that metal lines are
so much weaker than the {\la} line, since the inflowing material is mainly
unprocessed primordial gas with very low metallicity. The predicted metal line
absorption profiles appear tiny compared to the corresponding lines presented
in figures 6 or 10 of {\s10} having a line depth of 0.5 and $\eta \sim 1000$ km
s$^{-1}$ in metals. Detailed comparisons between our values and the observed
values by {\s10} are given in Table \ref{tab:cg}.

\begin{table}
\begin{center}
\setlength{\arrayrulewidth}{0.5mm}
\begin{tabular}{lccclrl}
\hline
    &             & {\s10}'s & {\s10}'s & our & our    & our \\
ion & $\lambda_0$ & MLD      & $W_0$    & MLD & $\eta$ & $W_0$ \\
    & [\AA]    &       & [\AA]  &   & [km s$^{-1}$] & [\AA] \\
\hline
{\la}   & 1216 &     &      & 0.90  & 1900 & 6.2  \\
{\CII}  & 1334 & 0.5 &  1.6 & 0.16  & 250 & 0.20  \\
{\SiII} & 1260 &     &      & 0.23  & 250 & 0.25  \\
{\OI}   & 1302 & 0.7 &  2.7 & 0.15  & 250 & 0.16  \\
{\SiII} & 1526 & 0.6 &  1.8 & 0.11  & 200 & 0.13  \\
{\SiII} & 1304 & 0.7 &  2.4 & 0.12  & 200 & 0.12  \\
{\CIV}  & 1548 & 0.5 &  3.1 & 0.013 & 350 & 0.024 \\
{\SiIV} & 1393 & 0.4 &  1.3 & 0.021 & 300 & 0.032 \\
{\CIV}  & 1550 & 0.5 &  3.4 & 0.015 & 350 & 0.027 \\
{\SiIV} & 1402 & 0.3 &  0.8 & 0.019 & 300 & 0.027 \\
{\MgII} & 2796 &     &      & 0.20  & 250 & 0.47  \\
{\FeII} & 2383 &     &      & 0.14  & 250 & 0.27  \\
\hline
\end{tabular}
\end{center}
\caption{Comparison of maximum line depths (MLD), FWHM $\eta$ and {\EW} $W_0$
in the central geometry between {\s10}'s observations (their figure 10) and our
predictions from the simulations (the ``$r_{\rm i} = 0.02 \ \Rv$'' line in our
figures \ref{fig:laline} and \ref{fig:metline}).}
\label{tab:cg}
\end{table}

It is hard to compare our {\la} absorption line profiles to the observations of
{\s10} as they actually show emission line profiles. One could still draw
interesting conclusions by comparing the underlying optical depths $\tau$ of
the {\s10} observations with our simulations. However, the {\la} emission is
not only an indicator of cold streams. Probably the most suitable lines for the
purpose of detecting cold streams in absorption are {\CII} and {\MgII} since
they have the strongest signal. Out of those {\MgII} is closer to being
observed with the needed sensitivity and resolution \citep[cf. for
example][]{matejek}. As {\s10} point out, the easiest way for a {\la} photon to
reach the observer is to acquire the velocity of the outflowing material on the
far side of the galaxy, and to be emitted in the observer's direction. In this
case the photon is redshifted by several hundred km s$^{-1}$ relative to the
bulk of the material through which it must pass to reach us. This picture would
explain qualitatively why the dominant component of {\la} emission always
appears redshifted relative to the galaxy systemic velocity \citep[see e.g,][]
{pettini02, adelberger03}, implying that their results for {\la} have to be
interpreted with caution. Our results for {\la} absorption in the simulations
are not subject to this complication, therefore one should be cautious when
comparing our {\la} results to the {\s10} data. Metal lines are less
problematic.

There are two main conclusions from our analysis of the central-source
geometry, which are discouraging concerning the potential for detecting cold
streams. First stacking of absorption line profiles does not strengthen but
weakens the signature of cold streams  this might be counter intuitive, but
compare figure \ref{fig:viewang} with figures \ref{fig:laline} and
\ref{fig:metline}). See figures \ref{fig:statlaline} and
\ref{fig:statmetallines} for the predicted strengths of a cold stream
absorption signal when observing without stacking. Second, the {\la} signal
detected by observations is naturally dominated by outflows which are expected
to have much higher covering factor and metallicity \citep{faucher2}.

\begin{figure}
\begin{center}
\includegraphics[width=8.45cm]{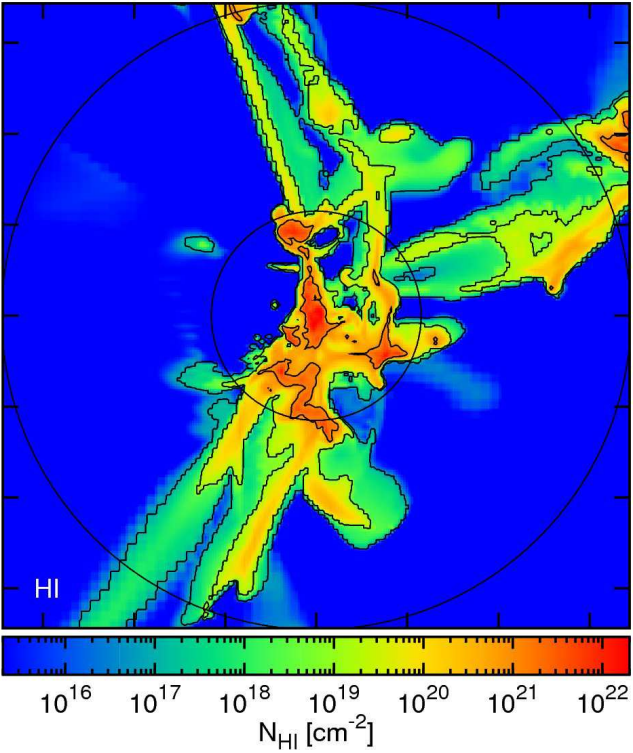}
\end{center}
\caption{Map of the projected neutral hydrogen column densities ($N_{\rm HI}$)
for the background geometry of our fiducial galaxy at $z = 2.3$. The box side
is $138\kpc$ (physical). The outer circle marks the virial radius and the inner
circle is at $1/3\ \Rv$. Contours are shown for $10^{17}$, $10^{19}$ and
$10^{21}\cmms$. One can see the three major streams outside $1/3 \ \Rv$ as well
as clumps which are actually satellites with dark matter haloes. Both of them
have $N_{\rm HI} > 10^{17}$ cm$^{-2}$. The background has $N_{\rm HI} < 10^{15}$
cm$^{-2}$. Inside $1/3 \ \Rv$ there is the messy region. It has column densities
with values between those of the stream and the background.}
\label{fig:NHIbg}
\end{figure}

\begin{figure}
\begin{center}
\psfrag{bRv}[B][B][1][0] {$b$ $[\Rv]$}
\psfrag{Ncm2}[B][B][1][0] {$N_{\rm Ai}$ [cm$^{-2}$]}
\psfrag{HI}[B][B][1][0] {\textcolor{red}{\HI}}
\psfrag{CII}[B][B][1][0] {\textcolor{green}{\CII}}
\psfrag{CIV}[B][B][1][0] {\textcolor{blue}{\CIV}}
\psfrag{OI}[B][B][1][0] {\textcolor{magenta}{\OI}}
\psfrag{SiII}[B][B][1][0] {\textcolor{cyan}{\SiII}}
\psfrag{SiIV}[B][B][1][0] {\textcolor{black}{\SiIV}}
\psfrag{bg}[Bl][Bl][1][0] {background source}
\includegraphics[width=8.45cm]{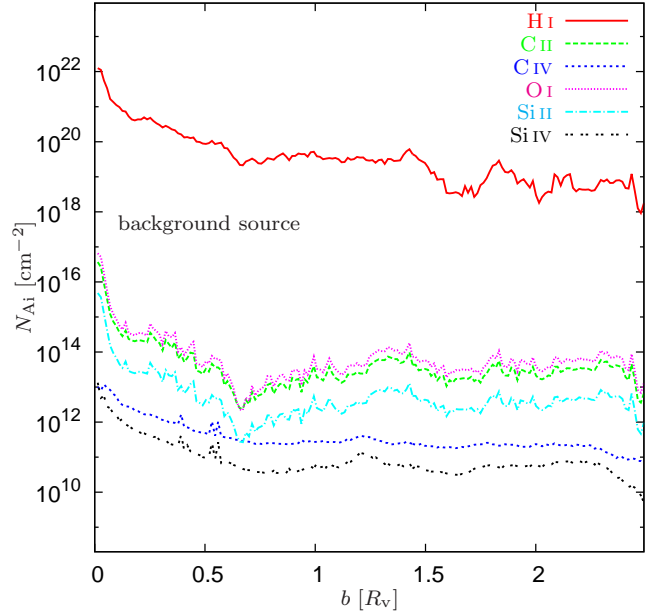}
\end{center}
\caption{Area weighted average column density $N_{\rm Ai}$ as a function of
impact parameter $b$ for {\HI} and five different metal lines in the background
source geometry. The {\HI} column density decreases through the interaction
region in the greater disc vicinity, from 10$^{22}$ cm$^{−2}$ near the centre of
the galaxy to 10$^{20}$ cm$^{−2}$ at 0.5 $\Rv$. It remains roughly constant at
10$^{19}$ cm$^{−2}$ in the stream regime out to 1.5 $\Rv$, and is not much lower
out to 2.5 $\Rv$.}
\label{fig:avcolden}
\end{figure}

\section{Background source}
\label{sec:bg}
In this section we consider the \lya and metal line absorption that occurs as
UV light emitted by a background galaxy or a quasar is absorbed by gas in the
circum-galactic environment of the foreground galaxy under investigation. In
this geometry one cannot distinguish between inflowing and outflowing gas.
However, the distribution of impact parameters allows us to explore the spatial
distribution of the circum-galactic gas surrounding the foreground galaxies.
{\s10} constructed a sample of 512 close angular pairs, $(1-15'')$, of galaxies
in the redshift range $z\sim 2-3$ with a large redshift difference such that
one is at the background of the other with no physical association. The pair
separations correspond to galactocentric impact parameters in the range $3-125$
kpc (physical) at $z = 2.2$, providing a map of cool gas as a function of
galactocentric distance for a well-characterised population of galaxies. The
discussion in this section will lead to a prediction of absorption line
profiles that are directly comparable to the ones observed and published in
section 6 of {\s10}.

\begin{figure*}
\begin{center}
\includegraphics[width=6.21cm]{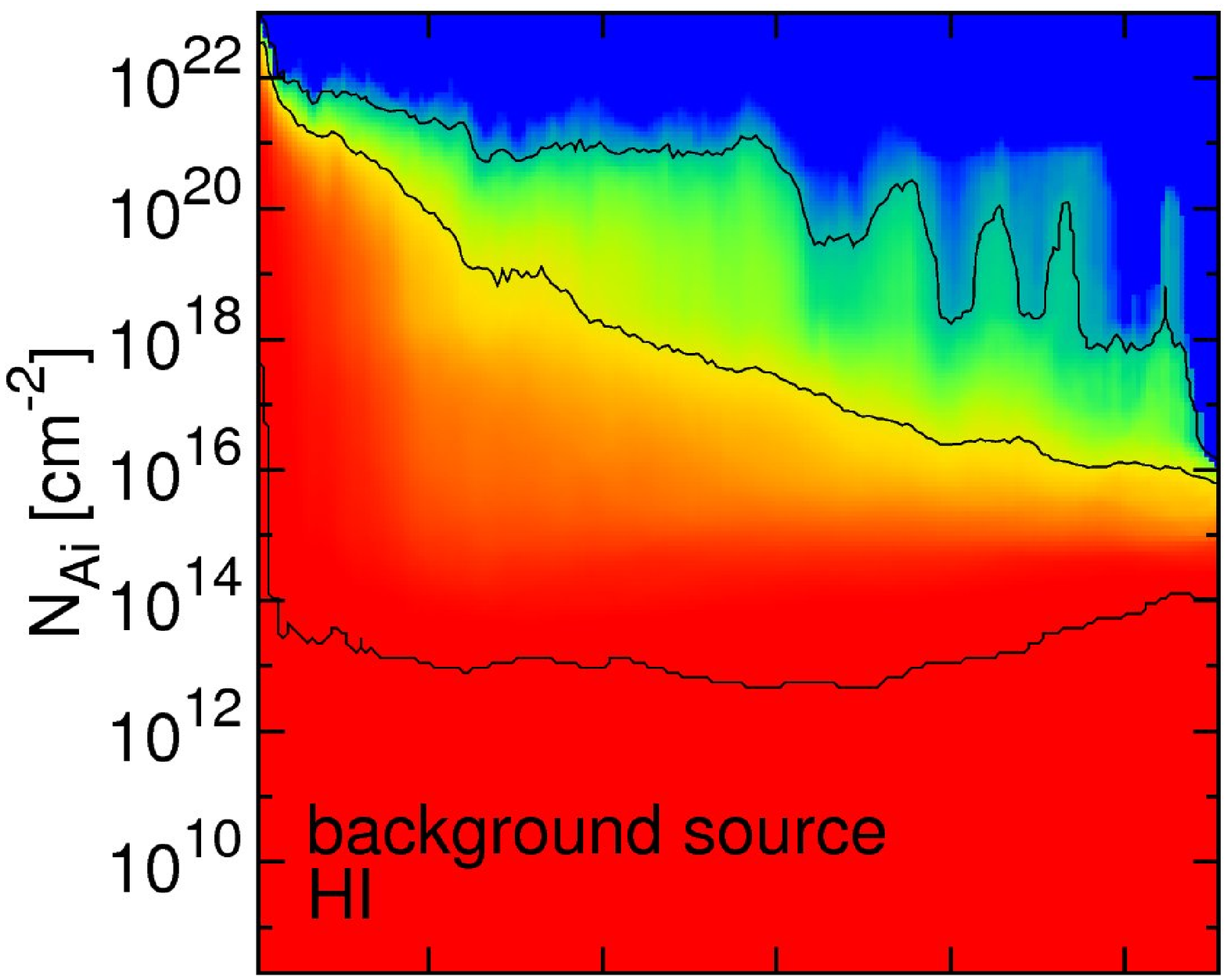}
\includegraphics[width=4.95cm]{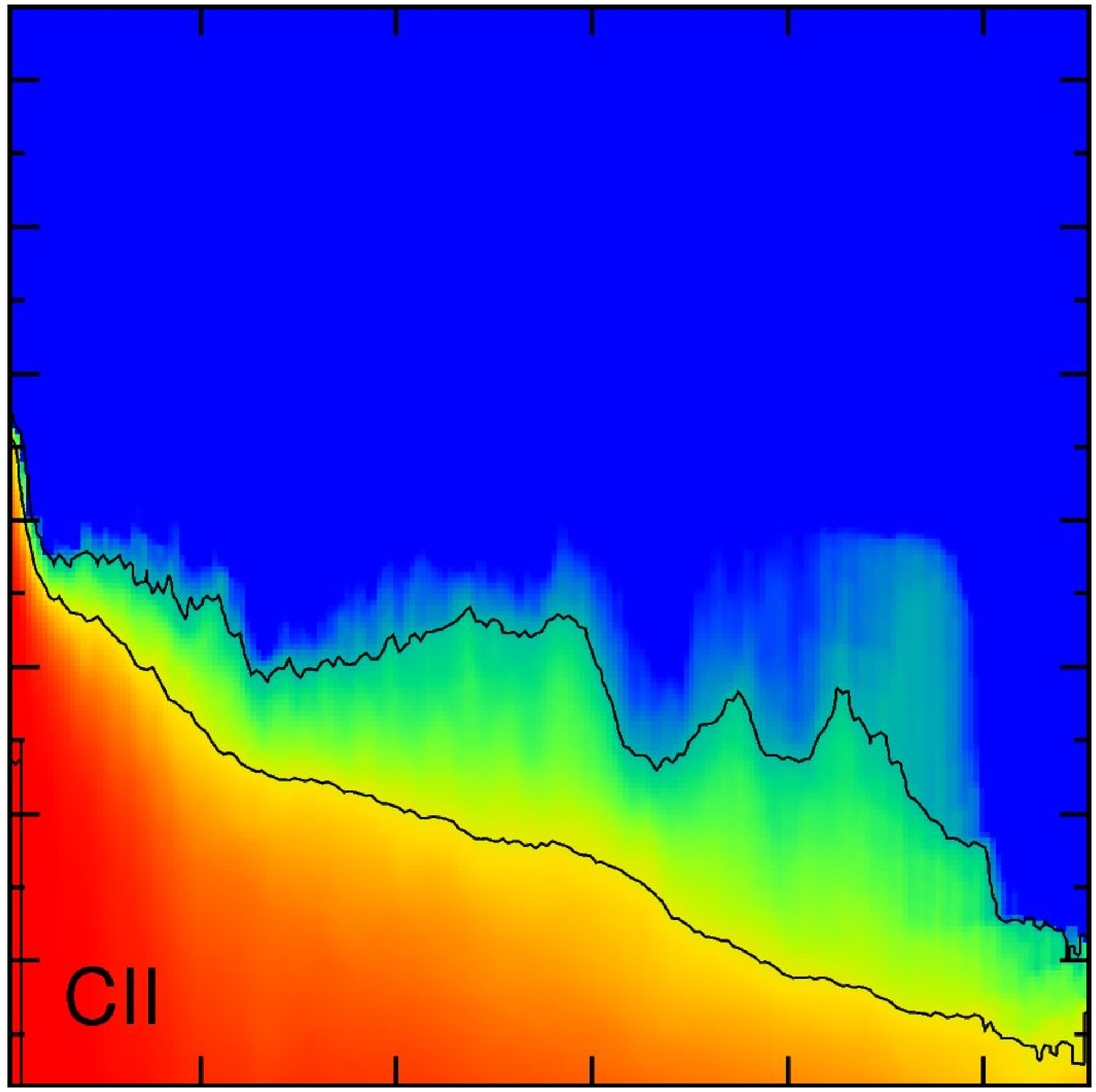}
\includegraphics[width=6.37cm]{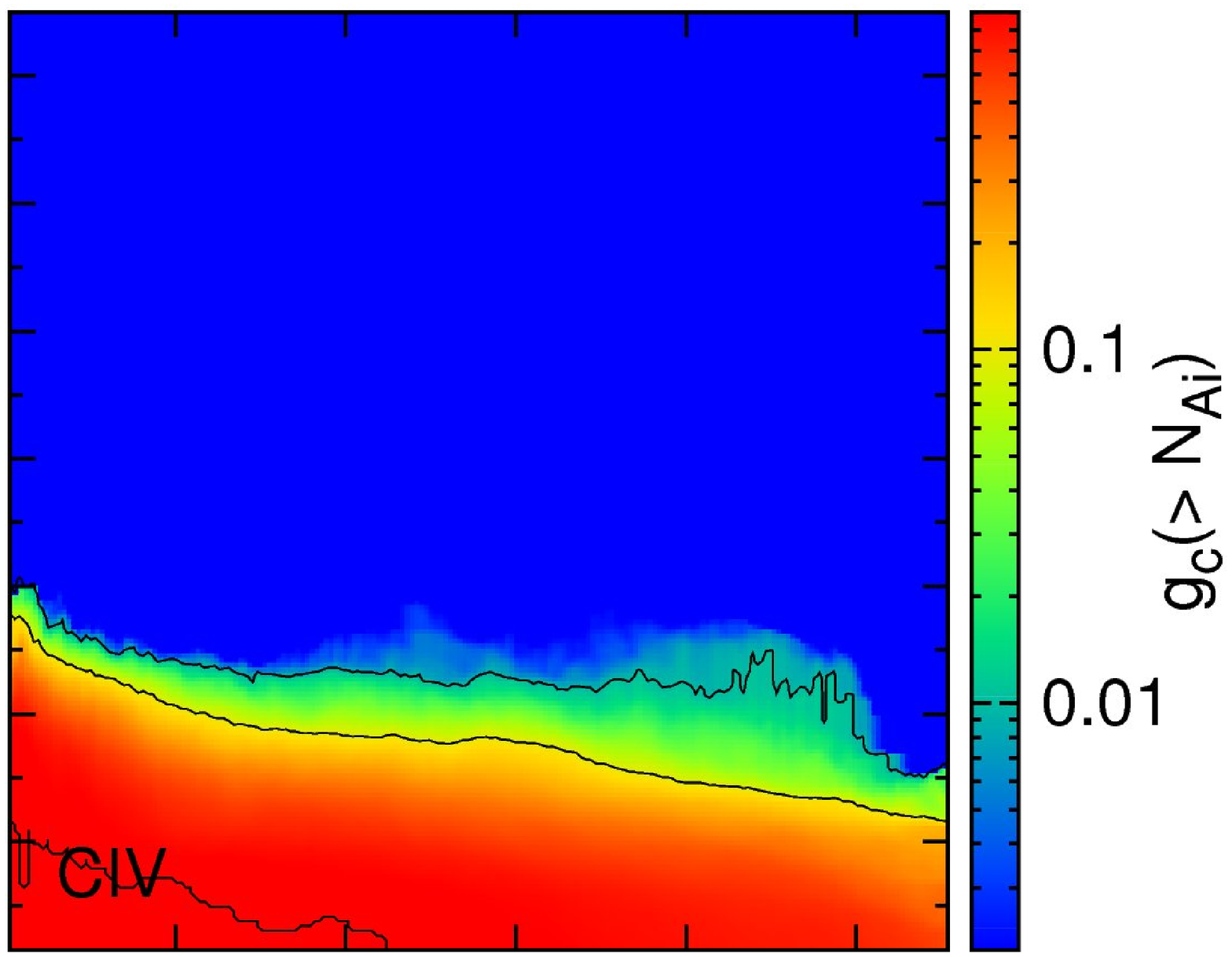}
\includegraphics[width=6.21cm]{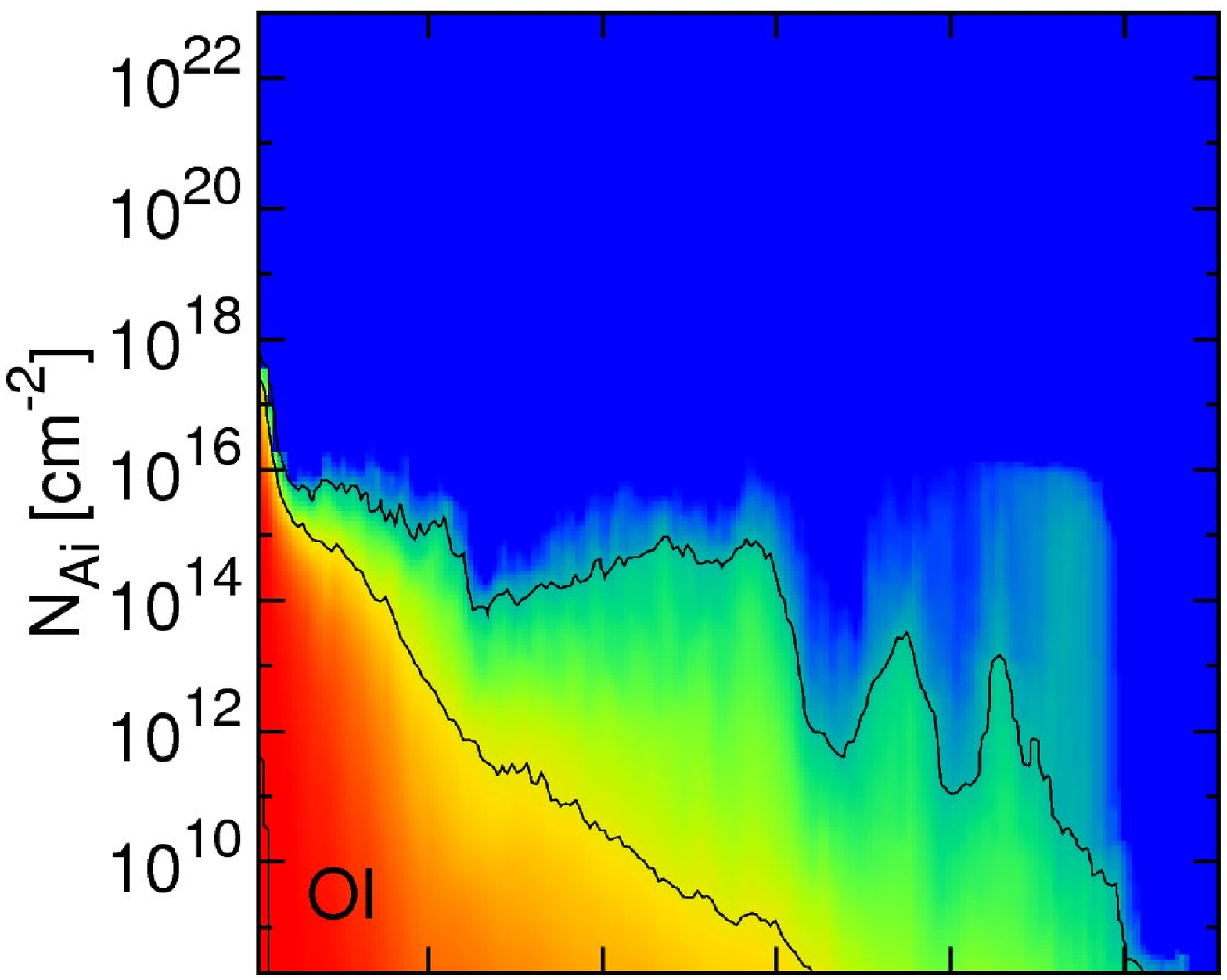}
\includegraphics[width=4.95cm]{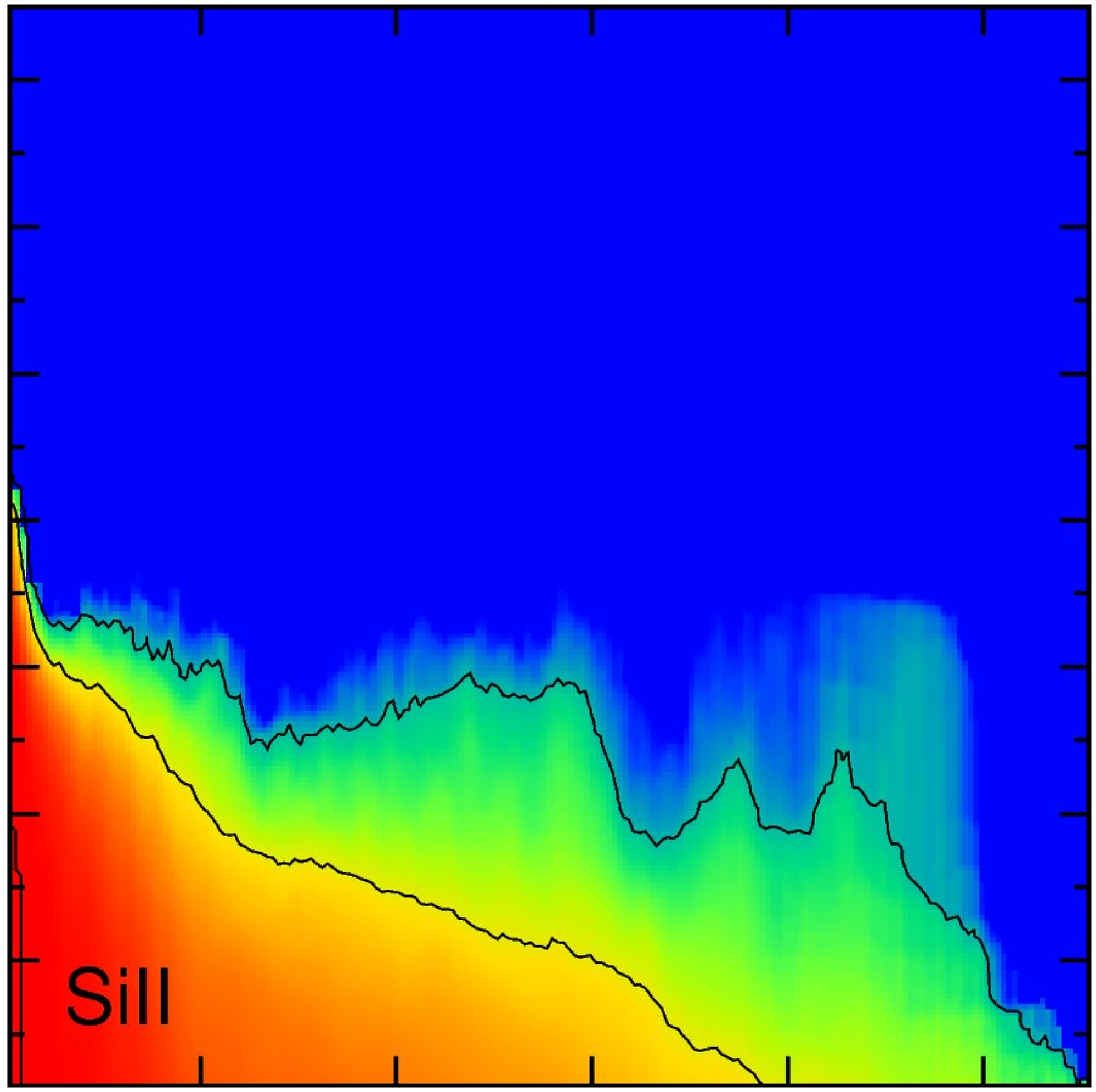}
\includegraphics[width=6.37cm]{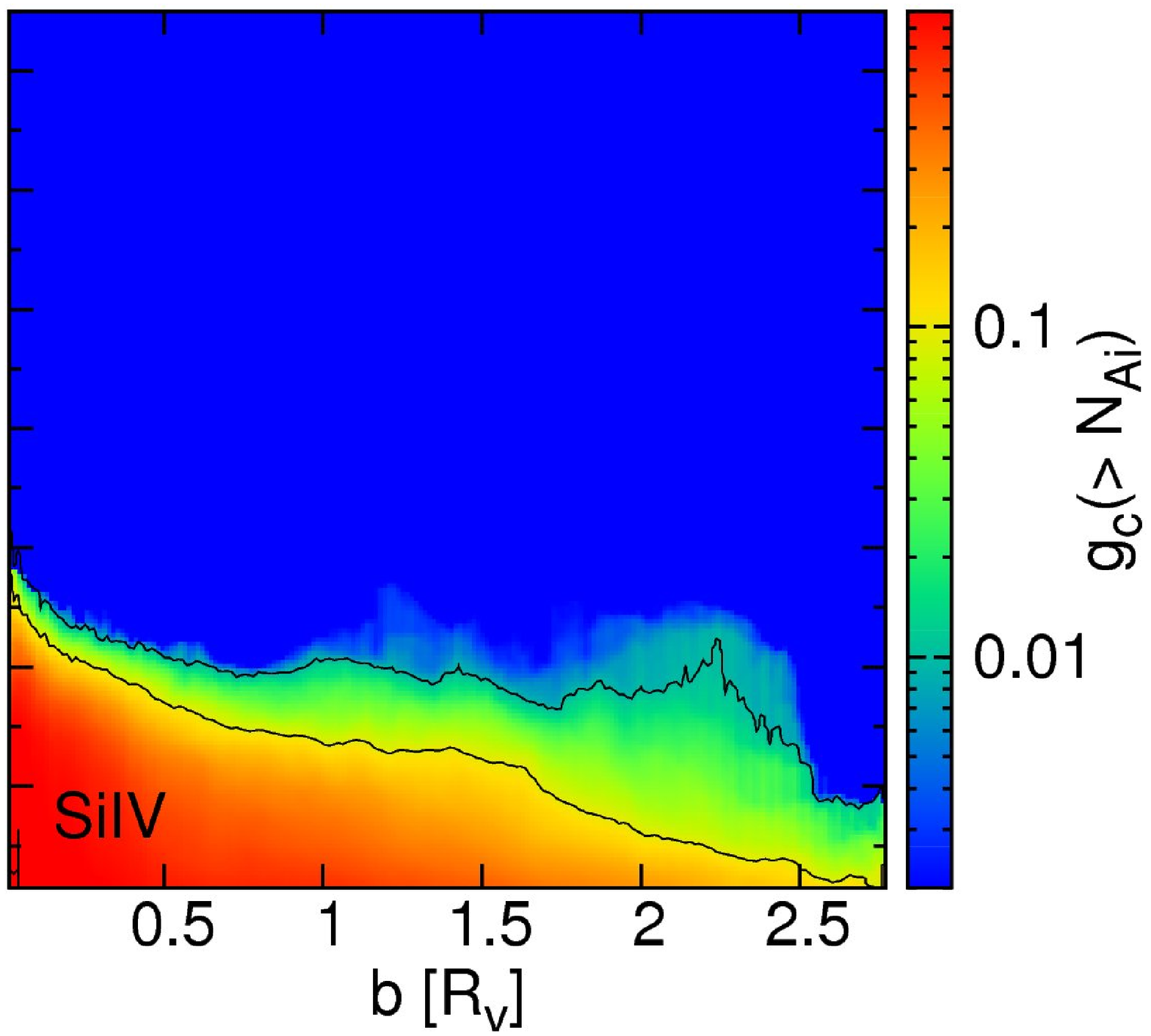}
\includegraphics[width=6.21cm]{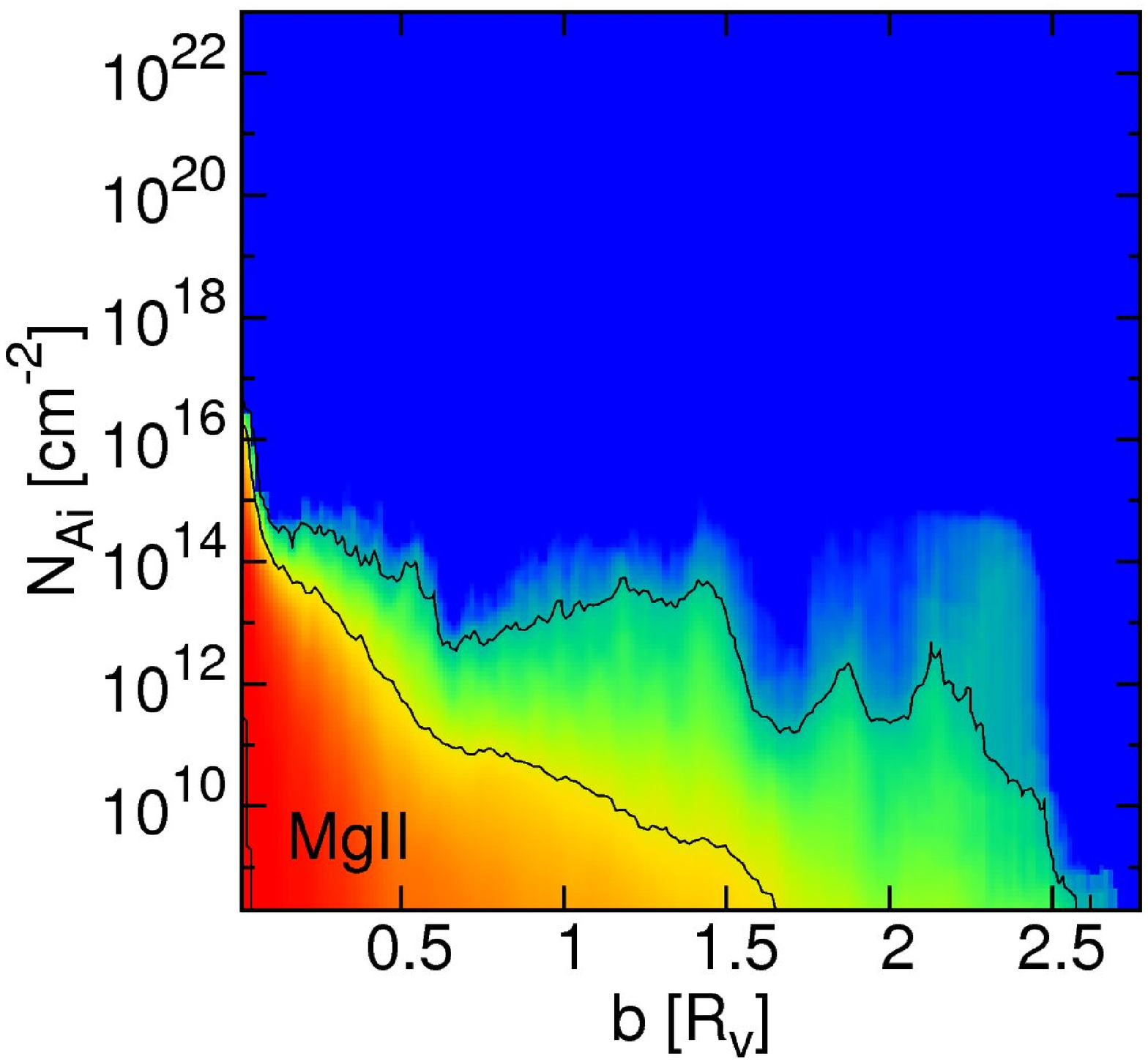}
\includegraphics[width=11.41cm]{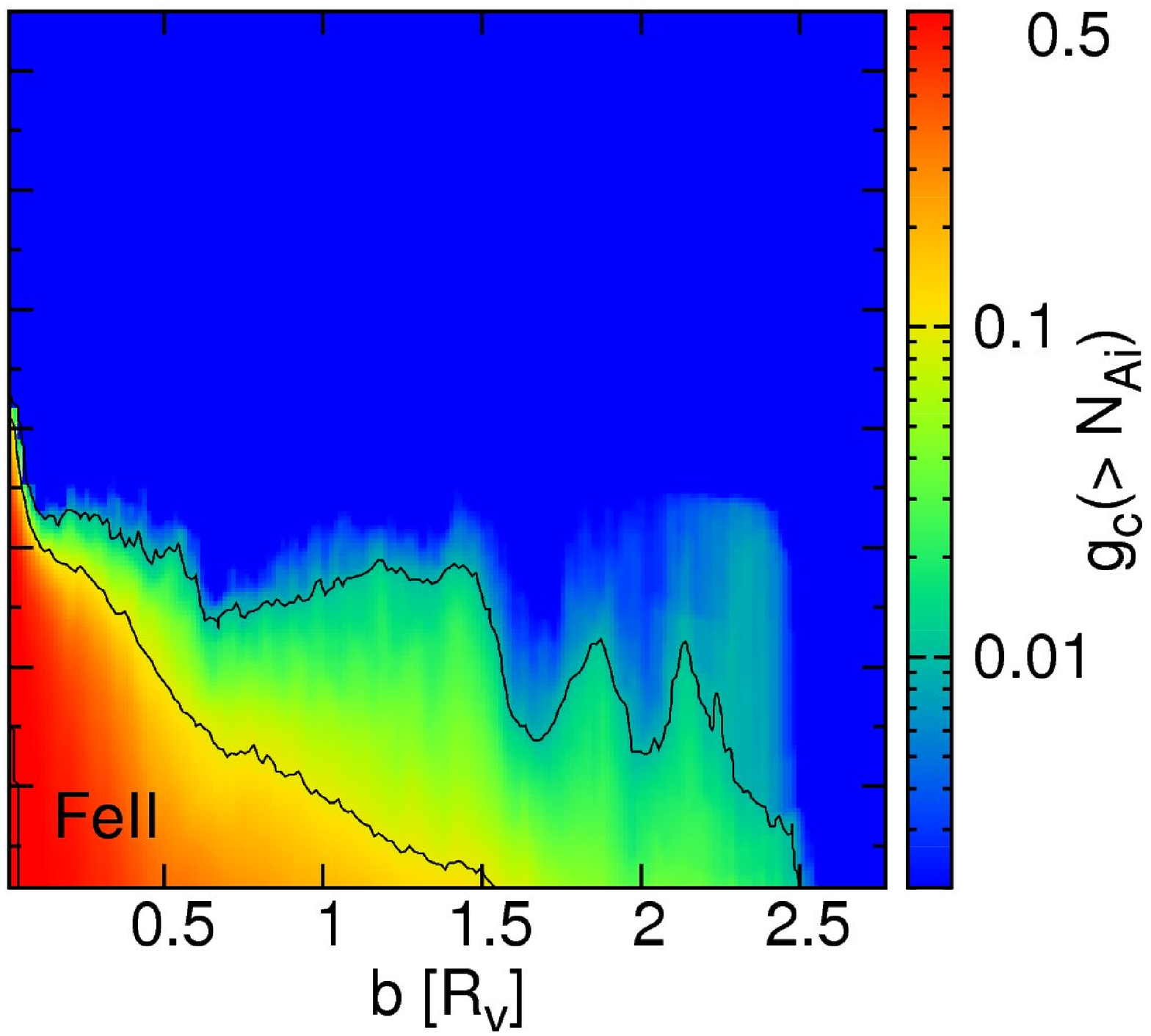}
\end{center}
\caption{Area covering fraction $g_{\rm c}$ as a function of impact parameter
$b$ and cumulative column density for different ionisation states $i$ of the
elements $A$ in the background source geometry. The panels show for every
impact parameter bin $b$ the fraction of the area within this impact parameter
bin which has a column density of $N_{\rm Ai}$ or above. The contours mark area
covering fractions $g_{\rm c}$ of 1.0, 0.1 and 0.01. The values are averaged
over three different galaxies and over three orthogonal directions. We see that
{\HI} shows the highest column densities over the whole radius range. It also
reaches a area covering fraction of 1.0 at $10^{13}$ cm$^{-2}$ over the whole
radius range. The 0.1 and 0.01 area covering fraction lines a decreasing
monotonically with radius for all panels. The metal ionisation states with the
highest area covering fractions are {\CII} and {\CIV}. The {\HI} panel agrees
with the results of {\f11} (their figure 4), \citet[][their Table 1]{faucher2}
as well as with the results of \citet{shen12} (their figure 10).}
\label{fig:areacovbg}
\end{figure*}

For a first impression of the simulation data in the background-source geometry
we show in figure \ref{fig:NHIbg} a map of projected neutral hydrogen column
density. The box of $140\kpc$ is centred on our fiducial galaxy of $\Mv \simeq
4\times 10^{11}\msun$ at $z = 2.3$, which serves as the foreground galaxy. The
column density $N_{\rm Ai}(x,y)$ of element $A$ and ionisation state $i$ at the
position $(x,y)$ itself is now calculated by
\begin{equation}
N_{\rm Ai}(x,y) = \int^{\rm bs}_{\rm -bs} x_{\rm Ai}(\vec{x}) \ n_{\rm A}(\vec{x}) \ dz,
\end{equation}
where $bs$ indicates the box size of twice the virial radius,
$n_{\rm A}(\vec{x})$ is the total gas density of element $A$ at position
$\vec{x} = (x,y,z)$ and $x_{\rm Ai}$ is the ionisation fraction of element $A$
in state $i$. This equation is the analog of equation \ref{eq:colden} for the
central geometry. The map in figure \ref{fig:NHIbg} corresponds to figure
\ref{fig:hammer} in the central geometry. The outer circle marks the virial
radius and the inner circle is at $1/3\ \Rv$. Contours are shown for $10^{17}$,
$10^{19}$ and $10^{21}\cmms$. One can see the three major streams which are very
pronounced outside $1/3 \ \Rv$. They have neutral hydrogen column densities of
$N_{\rm HI} > 10^{17}$ cm$^{-2}$. The background outside $1/3 \ \Rv$ has neutral
hydrogen column densities of $N_{\rm HI} < 10^{15}$ cm$^{-2}$. In the messy region
inside $1/3 \ \Rv$ one cannot distinguish between streams, galactic disk or
background. The neutral hydrogen column densities within this region vary
between $10^{15}$ cm$^{-2}$ and $10^{17}$ cm$^{-2}$. One should note that this
ideal map cannot be directly compared to any observed galaxy, because it
requires a background point source along the same line of sight of each pixel
in the map.

The impression from figure \ref{fig:NHIbg} is that the column density tends to
decrease with increasing distance from the galaxy centre. Figure
\ref{fig:avcolden} shows the area-weighted average column density, for {\HI}
and five different metal lines, as a function of the impact parameter $b$. The
{\HI} column density decreases through the interaction region in the greater
disc vicinity, from 10$^{22}$ cm$^{-2}$ near the centre of the galaxy to
10$^{20}$ cm$^{-2}$ at 0.5 $\Rv$. It remains roughly constant at 10$^{19}$
cm$^{-2}$ in the stream regime out to 1.5 $\Rv$, and is not much lower out to
2.5 $\Rv$. The metal ionisation states also decrease inside 0.5 $\Rv$ and are
roughly constant at values between 10$^{14}$\,cm$^{-2}$ (\OI) and 10$^{11}$
cm$^{-2}$ (\SiIV) in the outer halo and beyond.

\begin{figure*}
\begin{center}
\psfrag{relative intensity}[B][B][1][0] {relative intensity}
\psfrag{bg}[Bl][Bl][1][0] {background source}
\psfrag{corrfig}[Bl][Bl][1][0] {see figure \ref{fig:EWgalmap}}
\psfrag{vks}[B][B][1][0] {$\Delta w$ [km s$^{-1}$]}
\psfrag{lya}[Bl][Bl][1][0] {\la}
\psfrag{A}[B][B][1][0] {A}
\psfrag{B}[B][B][1][0] {B}
\psfrag{C}[B][B][1][0] {C}
\psfrag{D}[B][B][1][0] {D}
\psfrag{E}[B][B][1][0] {E}
\psfrag{F}[B][B][1][0] {F}
\psfrag{G}[B][B][1][0] {G}
\psfrag{H}[B][B][1][0] {H}
\psfrag{I}[B][B][1][0] {I}
\psfrag{J}[B][B][1][0] {J}
\psfrag{K}[B][B][1][0] {K}
\psfrag{L}[B][B][1][0] {L}
\psfrag{NHI=2.9e15}[Bl][Bl][1][0] {$N_{\rm HI} = 3.8 \times 10^{15}$ cm$^{-2}$}
\psfrag{NHI=4.3e19}[Bl][Bl][1][0] {$N_{\rm HI} = 2.6 \times 10^{19}$ cm$^{-2}$}
\psfrag{NHI=4.6e20}[Bl][Bl][1][0] {$N_{\rm HI} = 3.2 \times 10^{20}$ cm$^{-2}$}
\psfrag{NHI=3.0e19}[Bl][Bl][1][0] {$N_{\rm HI} = 9.3 \times 10^{19}$ cm$^{-2}$}
\psfrag{NHI=6.8e13}[Bl][Bl][1][0] {$N_{\rm HI} = 6.8 \times 10^{13}$ cm$^{-2}$}
\psfrag{NHI=5.4e13}[Bl][Bl][1][0] {$N_{\rm HI} = 6.1 \times 10^{13}$ cm$^{-2}$}
\psfrag{NHI=1.3e13}[Bl][Bl][1][0] {$N_{\rm HI} = 1.3 \times 10^{13}$ cm$^{-2}$}
\psfrag{NHI=3.3e14}[Bl][Bl][1][0] {$N_{\rm HI} = 3.0 \times 10^{14}$ cm$^{-2}$}
\psfrag{NHI=4.1e21}[Bl][Bl][1][0] {$N_{\rm HI} = 2.3 \times 10^{21}$ cm$^{-2}$}
\psfrag{NHI=7.3e19}[Bl][Bl][1][0] {$N_{\rm HI} = 1.2 \times 10^{20}$ cm$^{-2}$}
\psfrag{NHI=}[Bl][Bl][1][0] {$N_{\rm HI} =$}
\psfrag{1.4e18}[Bl][Bl][1][0] {$1.2 \times 10^{18}$ cm$^{-2}$}
\psfrag{3.8e18}[Bl][Bl][1][0] {$2.6 \times 10^{18}$ cm$^{-2}$}
\psfrag{W0=0.68A}[Bl][Bl][1][0] {$W_0 = 0.68$ \AA}
\psfrag{W0=1.2A}[Bl][Bl][1][0] {$W_0 = 1.2$ \AA}
\psfrag{W0=2.6A}[Bl][Bl][1][0] {$W_0 = 2.6$ \AA}
\psfrag{W0=6.1A}[Bl][Bl][1][0] {$W_0 = 6.1$ \AA}
\psfrag{W0=3.7A}[Bl][Bl][1][0] {$W_0 = 3.7$ \AA}
\psfrag{W0=1.0A}[Bl][Bl][1][0] {$W_0 = 1.0$ \AA}
\psfrag{W0=0.19A}[Bl][Bl][1][0] {$W_0 = 0.19$ \AA}
\psfrag{W0=0.17A}[Bl][Bl][1][0] {$W_0 = 0.17$ \AA}
\psfrag{W0=0.043A}[Bl][Bl][1][0] {$W_0 = 0.043$ \AA}
\psfrag{W0=0.50A}[Bl][Bl][1][0] {$W_0 = 0.50$ \AA}
\psfrag{W0=7.6A}[Bl][Bl][1][0] {$W_0 = 7.6$ \AA}
\psfrag{W0=3.3A}[Bl][Bl][1][0] {$W_0 = 3.3$ \AA}
\includegraphics[width=5.47cm]{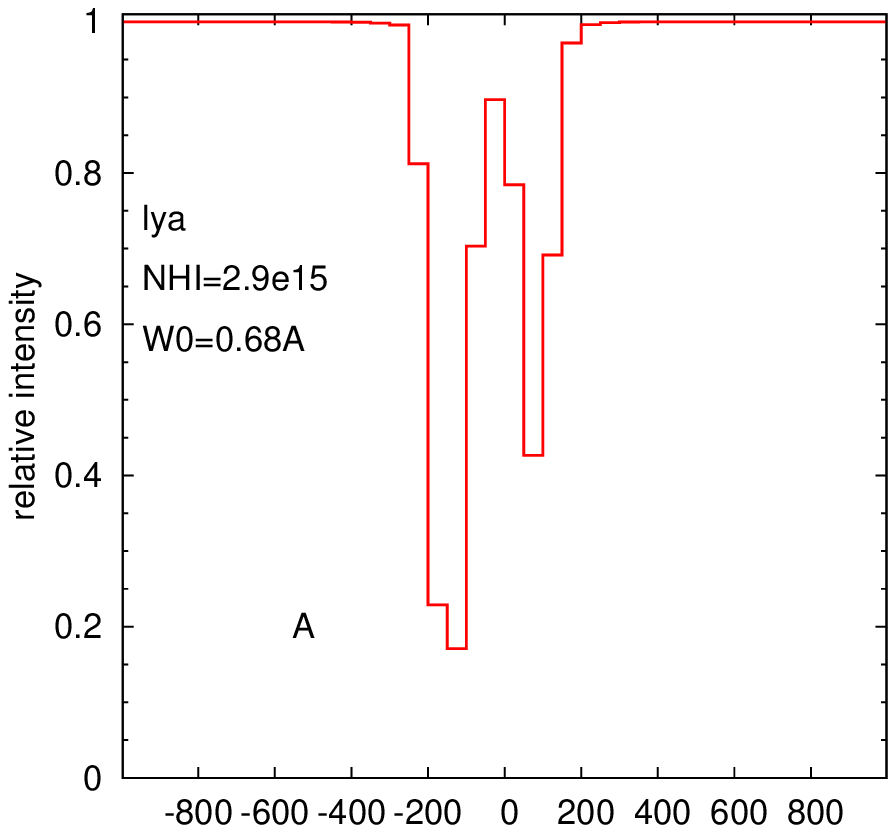}
\includegraphics[width=4.75cm]{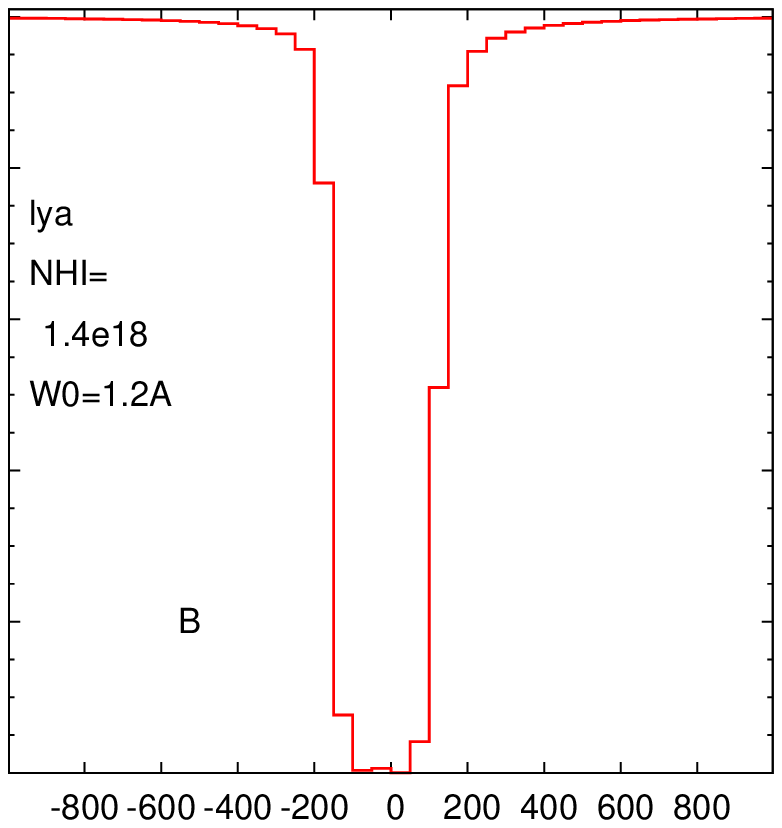}
\includegraphics[width=4.75cm]{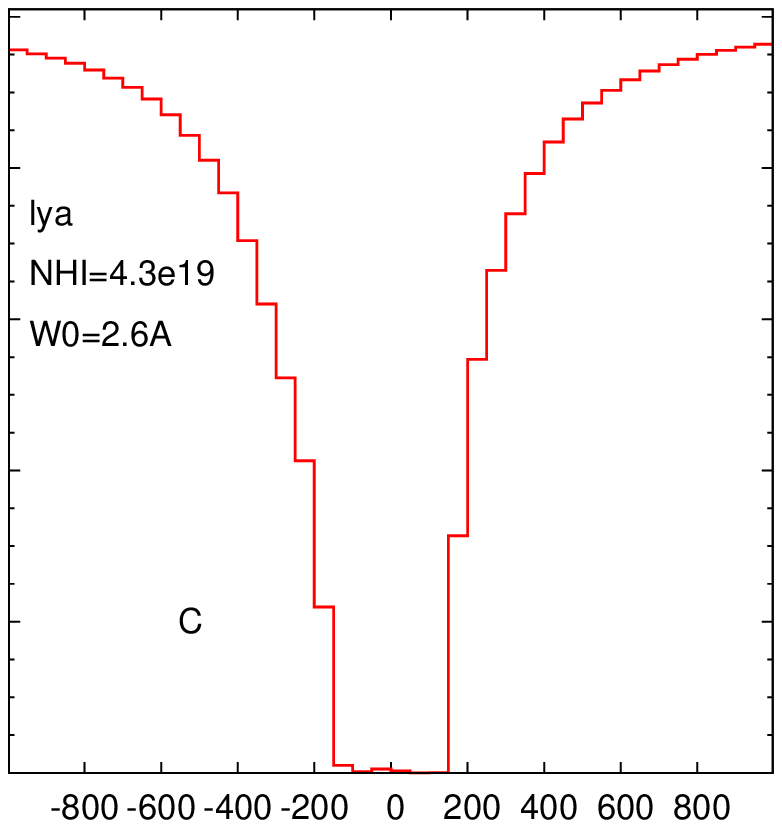}
\includegraphics[width=5.47cm]{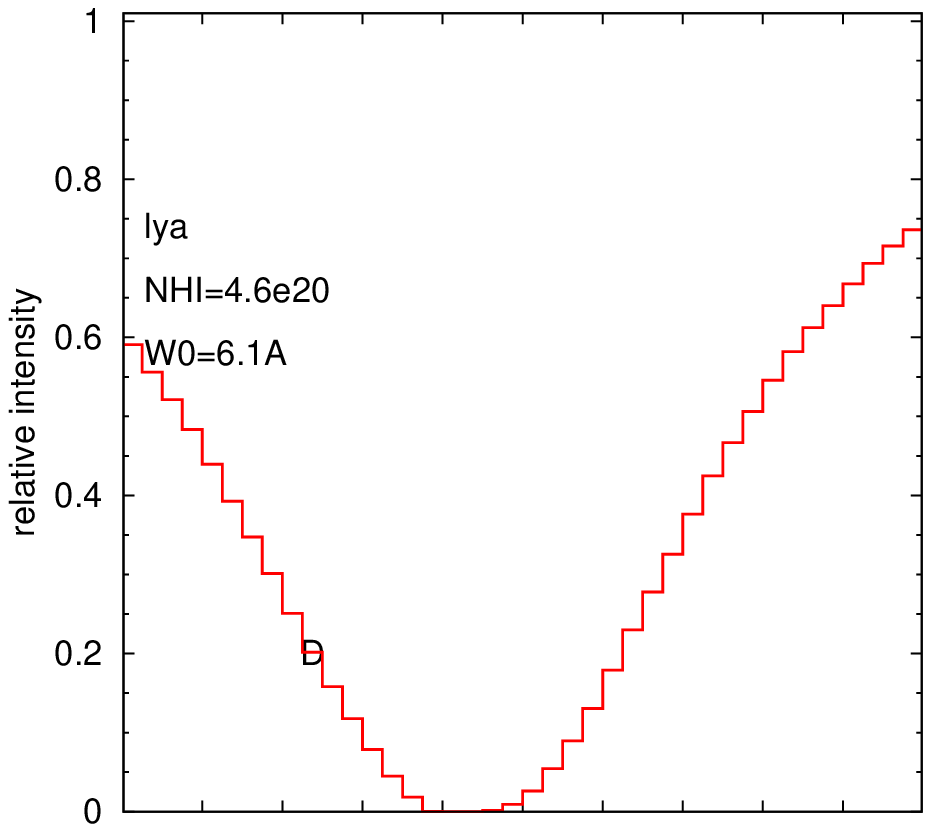}
\includegraphics[width=4.75cm]{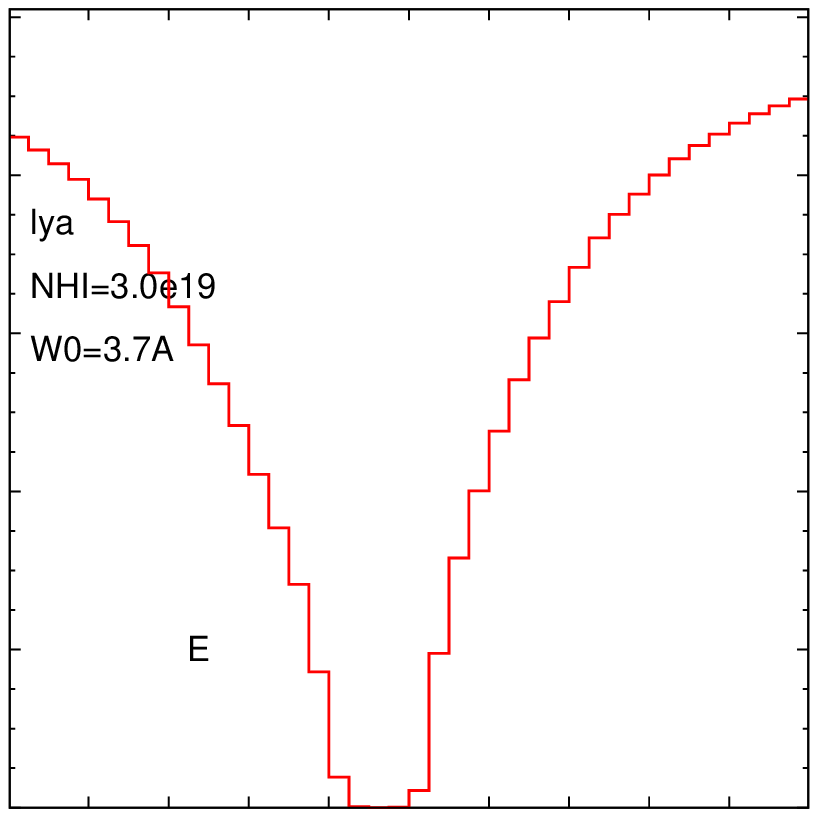}
\includegraphics[width=4.75cm]{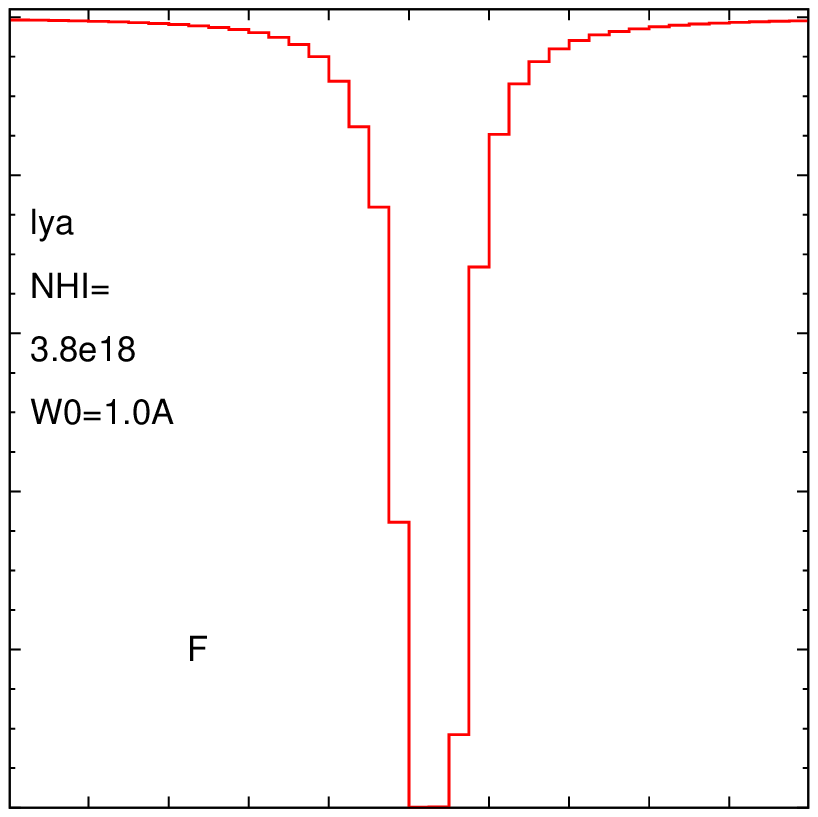}
\includegraphics[width=5.47cm]{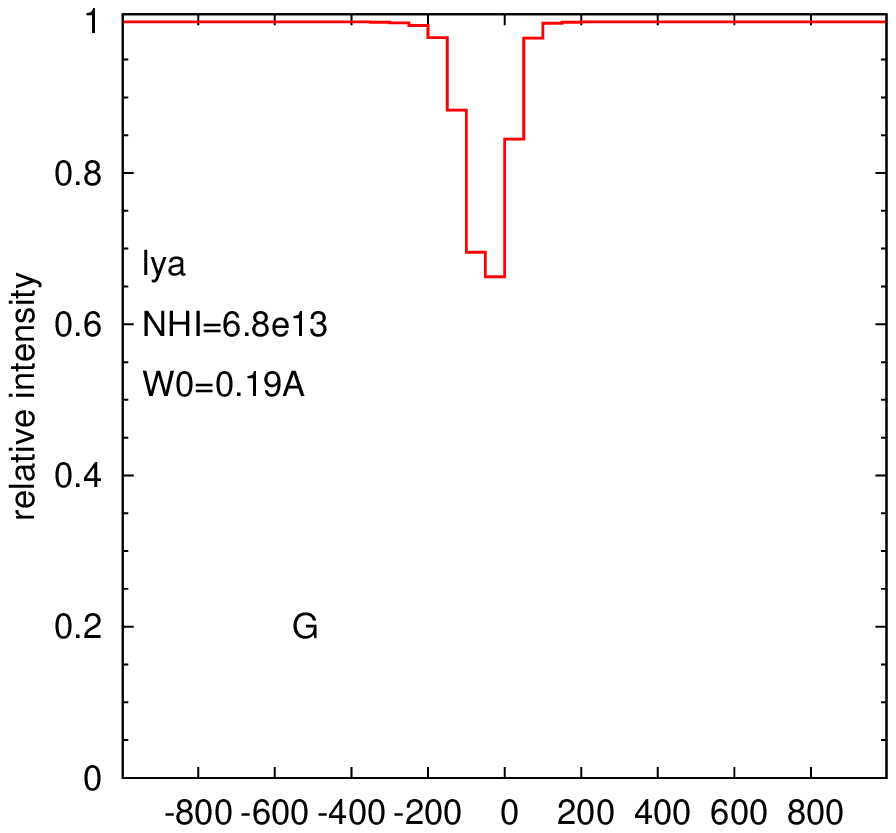}
\includegraphics[width=4.75cm]{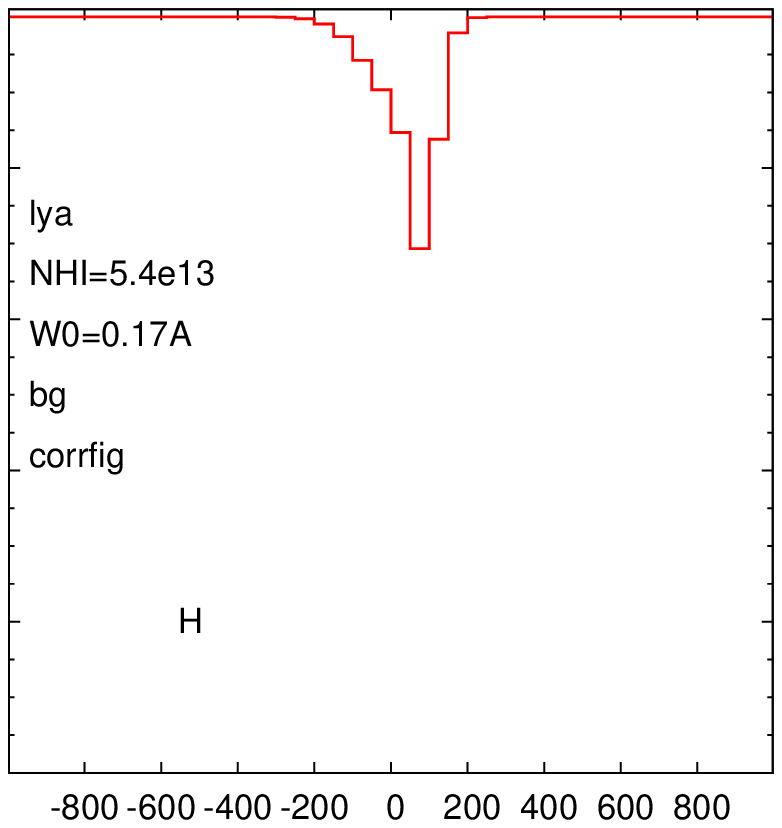}
\includegraphics[width=4.75cm]{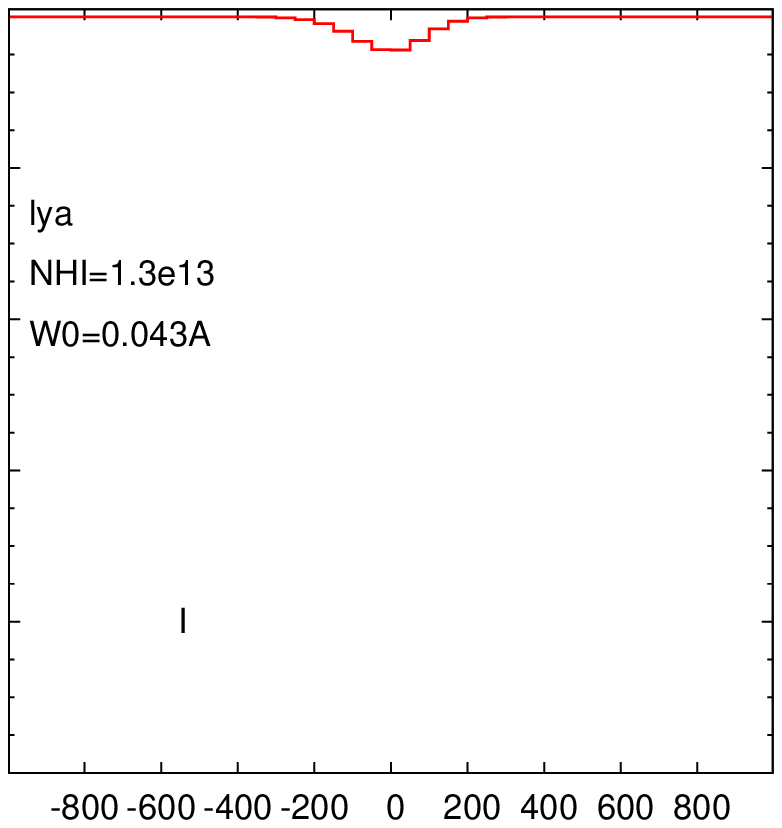}
\includegraphics[width=5.47cm]{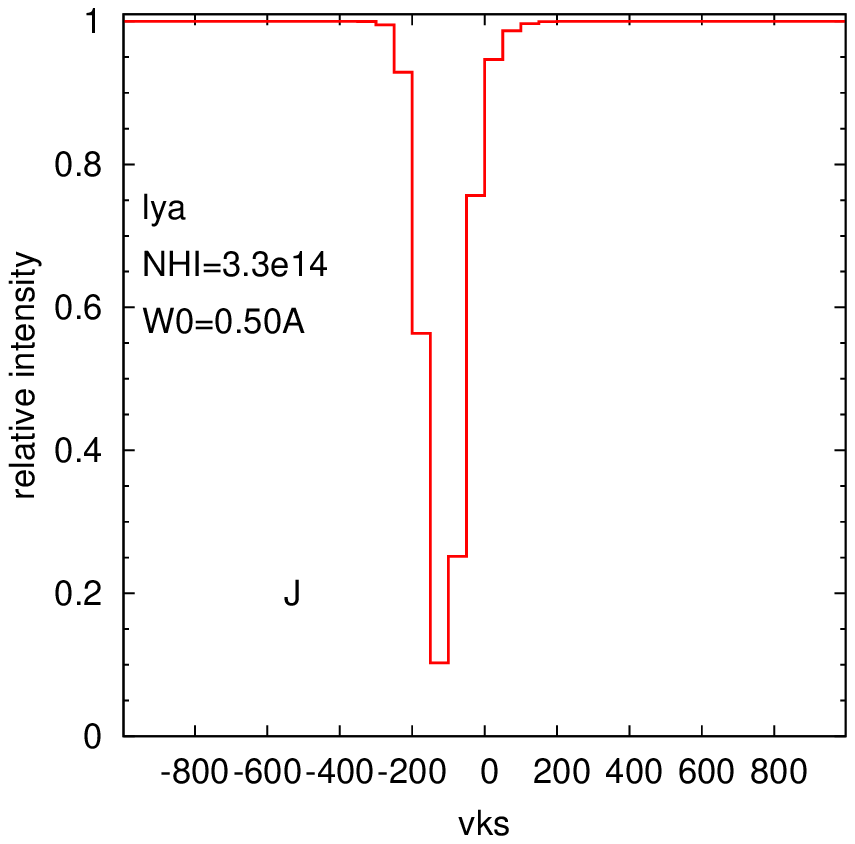}
\includegraphics[width=4.75cm]{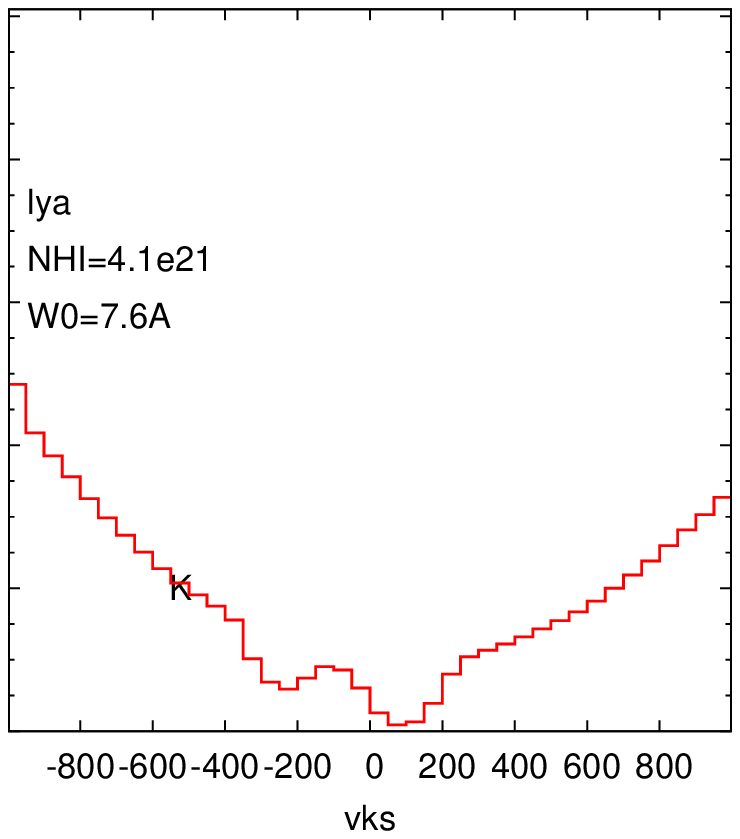}
\includegraphics[width=4.75cm]{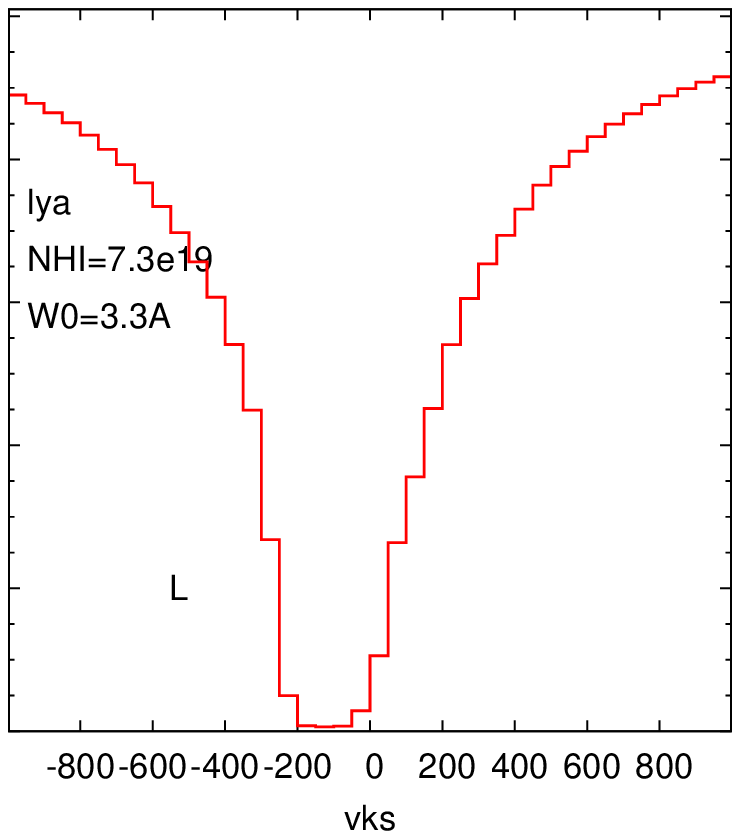}
\end{center}
\caption{Example {\la} absorption line profiles seen from a single direction in
the background geometry integrated from 0.3 to 1.0 $\Rv$. For a background
source one cannot tell from the redshift or blueshift of the line relative to
the central restframe whether it represents inflow or outflow with respect to
the galaxy centre. From inspection of the three-dimensional information in the
simulations, we know that almost in every case a significant coherent shift of
the line is associated with inflow. Values for neutral hydrogen column density
and {\EW} are quoted in each panel. Uppermost panel: various modes of bimodal
streaming, intermediate upper panel: different examples of strong
unidirectional streaming, intermediate lower panel: low absorption profiles and
lowermost panel: extraordinary cases. Some of the lines are saturated like B,
C, D, E, F, K and L, whereas others are not. The letters correspond to the
indicated position of the map from figure \ref{fig:EWgalmap}.}
\label{fig:backviewang}
\end{figure*}

\begin{figure*}
\begin{center}
\psfrag{relative intensity}[B][B][1][0] {relative intensity}
\psfrag{bg}[Bl][Bl][1][0] {background source}
\psfrag{corrfig}[Bl][Bl][1][0] {see figure \ref{fig:EWgalmap}}
\psfrag{vks}[B][B][1][0] {$\Delta w$ [km s$^{-1}$]}
\psfrag{K}[B][B][1][0] {K}
\psfrag{W0=0.17A}[Bl][Bl][1][0] {$W_0 = 0.17$ \AA}
\psfrag{W0=0.14A}[Bl][Bl][1][0] {$W_0 = 0.14$ \AA}
\psfrag{W0=0.25A}[Bl][Bl][1][0] {$W_0 = 0.25$ \AA}
\psfrag{W0=0.13A}[Bl][Bl][1][0] {$W_0 = 0.13$ \AA}
\psfrag{NCII=1.6e15}[Bl][Bl][1][0] {$N_{\rm CII} = 1.6 \times 10^{15}$ cm$^{-2}$}
\psfrag{NSiII=2.0e14}[Bl][Bl][1][0] {$N_{\rm SiII} = 2.0 \times 10^{14}$ cm$^{-2}$}
\psfrag{NMgII=1.9e14}[Bl][Bl][1][0] {$N_{\rm MgII} = 1.9 \times 10^{14}$ cm$^{-2}$}
\psfrag{NFeII=1.8e14}[Bl][Bl][1][0] {$N_{\rm FeII} = 1.8 \times 10^{14}$ cm$^{-2}$}
\psfrag{CII}[Bl][Bl][1][0] {{\CII} (1334 \AA)}
\psfrag{SiIIb}[Bl][Bl][1][0] {{\SiII} (1260 \AA)}
\psfrag{MgIIb}[Bl][Bl][1][0] {{\MgII} (2796 \AA)}
\psfrag{FeIIb}[Bl][Bl][1][0] {{\FeII} (2383 \AA)}
\includegraphics[width=4.94cm]{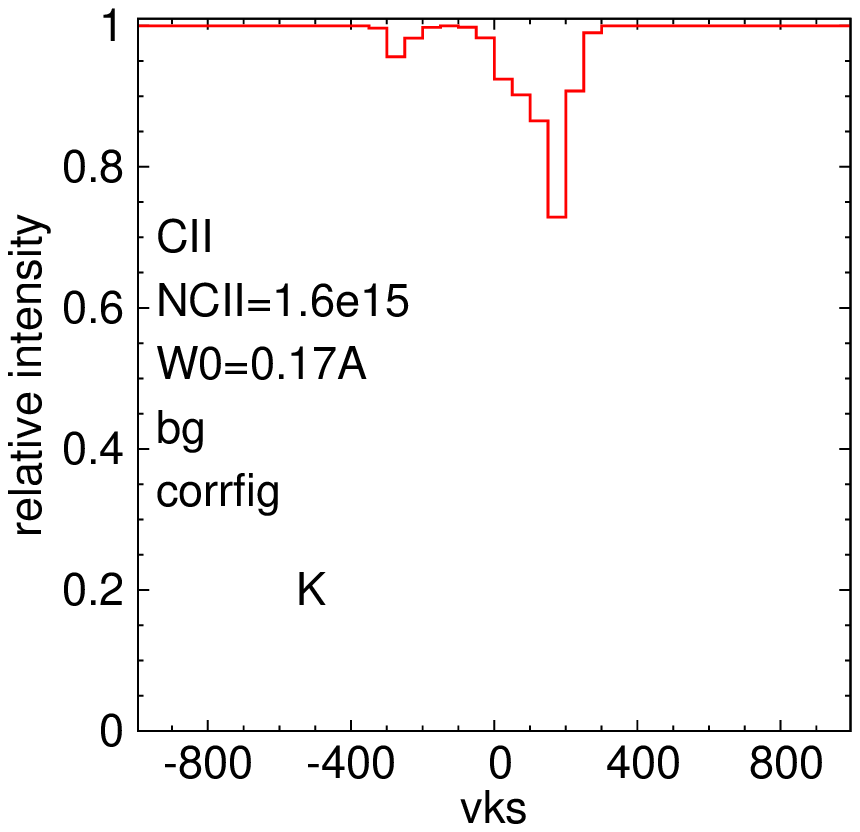}
\includegraphics[width=4.17cm]{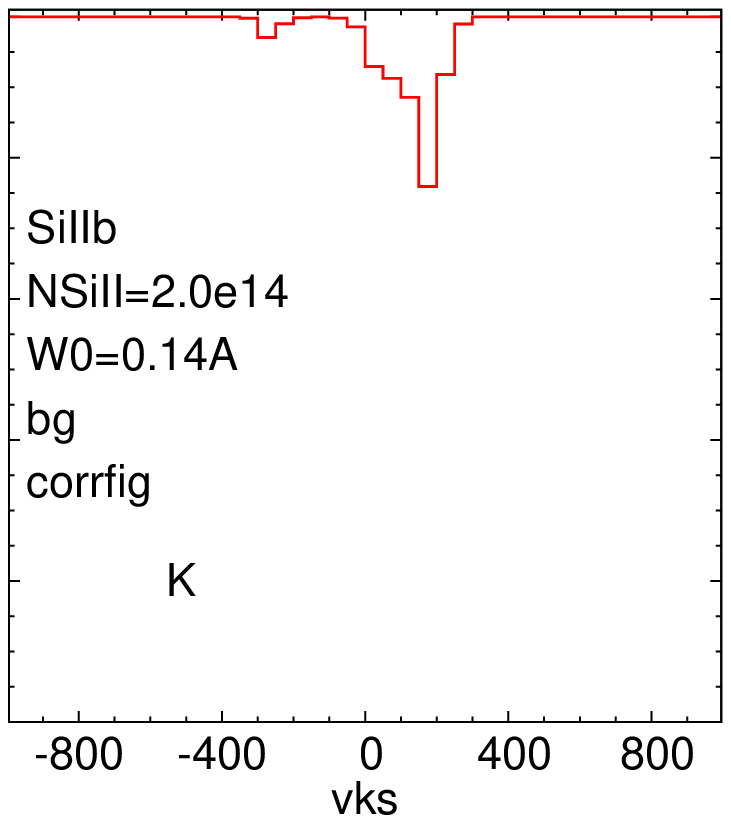}
\includegraphics[width=4.17cm]{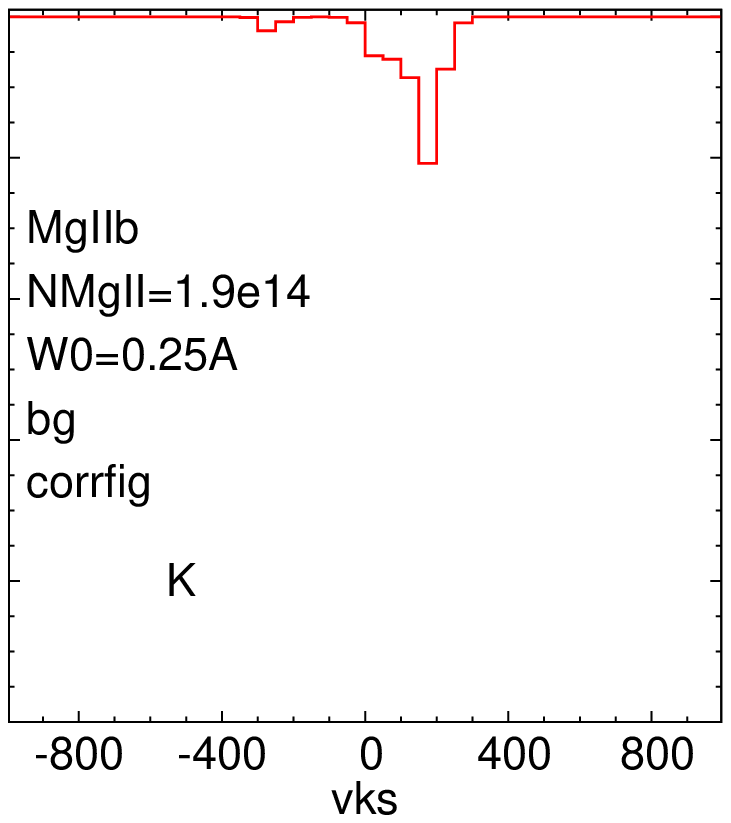}
\includegraphics[width=4.17cm]{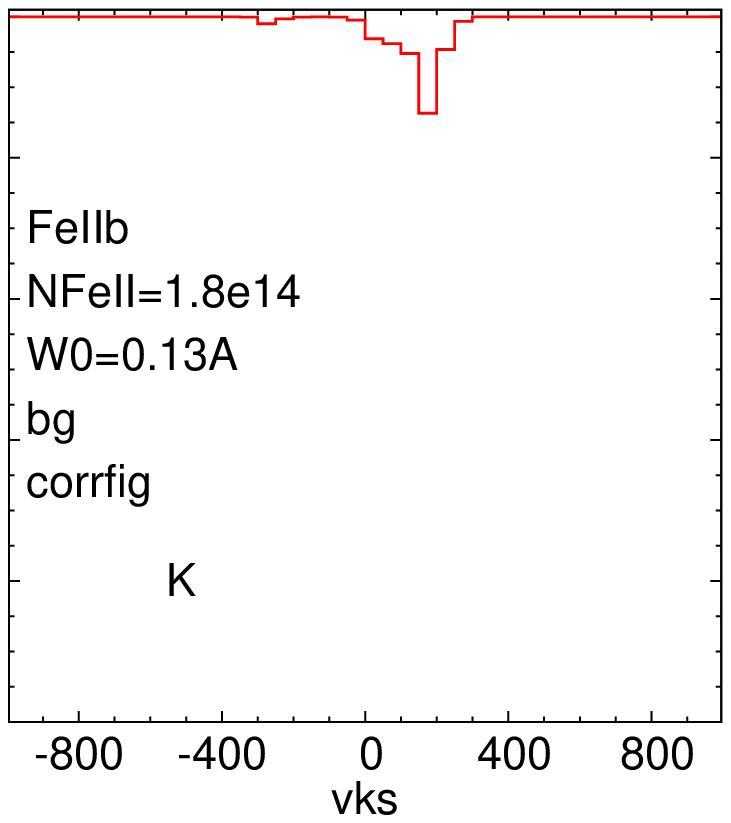}
\end{center}
\caption{Same figure as figure \ref{fig:backviewang}, panel K (the panel with
the deepest signal) this time for selected metal lines providing also the
deepest signal, namely: {\CII} (1334 \AA), {\SiII} (1260 \AA), {\MgII} (2796
\AA) and {\FeII} (2383 \AA). Panel K is fully saturated in {\la} but still
shows peak absorption line depths of 0.25 and {\EW} of $\sim 0.2$ {\AA} for
these metal lines. The {\CII} panel of this plot shows slight differences with
the results of \citet[][their figure 2, lower panel]{kimm} due to different
line computing algorithms.}
\label{fig:metalbackviewang}
\end{figure*}

Figure \ref{fig:NHIbg} is limited to the angle-averaged column density. Figure
\ref{fig:areacovbg} presents the distribution of column densities, for the
purpose of evaluating the likelihood of an observation through a given line of
sight. We plot area covering fractions $g_{\rm c}$ as a function of impact
parameter $b$ and cumulative column density $N_{\rm Ai}$. The values are
averaged over three different galaxies and three orthogonal projections per
galaxy. The panels show for every impact parameter $b$ the fraction of the area
within an annulus of radius $b$ and width $\delta b$ which has a column density
of $N_{\rm Ai}$ or above. One can see that there are obviously much higher
columns of {\HI} present than of the other metal ionisation state. {\HI} has
column densities which are several orders of magnitude higher, depending on the
actual radius and the area covering fraction one wants to look at. The columns
of the metal ionisation states lie closely to each other and decrease rather
smoothly with increasing radius. The $g_{\rm c} = 0.1$ lines of the two carbon
lines are fairly similar, both fall off almost linearly from $\sim 10^{15}$
cm$^{-2}$ at the left edge of the graph to $\sim 10^{10}$ cm$^{-2}$ at the right
edge of the graph. This is also true for the $g_{\rm c} = 0.1$ lines of the two
silicon lines. They fall off from around $10^{14}$ to below $10^{8}$. The
$g_{\rm c} = 0.01$ line, which lies much higher, on the other hand has more
differences between the low-ionised metals (like {\CII}, {\SiII} or {\FeII}) on
the higher ionised metals (like {\CIV} or {\SiIV}): In case of the low-ionised
metals the line has considerably more features like the extended bumps at impact
parameters of $1.0 \ \Rv$ or $2.0 \ \Rv$. The distribution for {\CII} indicates
the highest, the one for {\MgII} indicates the lowest overall column densities.

\begin{figure}
\begin{center}
\includegraphics[width=8.45cm]{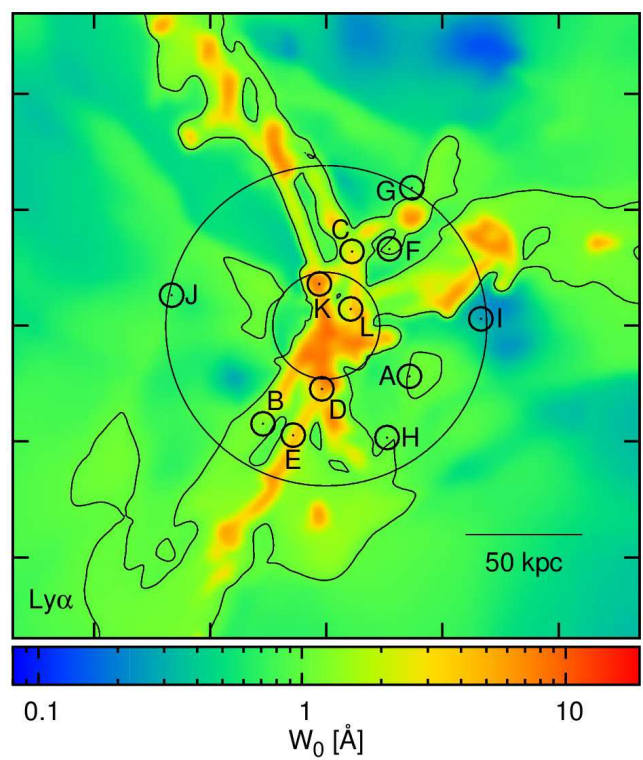}
\end{center}
\caption{Map of the {\la} absorption line {\EW}s for the background geometry of
one of our simulated galaxies. The box side is $276\kpc$ (physical). The outer
circle marks the virial radius and the inner circle is at $1/3\,\Rv$. The
contour lines indicate an {\EW} of 1 and 0.1\,{\AA} respectively. The circles
with the letters attached to it indicate the position of the example line
profiles from figure \ref{fig:backviewang}. The three major streams as well as
clumps (satellites with dark matter haloes) have $W_0 > 1$ {\AA}. The
background has $W_0 < 1$ {\AA}.}
\label{fig:EWgalmap}
\end{figure}

The {\HI} panel of figure \ref{fig:areacovbg} is comparable to figure 4 of
{\f11}, who implemented a full radiative transfer calculation. A comparison
between the two allows us to evaluate the accuracy of our simplified treatment
of self-shielding. For impact parameter other than $b = 2.0 \ \Rv$ the latter
show in their figure 4 values which are normalised to the value it would have
at $b = 2.0 \ \Rv$. In the following we will only quote the un-normalised
values. In our results the same covering factors correspond to column densities
that are usually smaller by a factor of $\sim 3$. For example at $b = 2.0 \
\Rv$ {\f11} have a covering factor of  0.01 at $3 \times 10^{20}$ cm$^{-2}$ and
of 0.1 at $10^{17}$ cm$^{-2}$. In our results these covering factors correspond
to column densities of $10^{20}$ cm$^{-2}$ and $3 \times 10^{16}$ cm$^{-2}$,
respectively. The {\HI} panel of figure \ref{fig:areacovbg} is also roughly
comparable to Table 1 of \citet{faucher2}. In our results we have covering
factors which are on average smaller by a factor of 1.5. For example at $b =
0.5 \ \Rv$ they have an average covering factor of $\sim 0.3$ at $10^{17.2}
\lsim N_{\rm HI} \lsim 2 \times 10^{20}$ and of $\sim 0.08$ at $> 2 \times
10^{20}$ cm$^{-2}$. Our corresponding covering factors are 0.2 and 0.1. In their
figure 10, \citet{shen12} quote a {\HI} covering factor of column densities
higher than $10^{17.2}$ cm$^{-2}$ for $r < \Rv$ ($r < 2 \Rv$) of 27\% (10\%). Our
corresponding covering factors are 0.2 and 0.5. We conclude that our results
based on the simplified treatment of self-shielding (section
\ref{sec:linestrength}) provide a rather good approximation to the results from
the full radiative transport analysis.

\begin{figure*}
\begin{center}
\psfrag{EW}[B][B][1][0] {$W_0$ [\AA]}
\psfrag{rRv}[B][B][1][0] {$b$ $[\Rv]$}
\psfrag{bg}[Bl][Bl][1][0] {background source}
\psfrag{La}[Bl][Bl][1][0] {\la}
\psfrag{CDB sim}[Br][Br][1][0] {\textcolor{red}{{\CD} simulation}}
\psfrag{S10 obs}[Br][Br][1][0] {\textcolor{blue}{{\s10} observation}}
\psfrag{CII}[Br][Br][1][0] {\textcolor{red}{{\CII} (1334 \AA)}}
\psfrag{CIVa}[Br][Br][1][0] {\textcolor{cyan}{{\CIV} (1548 \AA)}}
\psfrag{OI}[Br][Br][1][0] {\textcolor{green}{{\OI} (1302 \AA)}}
\psfrag{SiIIa}[Br][Br][1][0] {\textcolor{blue}{{\SiII} (1526 \AA)}}
\psfrag{SiIVa}[Br][Br][1][0] {\textcolor{black}{{\SiIV} (1393 \AA)}}
\psfrag{CIVb}[Br][Br][1][0] {\textcolor{red}{{\CIV} (1550 \AA)}}
\psfrag{SiIIb}[Br][Br][1][0] {\textcolor{magenta}{{\SiII} (1260 \AA)}}
\psfrag{SiIVb}[Br][Br][1][0] {\textcolor{green}{{\SiIV} (1402 \AA)}}
\psfrag{SiIIc}[Br][Br][1][0] {\textcolor{cyan}{{\SiII} (1304 \AA)}}
\includegraphics[width=5.55cm]{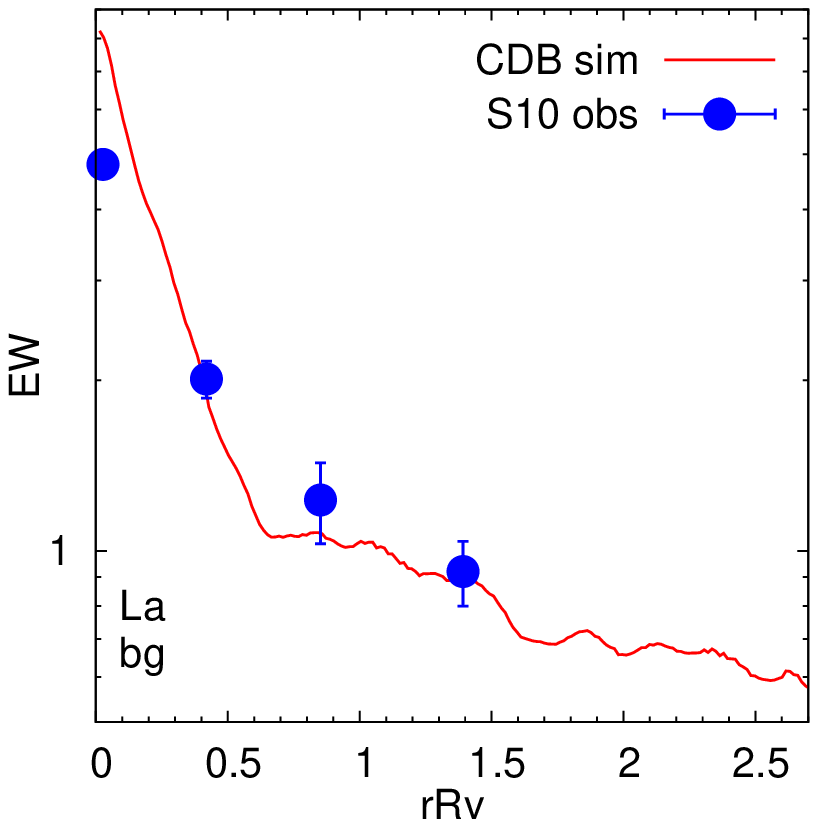}
\includegraphics[width=5.98cm]{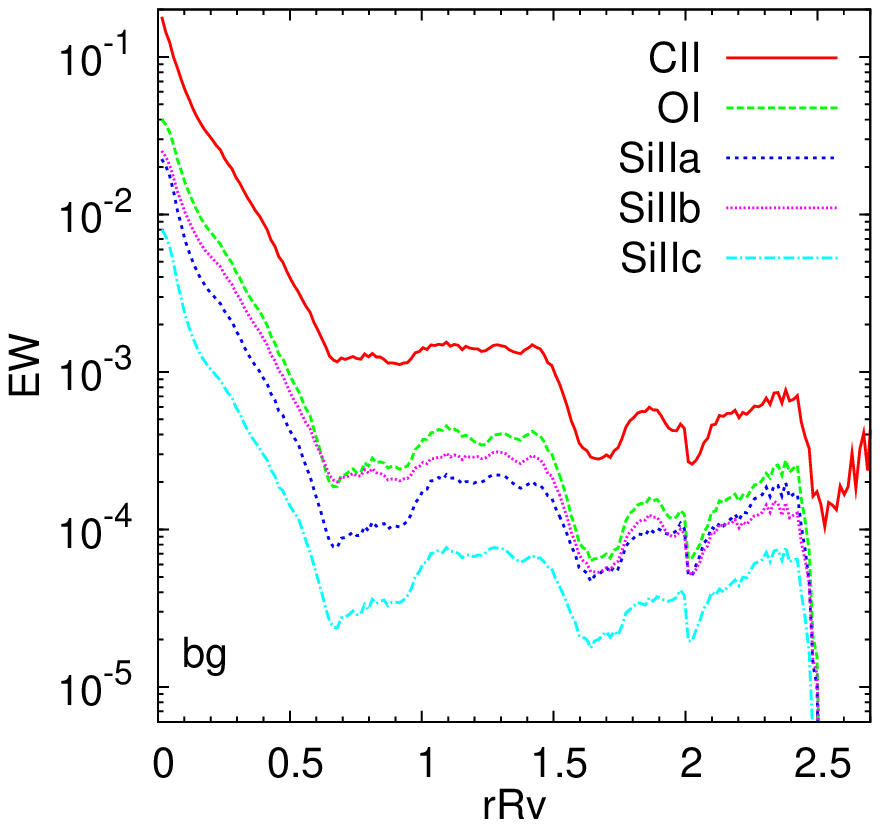}
\includegraphics[width=5.98cm]{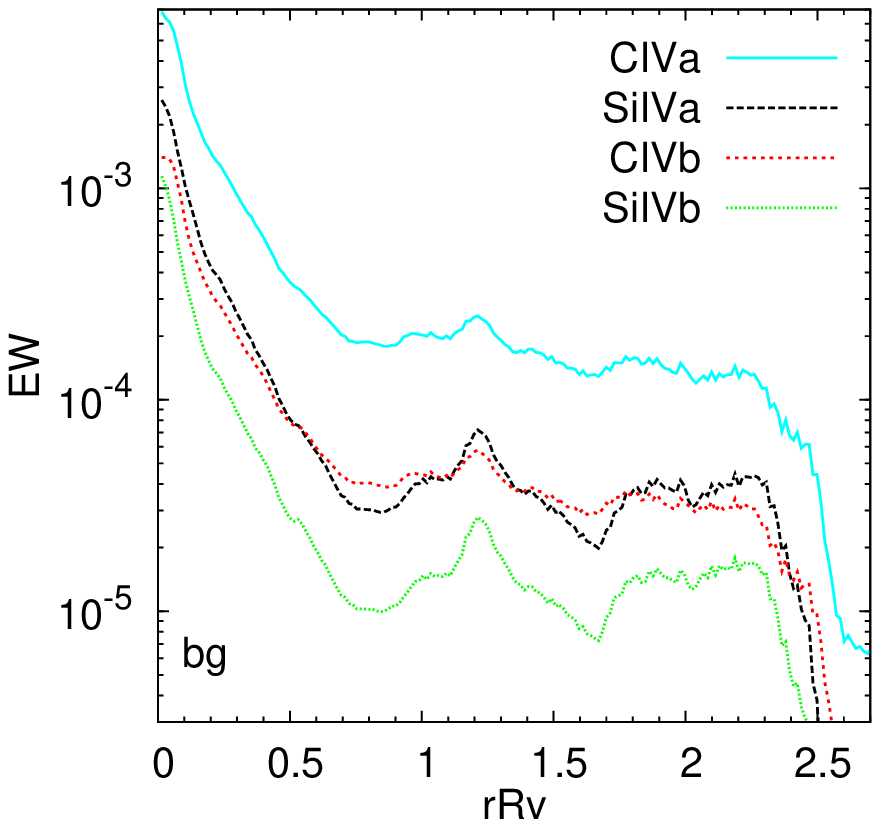}
\end{center}
\caption{Average {\EW} $W_0$ as a function of impact parameter $b$, averaged
over the three simulated galaxies. For {\la}, we show for comparison the
observational results of S10 (their Table 4). The inner galaxy is responsible
for a steep peak in {\EW}s in the range $b \le 0.5 \Rv$. Outside of this radius
the {\EW} is lower and it shows an overall gradually declining with $b$. As
expected there is a strong negative correlation between impact parameter $b$
and {\EW}. This behaviour is the same for all ten lines. The {\la} and the
{\SiII} (1260\,\AA) lines agree with the theoretical results of {\f11} (their
figure 14). The {\la} line also agrees with the theoretical results of
\citet{shen12} (their figure 9). The observational results of {\s10} for {\la}
(their Table 4 and their figure 21) are also in fairly good agreement.}
\label{fig:aEWvsradius}
\end{figure*}

We next compute example absorption line profiles for a background source, in
analogy to the line profiles for a central source, using equations
(\ref{eq:doppler}) to (\ref{eq:ablipro}). From now on, we apply a Gaussian
point spread function with a beamsize (= FWHM) of 4\,kpc, following the
algorithm described in section \ref{sec:cg}. In addition, the velocity
resolution is now downgraded to 50\,km\,s$^{-1}$, to match the velocity
resolution used in the corresponding figures of {\s10}. In figure
\ref{fig:backviewang} and \ref{fig:metalbackviewang} we show a selection of
absorption line profiles, for {\la} and metal lines, respectively,
demonstrating a large variety. These figures are the analogs of figures
\ref{fig:viewang} and \ref{fig:metalviewang} for a central source. Some line
profiles show a single dominant stream, others indicate two streams, some
display weak absorption and others show strong absorption. For a background
source one cannot tell from the redshift or blueshift of the line relative to
the central restframe whether it represents inflow or outflow with respect to
the galaxy centre. From inspection of the three-dimensional information in the
simulations, we know that almost in every case a significant coherent shift of
the line is associated with inflow. In the case of {\la}, 7 out of 12 lines in
the sample are saturated (B, C, D, E, F, K, L). Panel K, which shows the
strongest absorption in {\la}, has for selected metal lines a maximum
absorption depths of only 0.25 and an equivalent width {\EW} of only $\sim
0.2$\,{\AA}. Some of the weaker {\la} lines (e.g, panels I or even H) might be
completely erased by noise and the ISM component at $v = 0$. The leftmost panel
of figure \ref{fig:metalbackviewang} is analogous to the lower panel of figure
3 in \citet{kimm}. The latter obtain for {\CII} a maximum line depth of $\sim
0.9$ with a FWHM $\eta$ of 650 km s$^{-1}$. The difference from our result most
likely stems from the fact that they used a simulation with a resolution lower
by a factor of 20, the Horizon MareNostrum simulation \citep{ocvirk}, and
addressed haloes more massive by a factor of a few. Furthermore, they used a
different prescription for the Gaussian velocity distribution, where they
obtained their velocity dispersion from the neighbouring 26 cells instead of
the more appropriate algorithm described in equations (\ref{eq:doppler}) to
(\ref{eq:voigt}).

As for the central source, we compute the equivalent widths from the absorption
line profiles using equation (\ref{eq:EW}). The {\EW}s for a background source
are mapped for our fiducial galaxy in figure \ref{fig:EWgalmap}. The outer
circle marks the virial radius and the inner circle is at $1/3\,\Rv$. This plot
is the analog of figure \ref{fig:ewhammer} for a central source. The {\EW}s
shown in it span two and a half orders of magnitude. The three major streams
are fairly prominent, in addition to the central region in the greater vicinity
of the disc. Also very prominent are the parts of the IGM which are
stream-less.

Figure \ref{fig:EWgalmap} illustrates the dependence of the {\EW}s on the
projected distance from the centre, or impact parameter $b$. In the galaxy
vicinity at the halo centre, $b< 10\kpc$, say, we see the highest absorption,
which gradually declines toward larger radii, as the volume becomes dominated
by the dilute medium between the streams. Figure \ref{fig:aEWvsradius} shows
the average {\EW} as a function of impact parameter $b$, averaged over the
three simulated galaxies at $z = 2.3$, where the virial masses are $\Mv \simeq
4\times 10^{11} \msun$. For {\la}, we show for comparison the observational
results of {\s10} (their Table 4), assuming a virial radius of 74\,kpc. In our
simulation results for all the lines, the inner galaxy is responsible for a
steep peak in {\EW}s in the range $b \leq 0.5 \Rv$. Outside of this radius the
{\EW} is lower and it shows an overall gradually declining with $b$. A
comparison with the observations shown in figure 21 of {\s10} indicates that
the observational {\EW} for the {\la} line lies a factor of two or less below
our predicted value at low $b$, and is consistent with our prediction at larger
radii. The simulation results in figure 14 of {\f11} and in figure 9 of
\citet{shen12} for {\la} are in excellent agreement with our current results,
and so are the {\f11} predictions for {\SiII} (1260 \AA). The observational
results of {\s10} for the {\EW} of the metal lines are two orders of magnitude
above our predictions. These are likely to reflect massive, cold, dense and
metal-rich outflows that are not reproduced in an appropriate amplitude in our
simulations.

\begin{figure}
\begin{center}
\includegraphics[width=8.45cm]{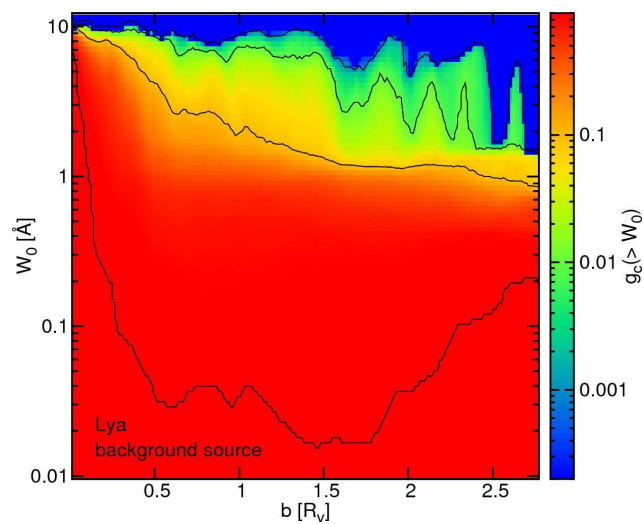}
\end{center}
\caption{Area covering fraction $g_{\rm c}$ as a function of impact parameter
$b$ and cumulative {\EW} $W_0$ for the {\la} absorption line in the background
source geometry. Shown is for every impact parameter bin $b$ the fraction of
the area within this impact parameter bin which has a {\EW} of $W_0$ or above.
The contours mark fractions of 1.0, 0.1, 0.01 and 0.001. The values are
averaged over three different galaxies and the six principal axis. We see that
we have an {\EW} $> 0.02$ {\AA} everywhere. The 0.1 area covering fraction line
is falling monotonically with $b$, whereas the 0.01 and 0.001 lines show noisy
distortions at $b > 1.5 \ R_{\rm v}$.}
\label{fig:laareacovewbg}
\end{figure}

\begin{figure*}
\begin{center}
\includegraphics[width=6.15cm]{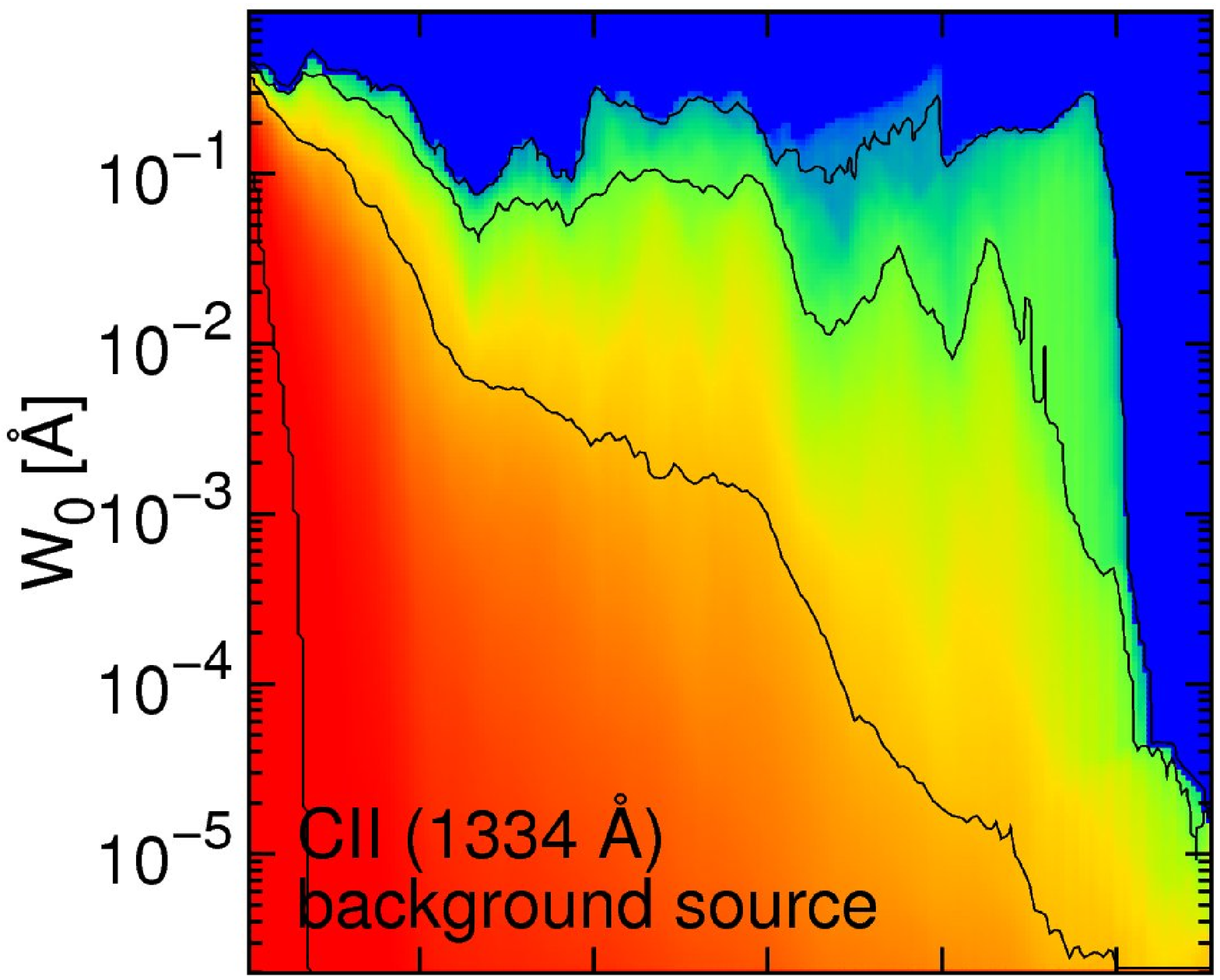}
\includegraphics[width=4.92cm]{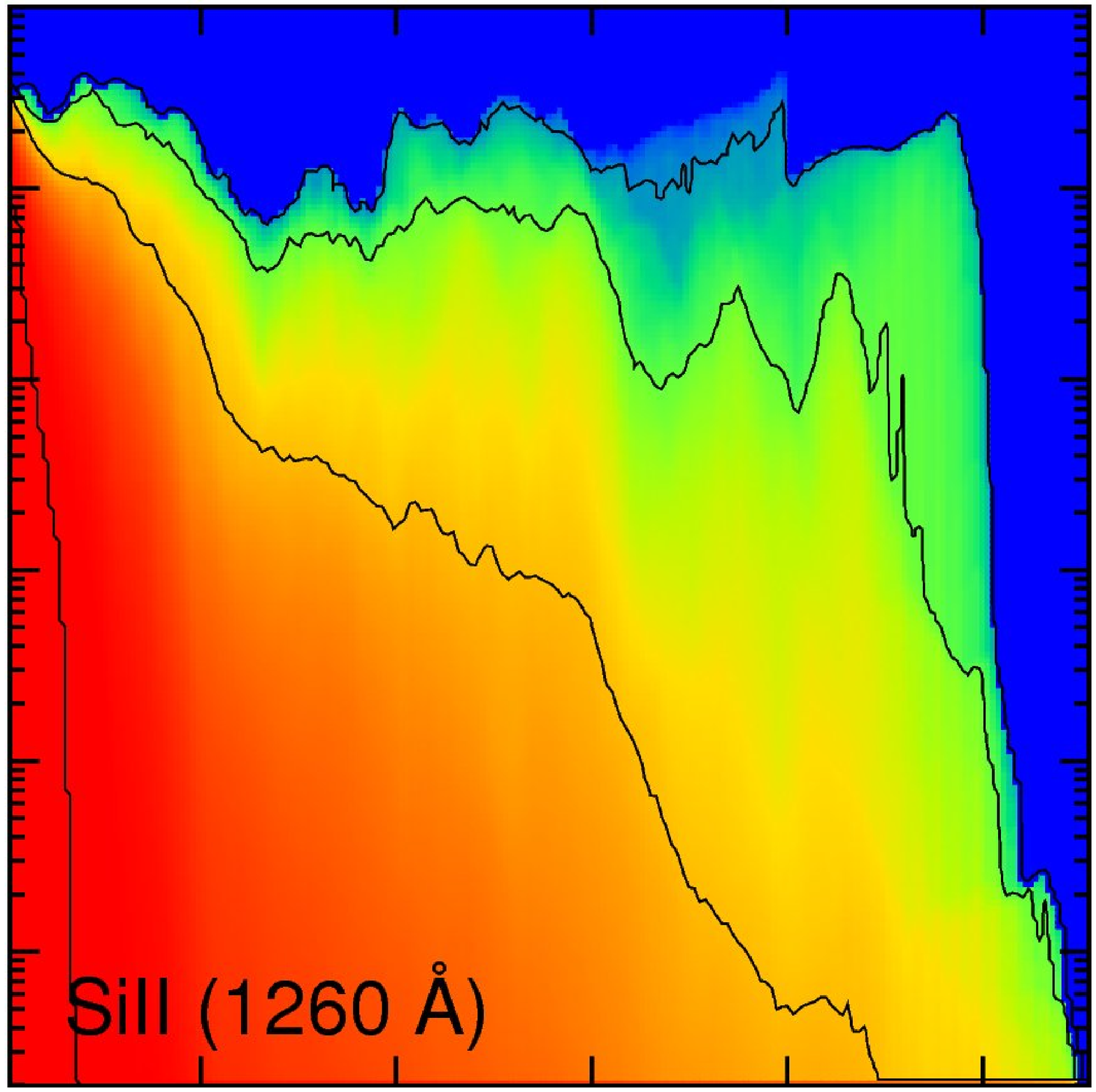}
\includegraphics[width=6.46cm]{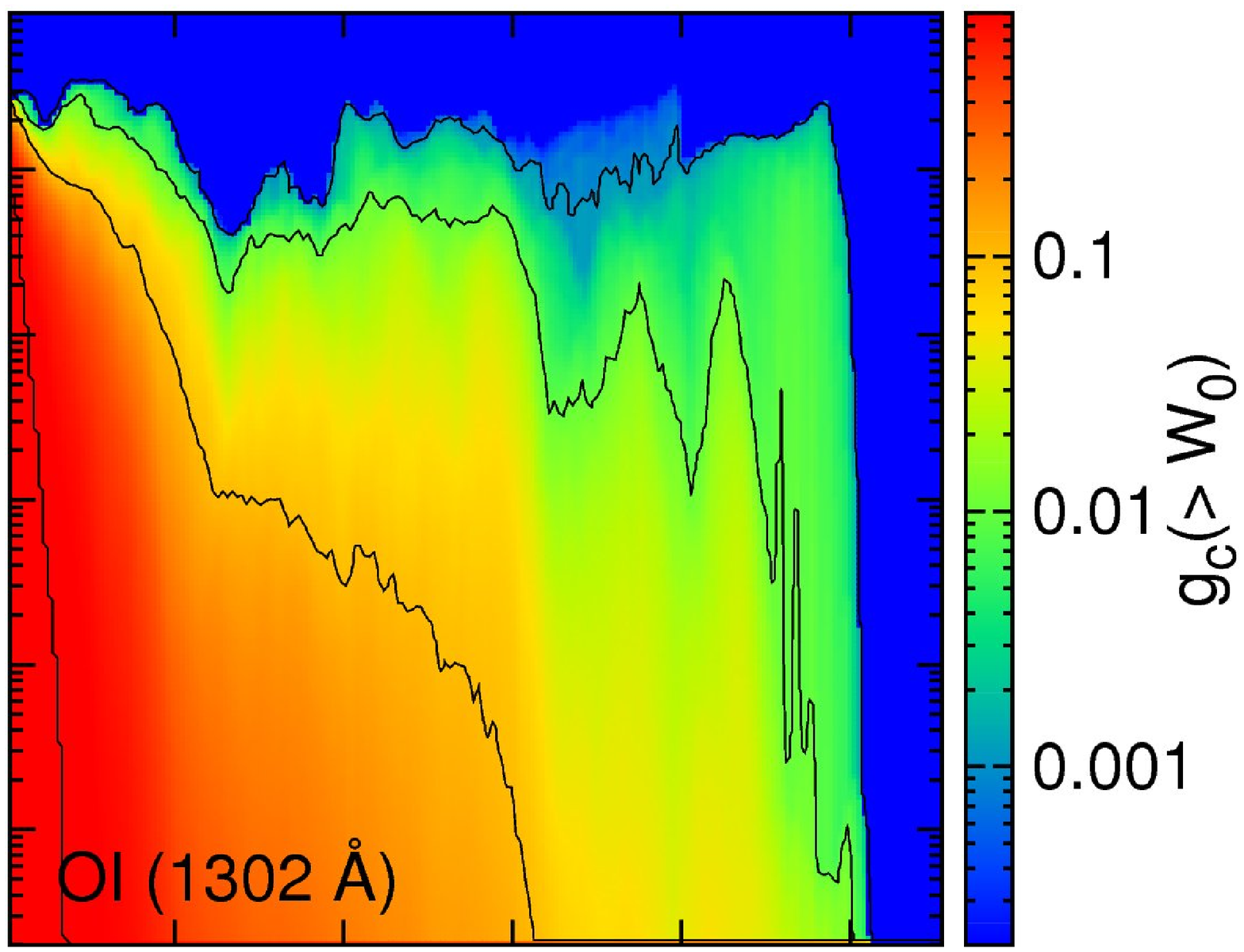}
\includegraphics[width=6.15cm]{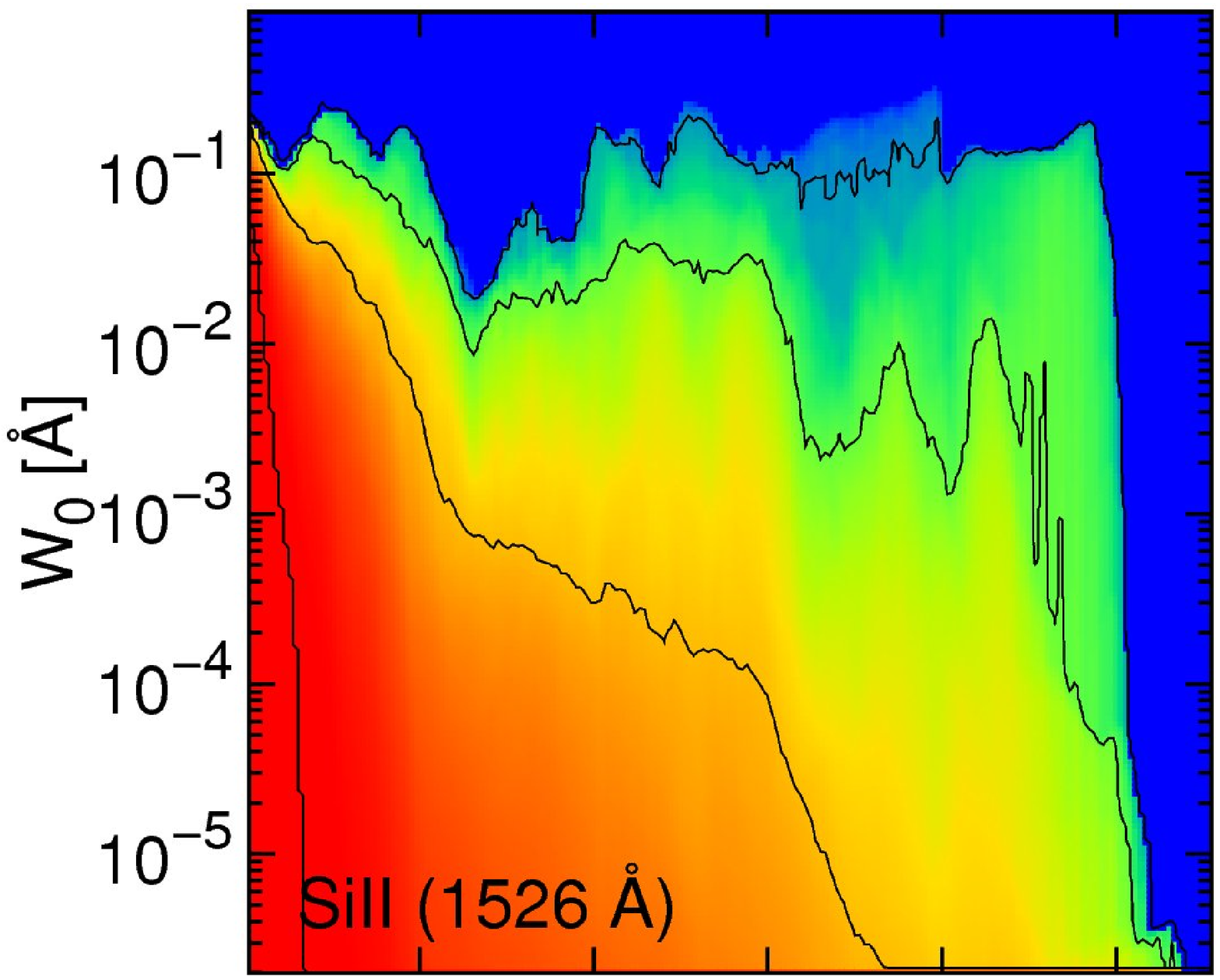}
\includegraphics[width=4.92cm]{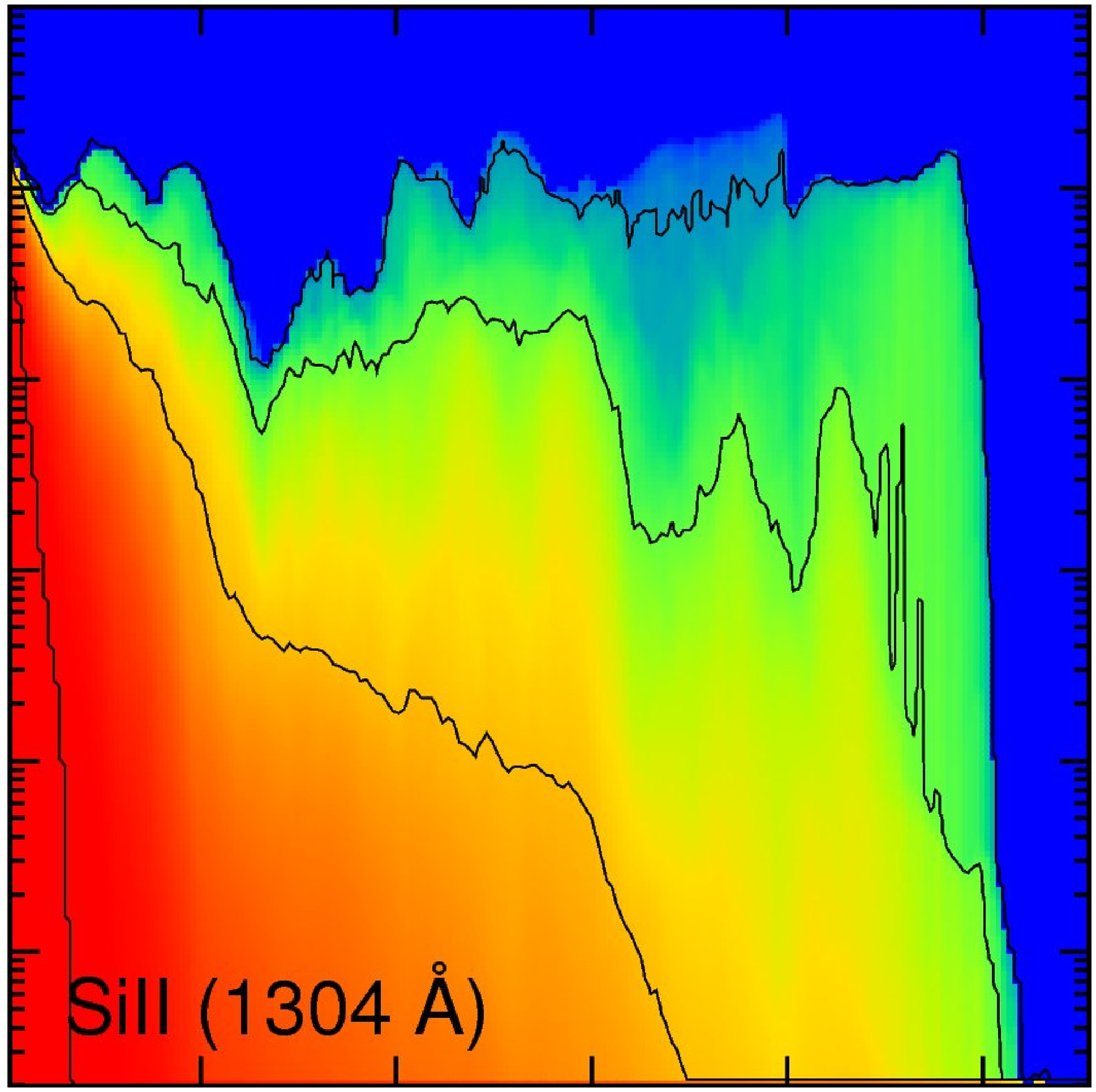}
\includegraphics[width=6.46cm]{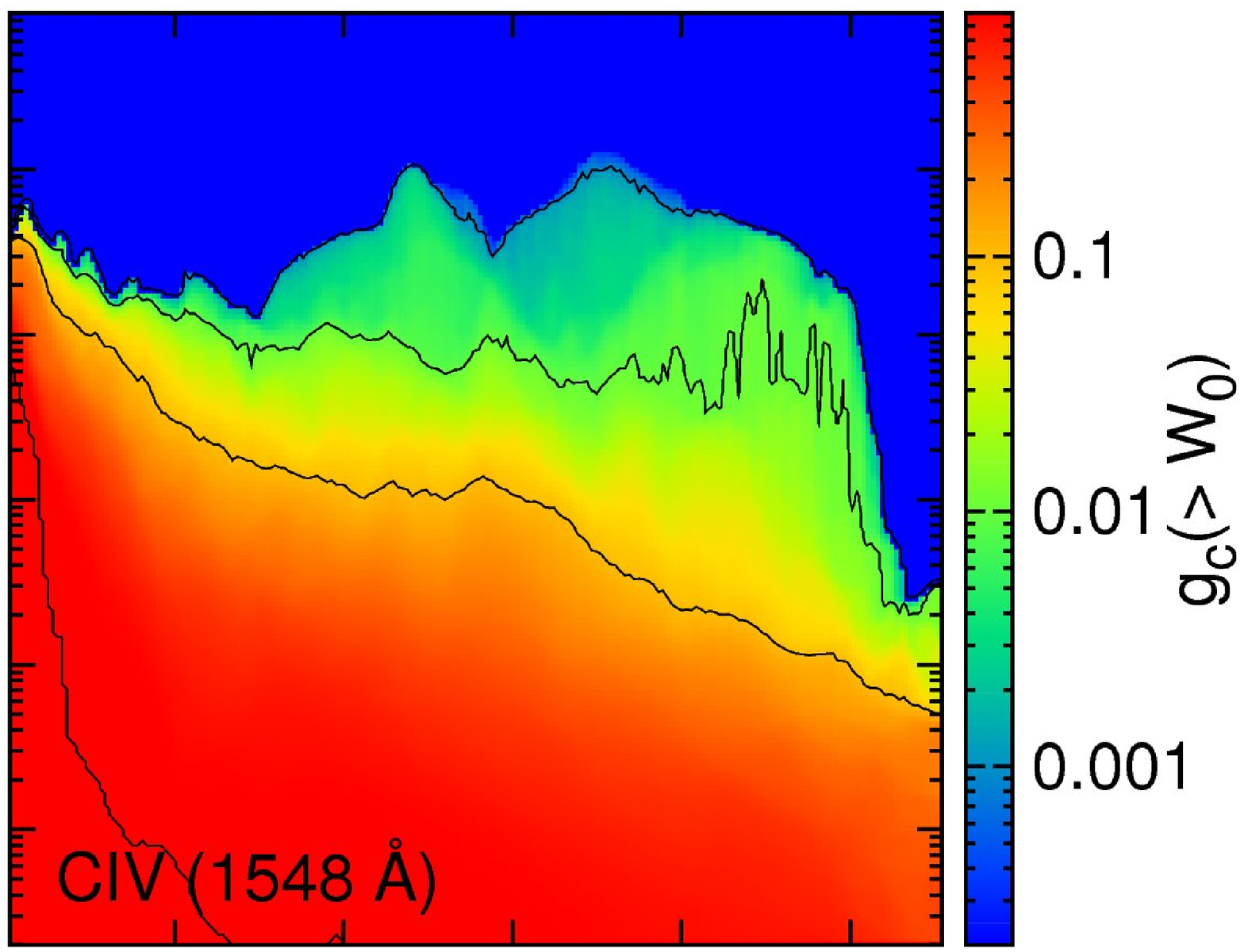}
\includegraphics[width=6.15cm]{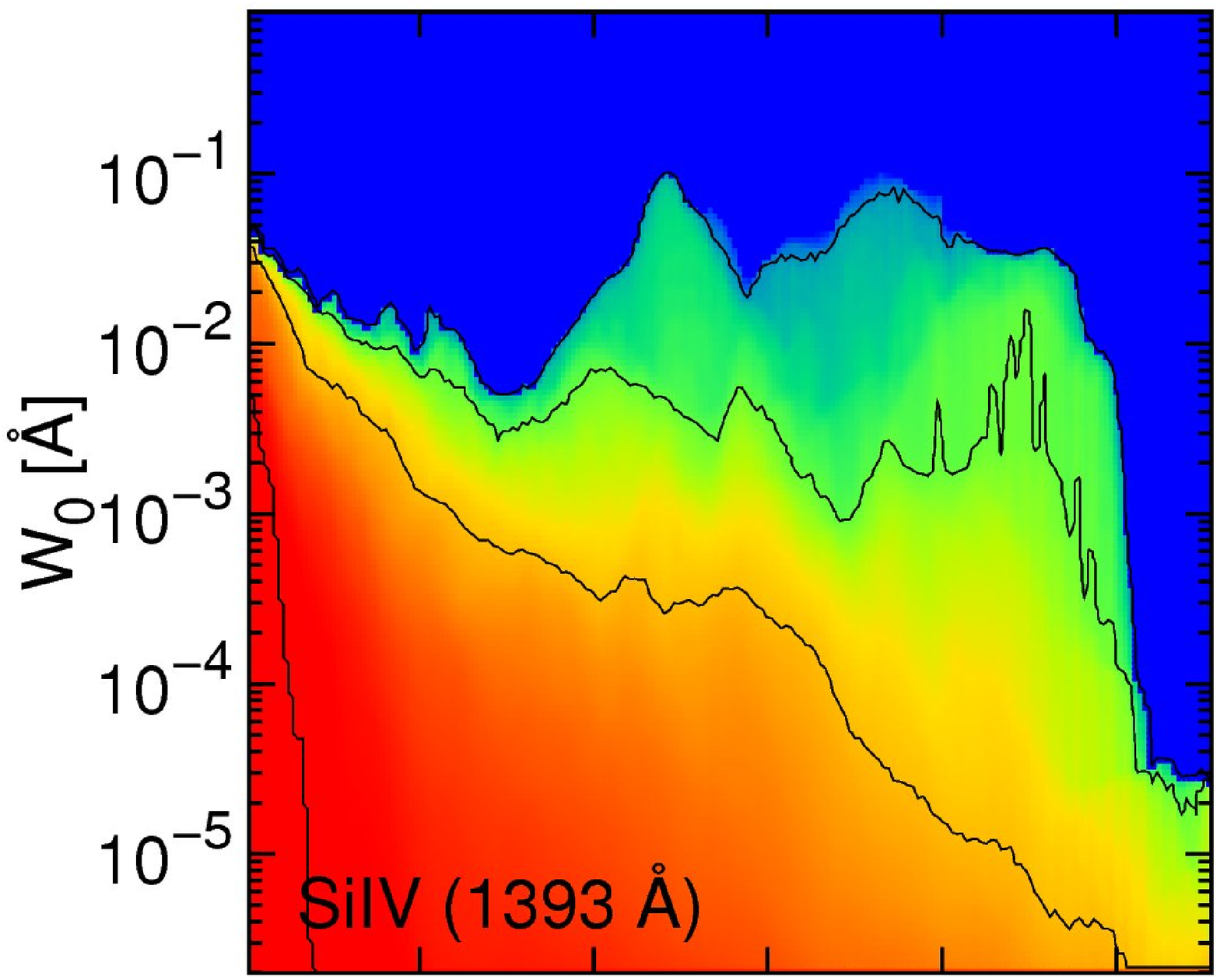}
\includegraphics[width=4.92cm]{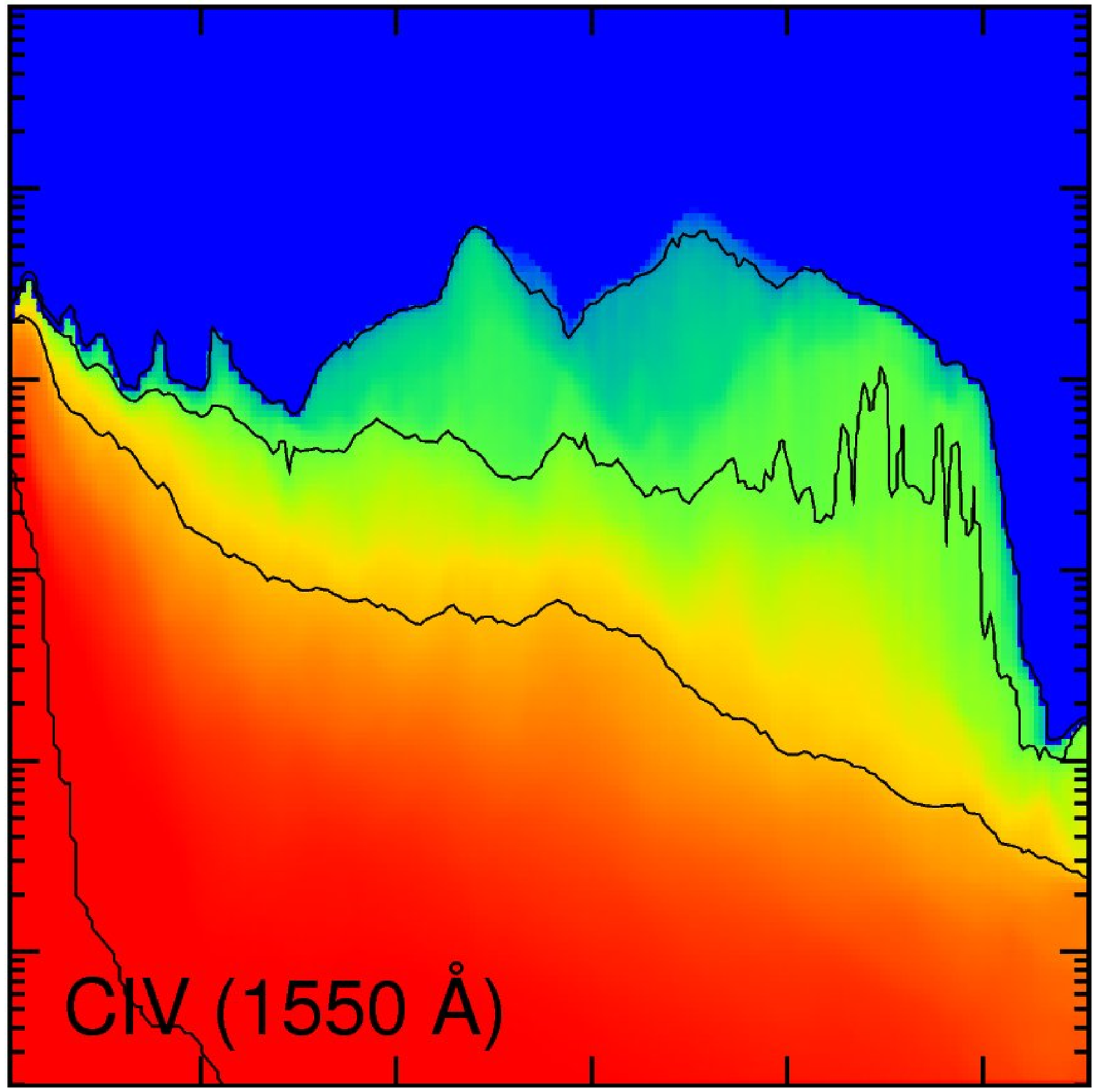}
\includegraphics[width=6.46cm]{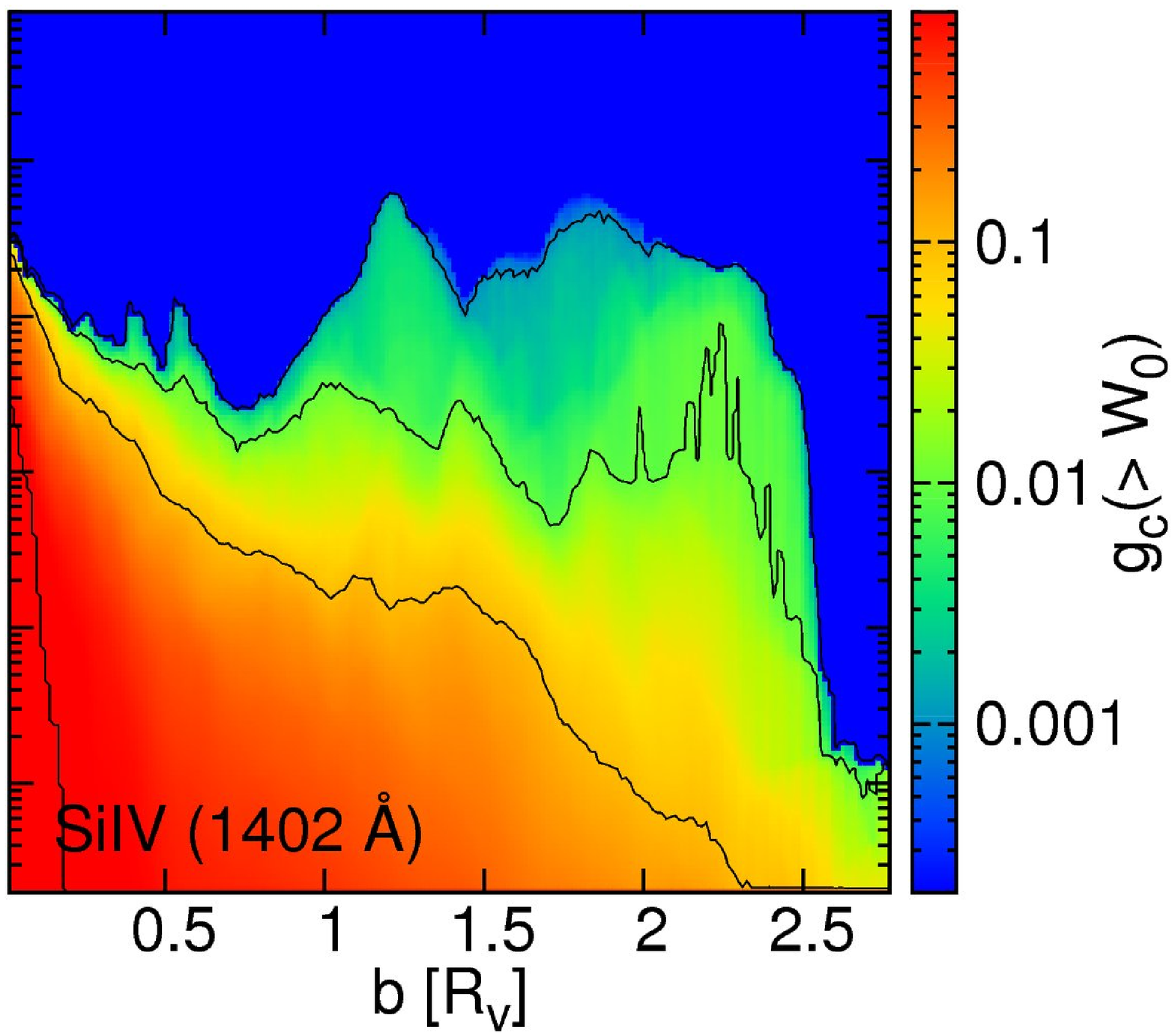}
\includegraphics[width=6.15cm]{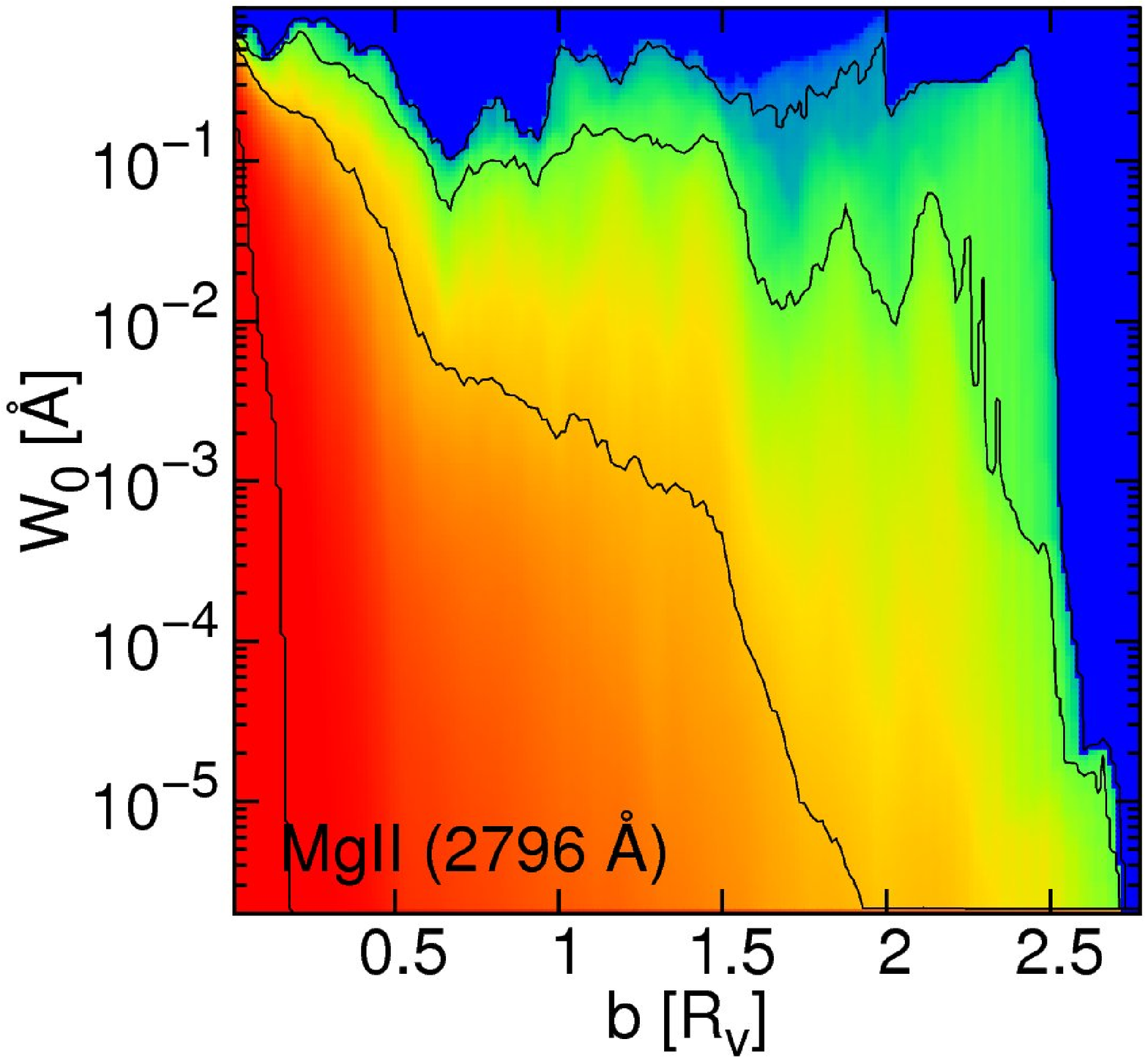}
\includegraphics[width=11.47cm]{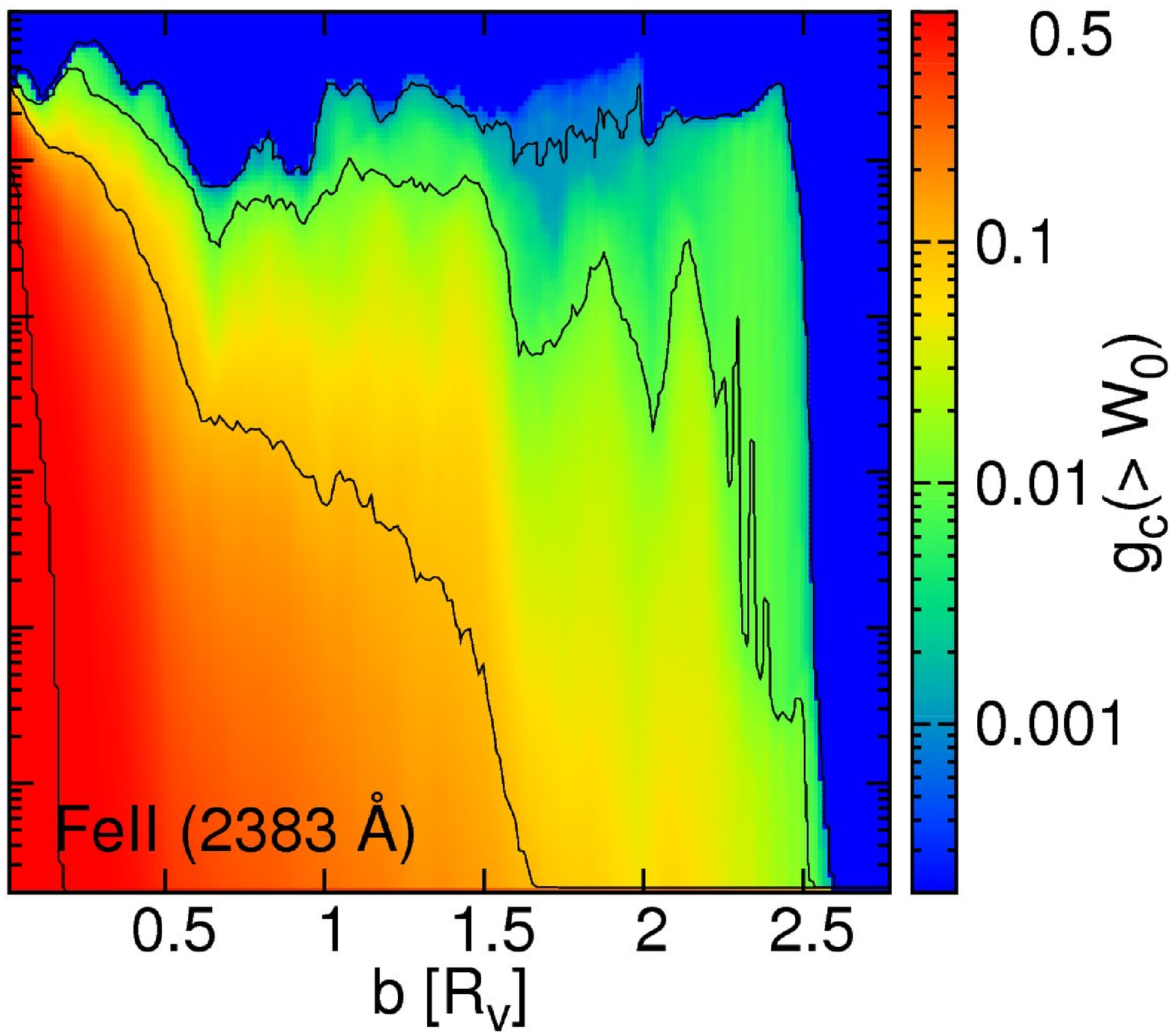}
\end{center}
\caption{Same area covering fraction $g_{\rm c}$ as in figure
\ref{fig:laareacovewbg}, this time for the metal line absorption lines. Peak
{\EW}s go as high as 0.1 {\AA} and the minimal values are well below 10$^{-6}$
{\AA}. In all cases the {\EW}s are decreasing mainly monotonic. The decrease
goes over five orders of magnitude and more. We see some noisy distortions in
the high $b$, low $g_{\rm c}$ regime. Over all the strongest lines are {\CII}
and {\MgII}.}
\label{fig:metalareacovewbg}
\end{figure*}

The probability to observe a given {\EW} along a given line of sight of impact
parameter $b$ is shown in figures \ref{fig:laareacovewbg} and
\ref{fig:metalareacovewbg}. Shown in colour is the area covering fraction
$g_{\rm c}$ as a function of impact parameter $b$ and cumulative {\EW} for
different absorption lines. The panels show for every impact parameter bin $b$
the fraction of the area within this impact parameter bin which has an {\EW} of
the indicated value $W_0$ or above. The contours mark fractions of 1.0, 0.1,
0.01 and 0.001. The values are averaged over three different galaxies and three
orthogonal projections, in both directions each. This figure is the analog of
figure \ref{fig:EWcovfrac} for a central source. We see that in {\la} we have
{\EW}s $> 0.01$ {\AA} everywhere. Peak {\EW}s go as high as 10 {\AA}. The 0.1
area covering fraction line is falling monotonically with $b$, whereas the 0.01
and 0.001 lines show noisy distortions at $b > 1.5 \ R_{\rm v}$. For the metal
lines the peak {\EW}s lie near 0.1 {\AA} and their minimum values are well
below 10$^{-6}$ {\AA}. In all cases the {\EW}s are declining mainly monotonic
with radius. For {\la} the decline is rather gradual, only by an order of
magnitude over the studied radius range. For the metal lines the decline is by
more than five orders of magnitude. We see some noisy distortions in the high
$b$, low $g_{\rm c}$ regime. Over all the strongest lines are {\CII} and
{\MgII}.

\begin{figure}
\begin{center}
\psfrag{relative intensity}[B][B][1][0] {relative intensity}
\psfrag{vks}[B][B][1][0]   {$\Delta w$ [km s$^{-1}$]}
\psfrag{bg}[Bl][Bl][1][0]  {background source}
\psfrag{Lya}[Bl][Bl][1][0] {\la}
\psfrag{1st}[Br][Br][1][0] {\textcolor{blue}{$10 - 41$ kpc}}
\psfrag{2nd}[Br][Br][1][0] {\textcolor{green}{$41 - 82$ kpc}}
\psfrag{3rd}[Br][Br][1][0] {\textcolor{red}{$82 - 125$ kpc}}
\includegraphics[width=8.45cm]{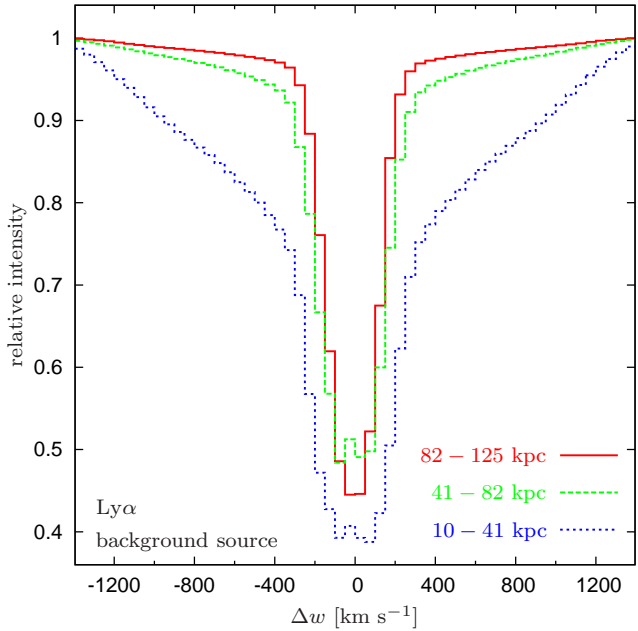}
\end{center}
\caption{Absorption line profiles for the background geometry for {\la}.
The data are averaged over three different impact parameter bins, three
different galaxies and six different viewing directions in the background
geometry. Note the different x- and y-scales in this figure in comparison to
Figure \ref{fig:metallinesbg}. This figure agrees with the results of
{\f11} (their figure 13). A qualitative comparison to the observation is
presented in table \ref{tab:bg}.}
\label{fig:lalinebg}
\end{figure}

Finally, we try to produce absorption line profiles that could be compared to 
those presented in figures 6 and 10 of \s10. They stacked spectra of several
galaxies in order to increase the signal to noise ratio. Their sample contains
512 close $(1'' -15'')$ angular pairs of $z\sim 2-3$ galaxies with redshift
differences that indicate no physical association. The sample explores the cool
gas as a function of galactocentric impact parameter in the range $3-125$ kpc
(physical) at $z = 2.2$. In figures \ref{fig:lalinebg} and
\ref{fig:metallinesbg} we show the resulting absorption line profiles for {\la}
and metal lines in our simulations, stacked over six principal different
viewing directions (three orthogonal axis in both directions). We show the line
profiles for the same three different impact parameter bins as in {\s10}: $10 -
41$ kpc, $41 - 82$ kpc and $82 - 125$ kpc. (Note the different scales in
figures \ref{fig:lalinebg}, \ref{fig:metallinesbg} and from row to row within
figure \ref{fig:metallinesbg}). The lines for $10 - 41$ kpc in those figures
correspond to figure 18 in {\s10}, the lines for $41 - 82$ kpc correspond to
figure 19 in {\s10}, and the lines for $82 - 125$ kpc correspond to figure 20
in {\s10}. For {\la} we have a FWHM $(= \eta)$ of up to $\sim 550$ km s$^{-1}$
and a relative intensity of 40-50\% at the line centre of the innermost radius
bin. It is not easy to compare our {\la} absorption line profiles to the
observations of {\s10} as they actually show emission line profiles. One could
still draw interesting conclusions by comparing the underlying optical depths
$\tau$ of the {\s10} observations with our simulations. However, due to the
special transfer behaviour of {\la} mentioned at the end of section
\ref{sec:cg} the {\la} emission is not only an indicator of cold streams.
Therefore the results have to be taken with caution. Our strongest metal
absorption lines, {\SiII} and {\CII}, have both a maximum line depth of $\sim
0.015$ with $\eta \sim 350$ km s$^{-1}$ for the $10 - 41$ kpc bin. This appears
tiny compared to the corresponding lines presented in figures 18 or 19 of
{\s10} having a line depth of 0.2 and $\eta \sim 1000$ km\,s$^{-1}$. Detailed
comparisons between our values and the observed values by {\s10} are given in
Table \ref{tab:bg}.

\begin{table}
\begin{center}
\setlength{\arrayrulewidth}{0.5mm}
\begin{tabular}{lcllcl}
\hline
 & & {\s10}'s & our & our & our \\
ion & $\lambda_0$ & $W_0$ & MLD & $\eta$ & $W_0$ \\
& [\AA] & [\AA] & &[km s$^{-1}$] & [\AA] \\
\hline
{\la}   & (1216) & 4.7  & 0.61    & 550 & 2.3 \\
{\CII}  & (1334) & 2.61 & 0.016   & 350 & 0.025 \\
{\SiII} & (1260) & 2.01 & 0.012   & 400 & 0.019 \\
{\OI}   & (1302) &      & 0.0080  & 400 & 0.013 \\
{\SiII} & (1526) & 1.96 & 0.0038  & 400 & 0.0072 \\
{\SiII} & (1304) &      & 0.0030  & 400 & 0.0047 \\
{\CIV}  & (1548) & 3.90 & 0.0013  & 350 & 0.0023 \\
{\SiIV} & (1393) & 2.04 & 0.00081 & 350 & 0.0013 \\
{\CIV}  & (1550) & 3.90 & 0.00068 & 350 & 0.0012 \\
{\SiIV} & (1402) &      & 0.00041 & 350 & 0.00064 \\
{\MgII} & (2796) &      & 0.0094  & 400 & 0.033 \\
{\FeII} & (2383) &      & 0.0062  & 400 & 0.019 \\
\hline
\end{tabular}
\end{center}
\caption{Comparison of maximum line depths (MLD), FWHM $\eta$ and {\EW} $W_0$
in the background geometry between {\s10}'s observations (their Table 4, second
line) and our predictions from the simulations (the ``$10 - 41$ kpc'' line in
our figures \ref{fig:lalinebg} and \ref{fig:metallinesbg}).}
\label{tab:bg}
\end{table}

Figure \ref{fig:lalinebg} as well as the upper middle panel of figure
\ref{fig:metallinesbg} can be compared to figure 13 of {\f11}. For {\la}, in
the $10 - 41$ kpc bin, the maximum line depth is 0.61 and 0.55 respectively,
and the FWHM $\eta$ is 550 and 400  km s$^{-1}$ respectively. For {\SiII} (1260
\AA), in the same $10 - 41$ kpc bin, the maximum depth is 0.013 and 0.05
respectively, and $\eta \sim 350$ km s$^{-1}$ in both. These results are
similar. The upper left panel of figure \ref{fig:metallinesbg} can be
compared to the right panel of figure 3 in \citet{kimm}, where for {\CII} the
maximum line depth is $\sim 0.65$ with a FWHM $\eta$ of 550 km s$^{-1}$. We
identified three possible reasons for these differences: First \citet{kimm}
used a simulation\footnote{the Horizon MareNostrum simulation \citep{ocvirk}
with a maximum resolution of 1 kpc} having a much lower resolution than we do.
Second, the haloes they analysed are in a very different (much higher) mass
range ($\Mv > 10^{12}$ M$_\odot$) than the haloes we use. Third and most
importantly they use a different prescription for the Gaussian velocity
distribution as described earlier. We conclude that stacking of the absorption
lines from a background source tends to wash out the cold filament absorption
signal (compare figure \ref{fig:backviewang} with figures \ref{fig:lalinebg} and
\ref{fig:metallinesbg}). Unlike the case of a central source, in the case of 
a background source the absorption by inflowing gas is not well separated 
from the absorption by outflowing gas.

\section{Conclusions}
\label{sec:con}
Theory, including hydrodynamical cosmological simulations, tells us that
massive galaxies at high redshifts were fed by cold gas streams, inflowing into
dark-matter haloes at high rates along the filaments of the cosmic web
\citep{keresa, db06, nature, danovich}, but observing this process is not
straightforward. Here we addressed the expected absorption signature from these
cold streams. We ``observed" simulated galaxies for {\la} absorption as well as
metal lines from low and medium ionisation ions. We focused on the absorption
line profiles as observed by \citet{steidel}, using their set of ten different
lines and mimicking their way of stacking data from several galaxies, for
sources that are either in the background or at the centre of the absorbing
halo itself. The simulations used are zoom-in cosmological simulations with a
maximum resolution of 35-70 pc \citep{cd}. Self-shielding was accounted for by
a simple density criterion.

\begin{figure*}
\begin{center}
\psfrag{relative intensity}[B][B][1][0] {relative intensity}
\psfrag{vks}[B][B][1][0]   {$\Delta w$ [km s$^{-1}$]}
\psfrag{bg}[Bl][Bl][1][0] {background source}
\psfrag{CII}[Bl][Bl][1][0]   {{\CII} (1334 \AA)}
\psfrag{OI}[Bl][Bl][1][0]    {{\OI} (1302 \AA)}
\psfrag{SiIIa}[Bl][Bl][1][0] {{\SiII} (1526 \AA)}
\psfrag{SiIIb}[Bl][Bl][1][0] {{\SiII} (1260 \AA)}
\psfrag{SiIIc}[Bl][Bl][1][0] {{\SiII} (1304 \AA)}
\psfrag{CIVa}[Bl][Bl][1][0]  {{\CIV} (1548 \AA)}
\psfrag{SiIVa}[Bl][Bl][1][0] {{\SiIV} (1393 \AA)}
\psfrag{CIVb}[Bl][Bl][1][0]  {{\CIV} (1550 \AA)}
\psfrag{SiIVb}[Bl][Bl][1][0] {{\SiIV} (1402 \AA)}
\psfrag{MgIIb}[Bl][Bl][1][0] {{\MgII} (2796 \AA)}
\psfrag{FeIIb}[Bl][Bl][1][0] {{\FeII} (2383 \AA)}
\psfrag{1st}[Br][Br][1][0] {\textcolor{blue}{$10 - 41$ kpc}}
\psfrag{2nd}[Br][Br][1][0] {\textcolor{green}{$41 - 82$ kpc}}
\psfrag{3rd}[Br][Br][1][0] {\textcolor{red}{$82 - 125$ kpc}}
\includegraphics[width=6.15cm]{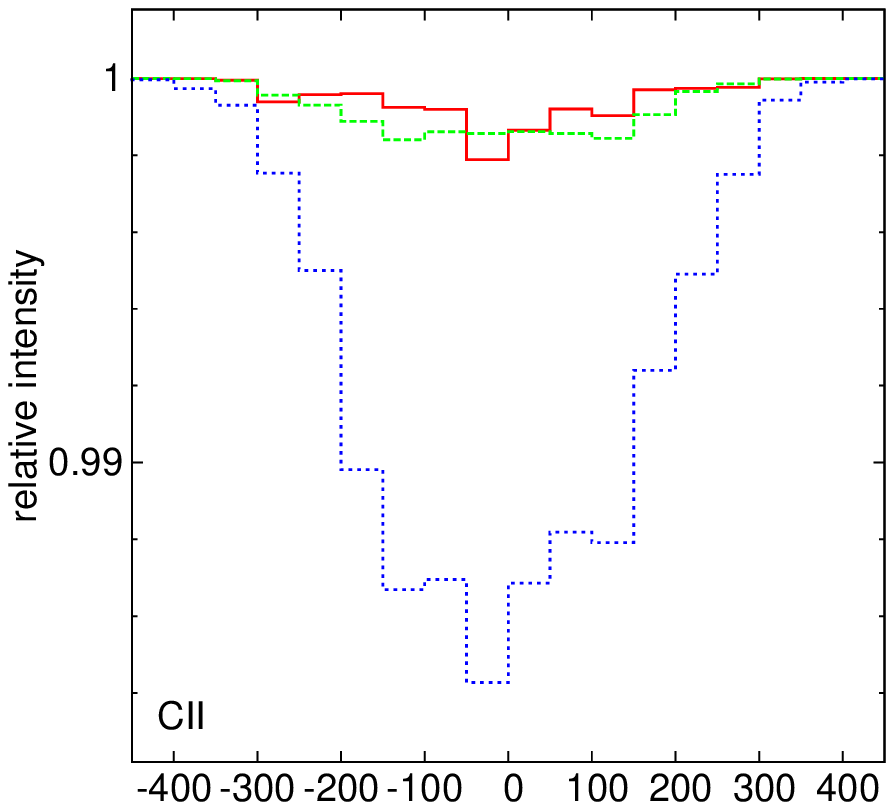}
\includegraphics[width=5.03cm]{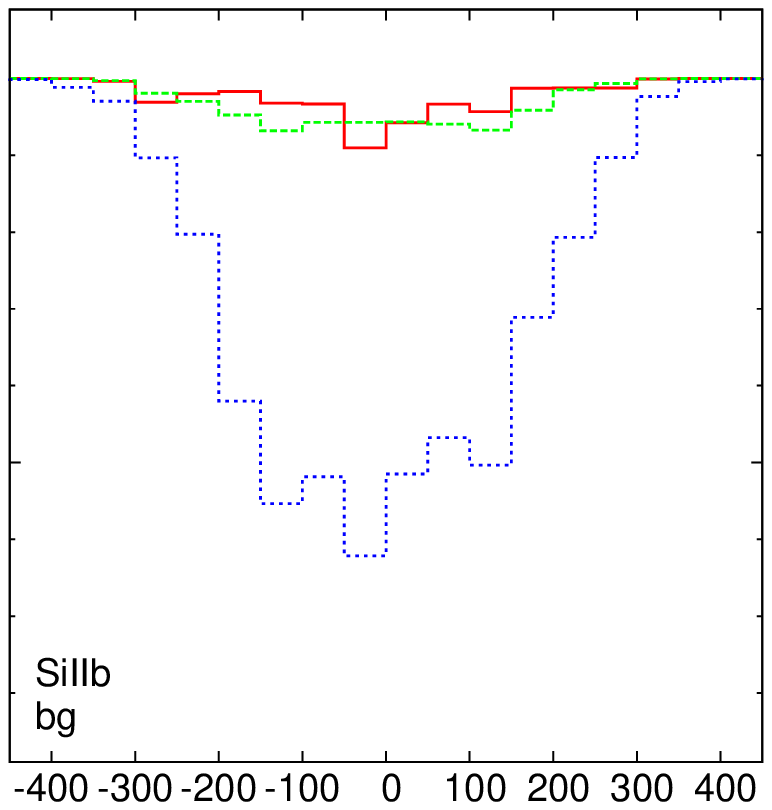}
\includegraphics[width=5.03cm]{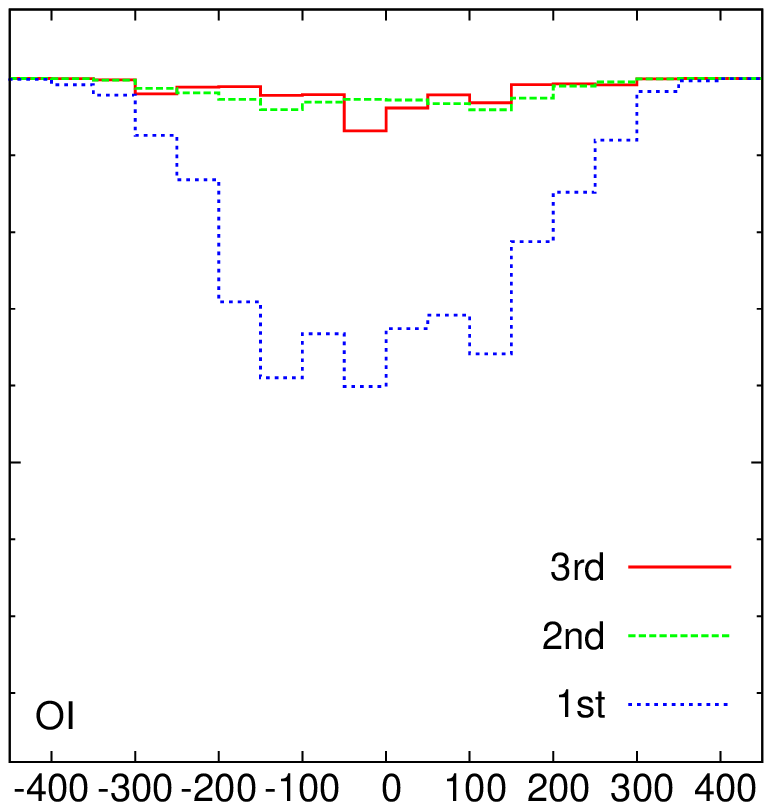}
\includegraphics[width=6.15cm]{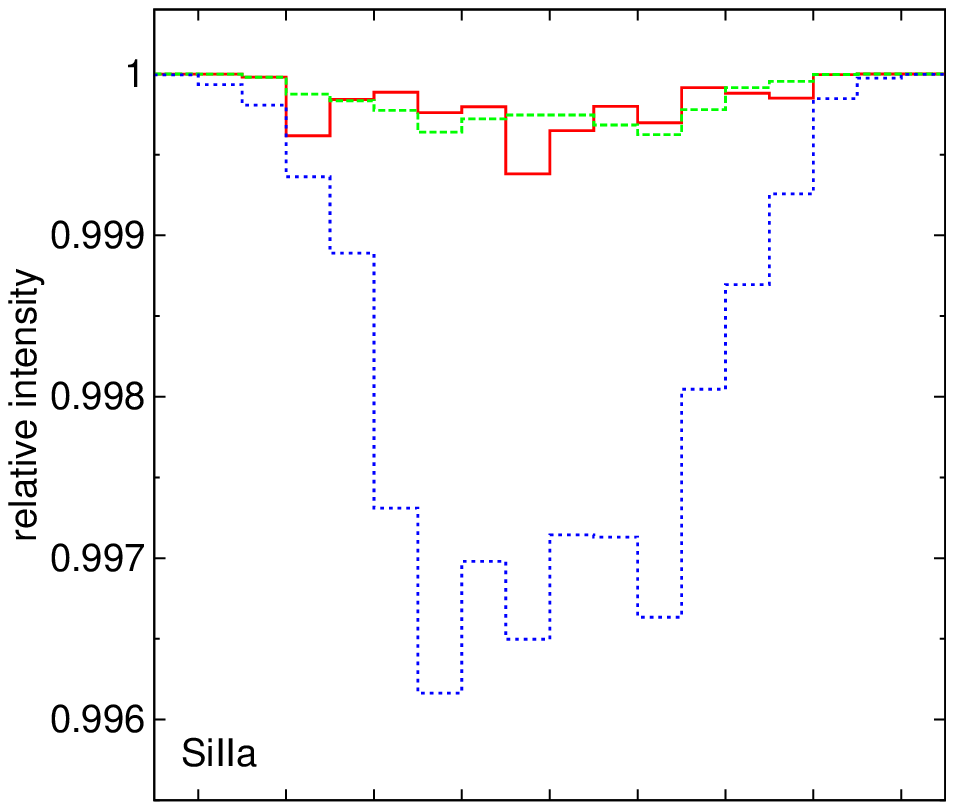}
\includegraphics[width=5.03cm]{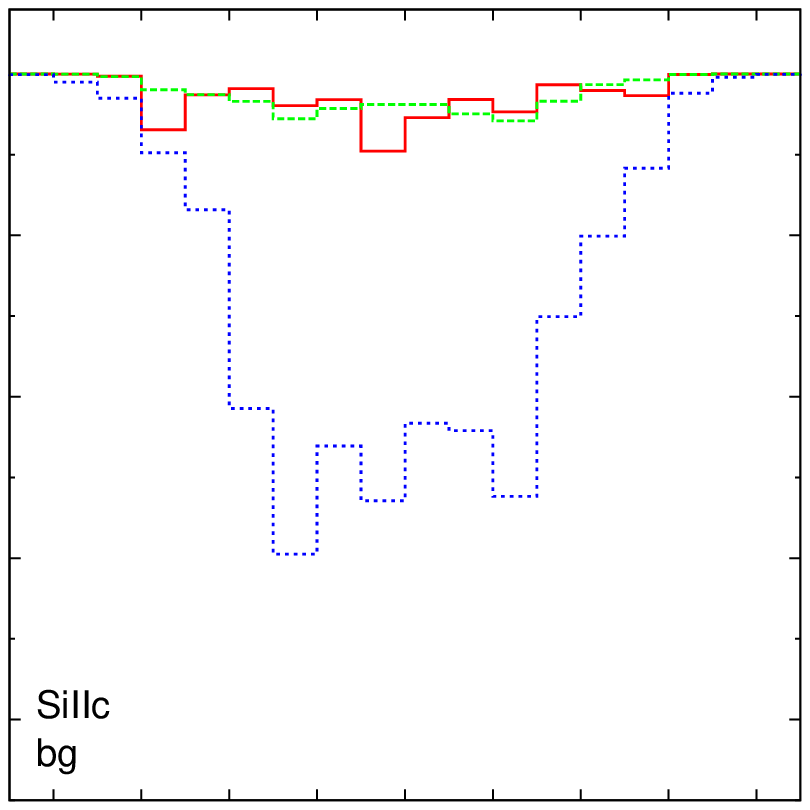}
\includegraphics[width=5.03cm]{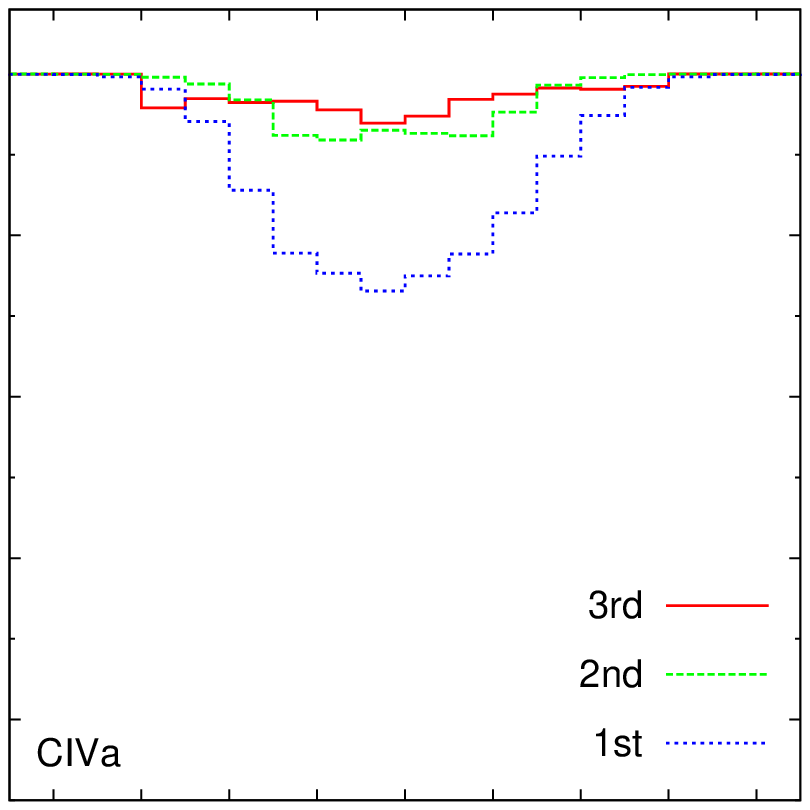}
\includegraphics[width=6.15cm]{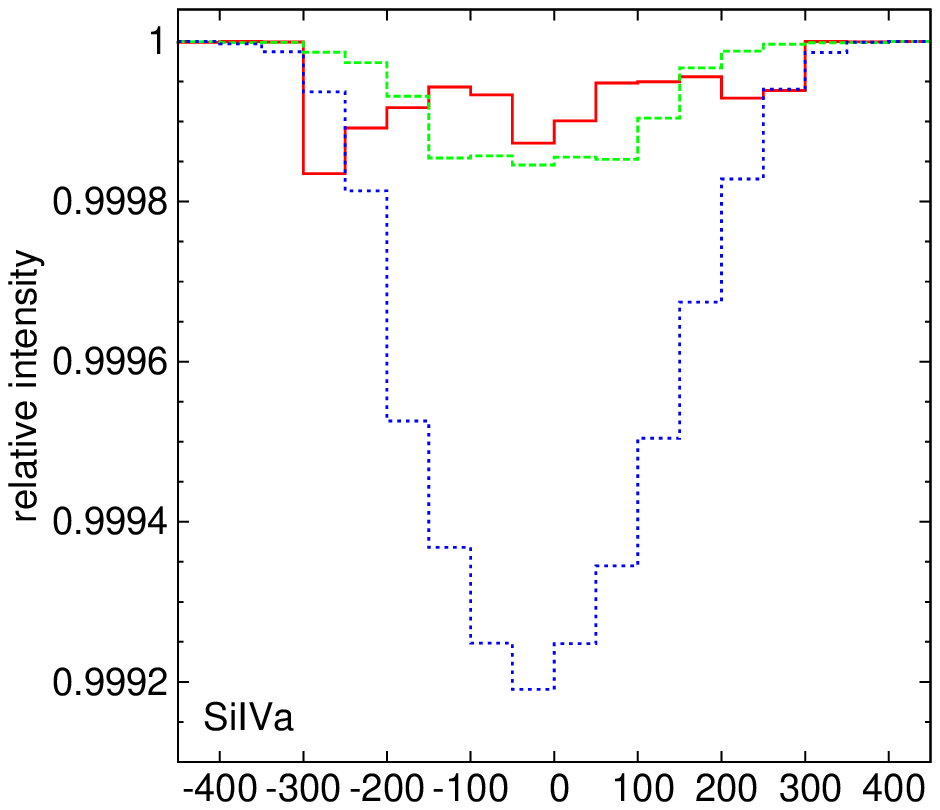}
\includegraphics[width=5.03cm]{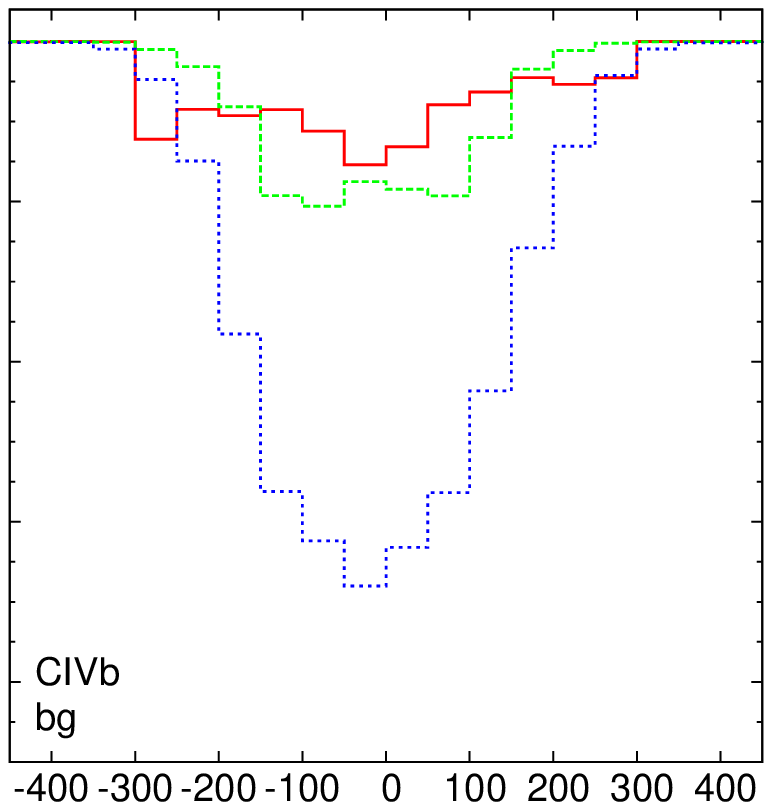}
\includegraphics[width=5.03cm]{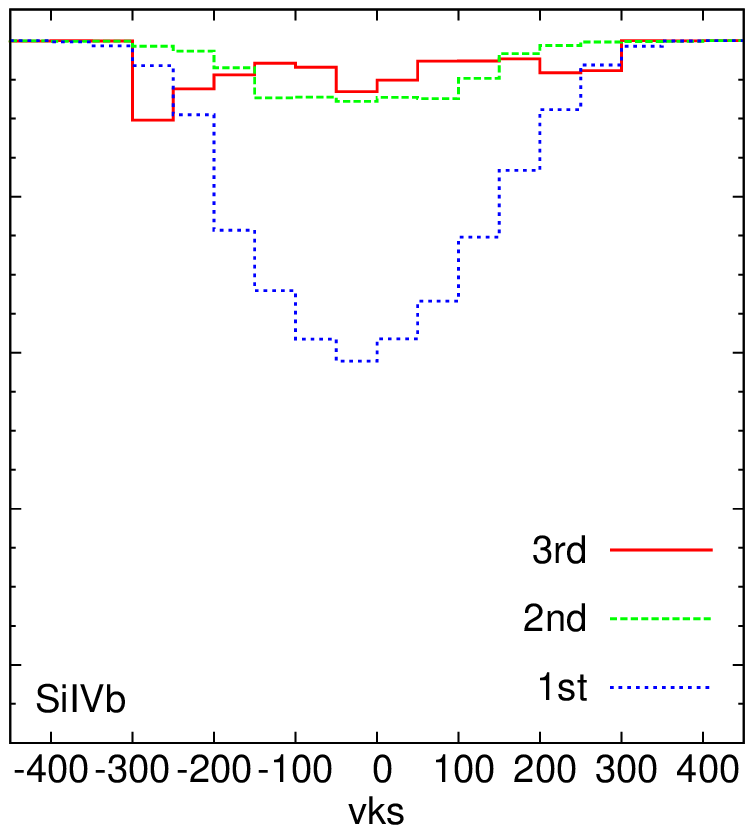}
\includegraphics[width=6.15cm]{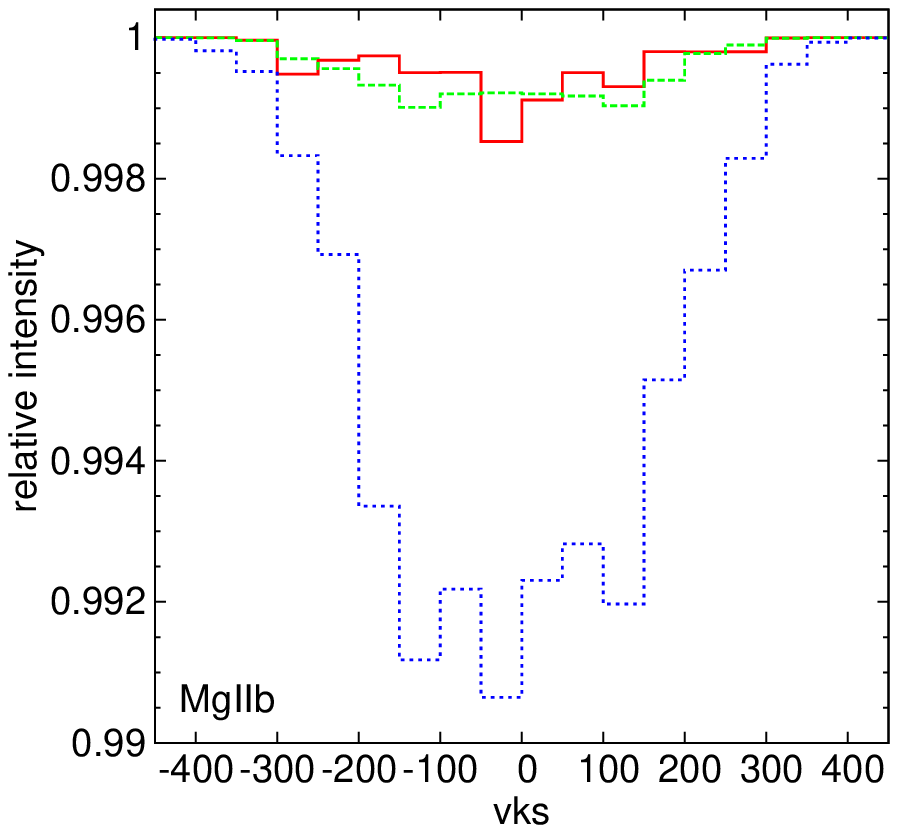}
\includegraphics[width=10.14cm]{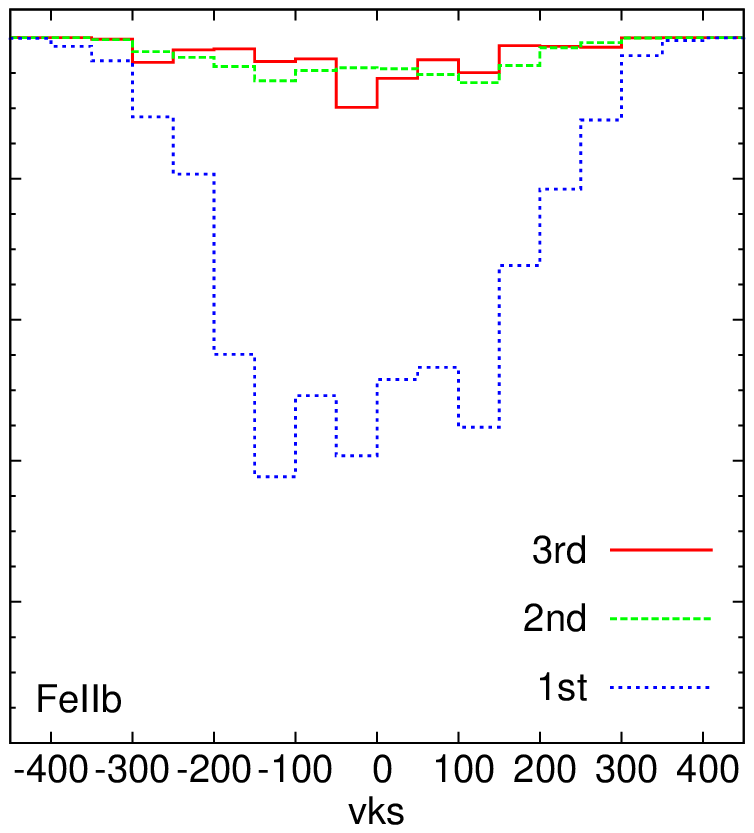}
\end{center}
\caption{Same absorption line profiles for the background geometry as in figure
\ref{fig:lalinebg}, this time for the metal lines. Note the changing y-scales
from row to row as well as the different x- and y-scales with respect to
figures \ref{fig:lalinebg}. The {\SiII} (1260\AA) panel of this plot agrees
with the results of {\f11} (their figure 13) the {\CII} panel of this plot
disagrees with the results of \citet[][their figure 3]{kimm} due to different
line computing algorithms. The lines with the deepest signals are {\CII} and
{\SiII} (1260 \AA). A more detailed qualitative comparison to the observation is
presented in table \ref{tab:bg}.}
\label{fig:metallinesbg}
\end{figure*}

We showed a variety of example absorption line profiles, as well as averaged 
absorption line profiles, both for a central source and a background source,
which are directly comparable to the observations of {\s10}. We computed the
resulting {\EW}s and their sky covering fractions for the two geometries. We
found in agreement with \citet{kimm} that when observations are averaged the
absorption signatures of the low-ionisation metals from the cold streams are
overwhelmed by the noise of other absorption processes in the host galaxy. This
has the following reasons: First, since the infalling gas has compared to the
ISM of the host galaxy so low densities and so low metallicities that
subsequently compared to other absorption processes in the host galaxy the
low-ionisation transitions of the cold streams also have tiny optical depths.
Second, the difference in redshift between the absorption signal of the cold
streams and that of the ISM is too small. The general absorption signature of
the ISM still overwhelms the one from the cold streams even in the very
redshifted part of the line. Last but not least, the flows have a radial,
narrow stream geometry, which makes them only very occasionally aligned with
the line of sight for a high optical depth. To show this we computed the
probabilities of detecting an inflow of a given speed with a given signal.

Similar to our work, {\f11} studied the absorption characteristics of the gas
in galaxies and streams, in order to compare with the statistics of observed
absorption-line systems. Like us they postprocessed high-resolution simulated
galaxies for determination of the ionisation states of the gas. To compute
those however we used the combined effects of electron-impact collisional
ionisation and photoionisation together with a distinction between UV-shielded
gas (where collision-ionisation equilibrium was assumed) and UV-unshielded gas
(where photo-ionisation equilibrium was assumed). {\f11} on the other hand used
a full radiative transfer approach. Since both analyses use the same data, the
high resolution {\CD} simulations, direct comparisons to {\f11} gives us a good
handle to estimate the strengths and weaknesses of our simplifying procedure.
We compared the distributions by volume of the neutral hydrogen fractions
$x_{\rm HI}$ and total hydrogen gas densities $n_{\rm H}$ in the circumgalactic
environment (our figure \ref{fig:danieltest}). We also compared the {\HI} area
covering fractions (our figure \ref{fig:areacovbg}) as well as the average
{\EW} as a function of impact parameter $b$ for {\la} and the {\SiII} (1260
\AA) lines (our figure \ref{fig:aEWvsradius}) both for the background geometry.
We finally compared the absorption line profiles for the background geometry in
{\la} and {\SiII} (1260 \AA) (our figures \ref{fig:lalinebg} and
\ref{fig:metallinesbg}). The differences from our results are small, justifying
the assumptions made in our simplifying model.

Both {\f11} and \citet{faucher2} analysed the sky covering fractions of neutral
hydrogen column densities, and came up with similar results to ours.
\citet{kimm} predicted {\CII} absorption line profiles similar to ours. Certain
differences emerge from the fact that \citet{kimm} deploy a Gaussian profile
instead of a Voigt profile. Also, while we used thermal Doppler broadening that
depends on the temperature of the cell, they used turbulent Doppler broadening
that depends on the velocity differences with respect to the 26 neighbouring
cells. {\f11}, like us, concluded that cold streams are unlikely to produce the
large {\EW}s of low-ion metal absorption around massive galaxies. If the average
equivalent widths reported by {\s10} are confirmed (e.g. with higher resolution
observations), more precise feedback models must be included in the
simulations, which are capable of causing winds with the right densities and
temperature range, since only such winds are able to produce these high
equivalent widths of low-ion metal absorption lines.

We mention three potential limitations of our analysis. A limitation of our
simulations may arise from the artificial pressure floor imposed in order to
properly resolve the Jeans mass. This may have an effect on the temperature and
density of the very dense and cold parts of the streams, with potential
implications on the computed absorption. Still, the AMR code is the best
available tool for recovering the stream properties. With 35-70 pc resolution,
and with proper cooling below $10^4$K, these simulations provide the most
reliable description of the cold streams so far.

{\s10} and others observed strong signatures of metals at distances as large as
200 kpc from the galaxy centre, that are interpreted as massive outflows. Such
massive and cold outflows are not reproduced in our current simulations, which
incorporate supernova feedback, but not yet strong radiative feedback and
other processes that are capable of driving sufficiently massive outflows. Our
studies of absorption from the simulated inflows are valid only if one could
assume that the outflows do not drastically affect the inflowing streams. The
strong outflows produced in the simulations of \citet{shen12} indeed avoid the
incoming dense narrow streams, and find their way out in a wide solid angle
through the dilute medium between the streams. The same is true for the weaker
and hot outflows that are produced in our simulations, with a mass loading
factor of about a third. Therefore our analysis is reliable.

The ionisation states should ideally be coupled to the hydrodynamical
calculations using radiative transfer, but in our current analysis they are
computed in post-processing. The analysis distinguished between UV-shielded
gas, where collision-ionisation equilibrium was assumed, and UV-unshielded gas,
where photo-ionisation equilibrium was assumed. An additional source of local
ionising radiation is the Lyman continuum from newly born stars. This could be
a non-negligible source of ionisation, which is neglected in our calculations
so far.

We conclude that the signatures of cold inflows are subtle, and when stacked
are overwhelmed by the outflow signatures. Our predicted {\la} line absorption
profiles agree with the observations, while the stacked metal line absorption
from the inflows is much weaker than observed in the outflows. The
single-galaxy line profiles predicted here will serve to compare to
single-galaxy observations.

\section*{Acknowledgements}
The computer simulations were performed in the astro cluster at HU Jerusalem,
at the National Energy Research Scientific Computing Centre (NERSC), Lawrence
Berkeley National Laboratory, at NASA Advanced Supercomputing (NAS), at NASA
Ames Research Centre and at CCRT and TGCC under GENCI allocation 2011-GEN2192.
The analysis was performed on TGCC Curie under Prace project number ra0317. We
acknowledge stimulating discussions with Michele Fumagalli, Alexander Knebe,
Steffen Knollmann, Jonty Marshall, Crystal Martin and Chuck Steidel. We thank
the DFG for support via German-Israeli project cooperation grant
STE1869/1-1.GE625/15-1 and the Spanish {\it Ministerio de Ciencia e
Innovaci{\'o}n (MICINN)} for support via the project AYA 2009-13875-C03-02.
This work was also supported by ISF grant 6/08, by GIF grant G-1052-104.7/2009
and by NSF grant AST-1010033 at UCSC. Tobias Goerdt and Daniel Ceverino are
Juan de la Cierva fellows.

\bibliographystyle{mn2e}
\bibliography{absor23.bbl}

\label{lastpage}
\end{document}